\documentclass{bmcart}

\usepackage{amsthm,amsmath}

\usepackage[utf8]{inputenc}
\usepackage{xurl}
\usepackage{hyperref}
\usepackage{float}
\usepackage{lscape}
\usepackage{ulem}

\usepackage[]{graphicx}
\startlocaldefs

\endlocaldefs

\graphicspath{{fig/}{\main/fig/}}

\begin{document}

\begin{frontmatter}
\begin{fmbox}
\dochead{Research}
\title{Towards complete representation of 
bacterial contents in metagenomic samples}

\author[
addressref={aff1,aff2},
]{\inits{X.F.}\fnm{Xiaowen} \snm{Feng}}

\author[
addressref={aff1,aff2},
email={hli@ds.dfci.harvard.edu}
]{\inits{H.L.}\fnm{Heng} \snm{Li}}

\address[id=aff1]{
\orgdiv{Department of Data Sciences},
\orgname{Dana-Farber Cancer Institute},
\city{Boston},
\cny{US}
}
\address[id=aff2]{
\orgdiv{Department of Biomedical Informatics},
\orgname{Harvard Medical School},
\city{Boston},
\cny{US}
}

\end{fmbox}% comment this for two column layout

\begin{abstractbox}

\begin{abstract}  % max: 250 words, do no cite, minimum abbrev
\paragraph{Background:}
In the metagenome assembly of a microbiome community, 
we may think abundant species would be easier to assemble due to their 
deeper coverage. However, this conjucture is rarely tested. 
We often do not know how many abundant species we are missing and 
do not have an approach to recover these species.

\paragraph{Results:}
Here we proposed k-mer based and 16S RNA based methods to 
measure the completeness of metagenome assembly. 
We showed that even with PacBio High-Fidelity (HiFi) reads, 
abundant species are often not assembled as high strain diversity may 
lead to fragmented contigs. 
We developed a novel algorithm to recover abundant 
metagenome-assembled genomes (MAGs) by identifying circular assembly subgraphs. 
Our algorithm is reference-free and complement to standard metagenome binning. 
Evaluated on 14 real datasets, it rescued many abundant species that 
would be missing with existing methods.

\paragraph{Conclusions:}
Our work stresses the importance of metagenome completeness which is 
often overlooked before. Our algorithm generates more circular MAGs and 
moves a step closer to the complete representation of microbiome communities. 


\end{abstract}

\end{abstractbox}

\end{frontmatter}

\section*{Background}

{\it De novo} metagenome assembly has been an allusive promise of
unbiased and comprehensive snapshot of microbial communities of interest, 
independent to isolation and cultivation~\cite{tully2018reconstruction,howe2014tackling,kroeger2018new}.
Neither of the two aspects has been close to realization.
\underline{First}, most past metagenome sequencing projects were based on 
short read sequencing and produced short contigs of 
tens of kilobases (kb) long, which 
need to be clustered to form metagenome-assembled genomes (MAGs).
Most {\it de novo}genome-complete~\cite{Parks2015ua,bowers2017minimum} MAGs still 
contain an average of 87 assembly gaps with median length about 1.3kb.
We manually checked some and found that these gaps either 
have no presence in the BLAST nr/nt database using BLASTn,
or were homologous to shared genes such as ribosomal RNA (rRNA) operons. 
\underline{Second}, MAGs are rarely checked for their representation-completeness. 
Studies often assumed that sufficiently abundant or the most abundant species 
will be reconstructed~\cite{albertsen2013genome,vicedomini2021strainberry,nayfach2019new}.
There is no such guarantee despite efforts to approve it ~\cite{luo2012individual}.
A distinct species with low coverage can be easier to recover than
abundant but highly similar species or strains.
Horizontal gene transfer (HGT) or large duplication events 
are even more difficult to resolve.
One major obstacle for improving the situation was that 16S rRNA sequences,
which is a proxy of species definition~\cite{stackebrandt1994taxonomic,brumfield2020microbial,edgar2018updating}, 
were no easier to assemble than the whole genome~\cite{yuan2015reconstructing}
and will require amplicon sequencing~\cite{poretsky2014strengths,kai2019rapid,johnson2019evaluation}.
Therefore it is hard to  
cross validate between gene-based composition inference 
and assembly~\cite{nayfach2019new,wang2019metagenomic,almeida2019new}.

Pooling short read MAGs to form a complete reference catalogs~\cite{
kim2021human,almeida2019new,pasolli2019extensive}
is also flawed. 
The human gut microbiome is the most genome-sequenced metagenome context.
SRA has near 83k results assigned to this category as of July 2022.
However, recent large-scale studies still found 42\% of their 
quality MAGs
missing from major public repositories~\cite{almeida2019new}. 
For gut microbiome of other species, 
this number can be as high as 86\%~\cite{chen2021expanded}.
Moreover, 
taxonomic profiling of 1004 faecal samples from TwinsUK registry
found each species was observed in a median of 2.7\% samples, 
with 12\% of species being sample-specific
and 50\% species found in less than 1\% samples~\cite{visconti2019interplay},
demonstrating vast diversities.
To make the situation more complicated, 
there is no consensus on the optimal library size. 
Most sequencing projects prioritized sample size over high-coverage,
perhaps due to the observation that 
MAG yield per gigabases sequenced per sample peaks 
at the lower end of library sizes (Figure S1-2).
Some of the most deeply sequenced~\cite{liu2022towards,hillmann2018evaluating}
or unusual~\cite{SRwetland}
datasets were analyzed unassembled.
The endgame of the reference MAG accumulation process is unclear.
On the other hand,
the drawbacks of short read MAGs are still relevant even if MAGs are just 
treated as bags of genes~\cite{kanehisa2016blastkoala,frioux2020bag},
or used for composition inference~\cite{brumfield2020microbial,laudadio2018quantitative}.

Metagenome assembly using accurate long reads, such as Pacbio HiFi reads,
can now recover haplotype-resolved near-complete closed MAGs
and separate species with sequence divergence 
as low as 1\%~\cite{Kolmogorov2020-gu,Nurk2020-zh,feng2022metagenome}.
This somewhat resolves the first issue of short read MAGs, 
although the quality evaluation remains reference-dependent.
Binning re-emerged to be problemetic.
Traditional binners tend to collected contigs from multiple haplotypes, 
despite they are well capable of creating high quality short read MAGs 
containing tens of hundreds of short contigs.
In HiFi assemblies, only when the binner pulls less than ten contigs would a bin 
likely to pass quality checks, basically relying on the recruitment of
a few long contigs and nothing more.
The second issue about representation-completeness is an open question. 
However, with circular contigs no longer relying on binning
and full length 16S rRNA sequences readily available from the HiFi reads, 
we think it is possible to approach it from both k-mer specturm,
leanring from evaluations in single sample large genome assemblies, 
and 16S-based species-level operational taxonomic units (OTUs). 
It is no less important about learning what is absent
from the assemblies than examining what has been recovered
in order to properly compare across samples and 
different microbiotas, and further, to improve the assembly method.

\uline{This manuscript has three major components.}
We \uline{first} introduce a graph topology-based, reference-free postprocessing step 
for the hifiasm-meta~\cite{feng2022metagenome} assembler.
The core is a circle-finding depth-first search.
Although simple, this could replace about half of regularly binned MAGs
with pseudo circular paths of identical or better quality,
as well as adding in MAGs that are not recoverable by regular binners.
There is no external data needed,
and the end results were very close to 
the almost-best-possible MAG recovery (i.e. if assisted by checkM).
We \uline{then} propose two evaluations as mentioned above 
and apply them to the HiFi MAGs.
\uline{We demonstrate} that in the best cases, near-complete MAGs alone could offer 
proper sample representations comparable to that of 
a comprehensive reference catalog and delegate almost 90\% 
abundant OTUs (more than 30 16S copies, using 99\% boundary).

\section*{Results}
\subsection*{Topology-based cycle-finding on the contig graph as a binning method 
and its merging with traditional binning}

Established binners were developed for 
binning short read metagenome assemblies.
The process relies on tetranucleotide profiles and coverage 
estimates~\cite{kang2019metabat,nissen2021improved}, % metabat2 and vamb 
and sometimes consults assembly graph for 
further refinements~\cite{mallawaarachchi2020graphbin}.
Single-copy marker genes and other prior knowledge 
can also be 
exploited~\cite{wu2014maxbin,sieber2018recovery},  % maxbin and das tools
though such binning complicates the final evaluations, 
as assessing assemblies of empirical datasets heavily relies on 
checkM~\cite{Parks2015ua} or similar methods
that also utilize marker gene sets and phylogenetic placement.

In short read assemblies, the binners are capable of pooling 
hundreds of short contigs to form very high quality MAGs.
However, they do not perform as good in HiFi assemblies.
We found most near-complete or high-quality MAGs generated by 
binners contain less than 10 contigs. 
The completeness of these bins mostly come from only one or two 
long linear contig(s).
While this does not defeat the purpose of binning on HiFi assemblies, 
haplotypes that do not assemble well would remain missing.
Additionally, long contigs of hundreds or more kilobases 
from closely related haplotypes are 
prone to be clustered together, 
which results in complete but highly contaminated bins.
The first issue was especially hard to be improved.
The second can be somewhat mitigated by arbitrarily splitting bins
based on intra-bin sequence similarity.
We noticed that the assembly graph of hifiasm-meta 
was simple enough to provide binning hints when the assembly was 
visually~\cite{wick2015bandage} reasonably not bad (Figure 1).
We propose a simple topology-based assembly postprocessing 
in complement to binning to achieve a reasonable complete representation
of the library.
The method also has bonus contig ordering and circularization.
We implemented this approach in the hifiasm-meta assembler.
The other two capable assemblers, metaFlye and HiCanu, produces more tangled
graphs or does not produce a graph, respectively.

%%%%%%figure 1
\begin{figure}[bp!]
\includegraphics[width=0.95\linewidth]{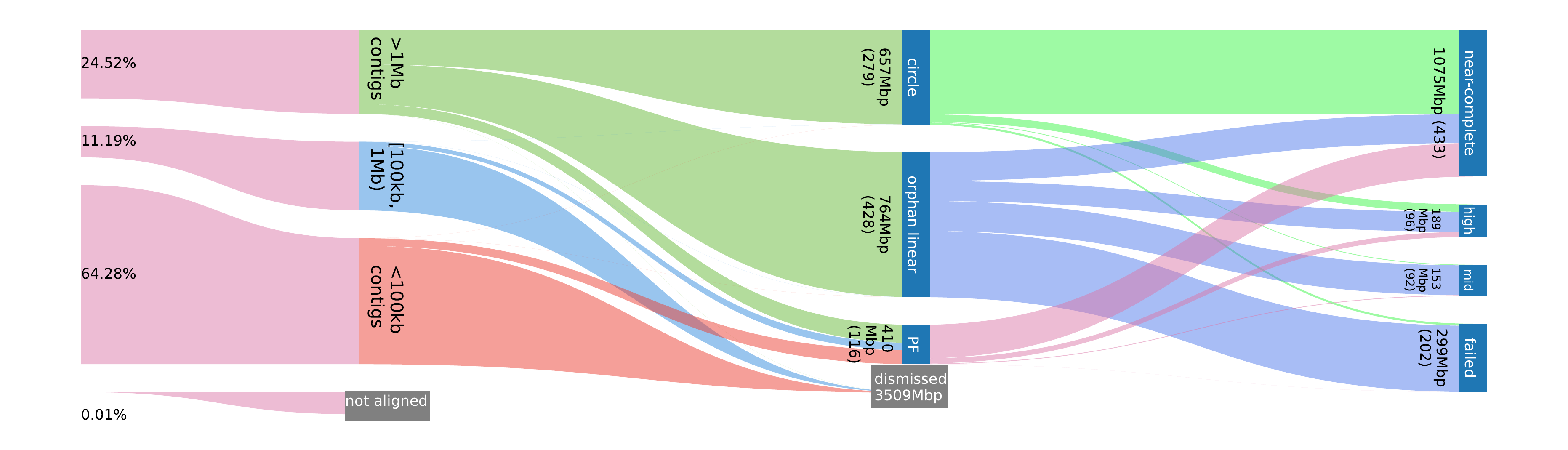}
\includegraphics[width=0.95\linewidth]{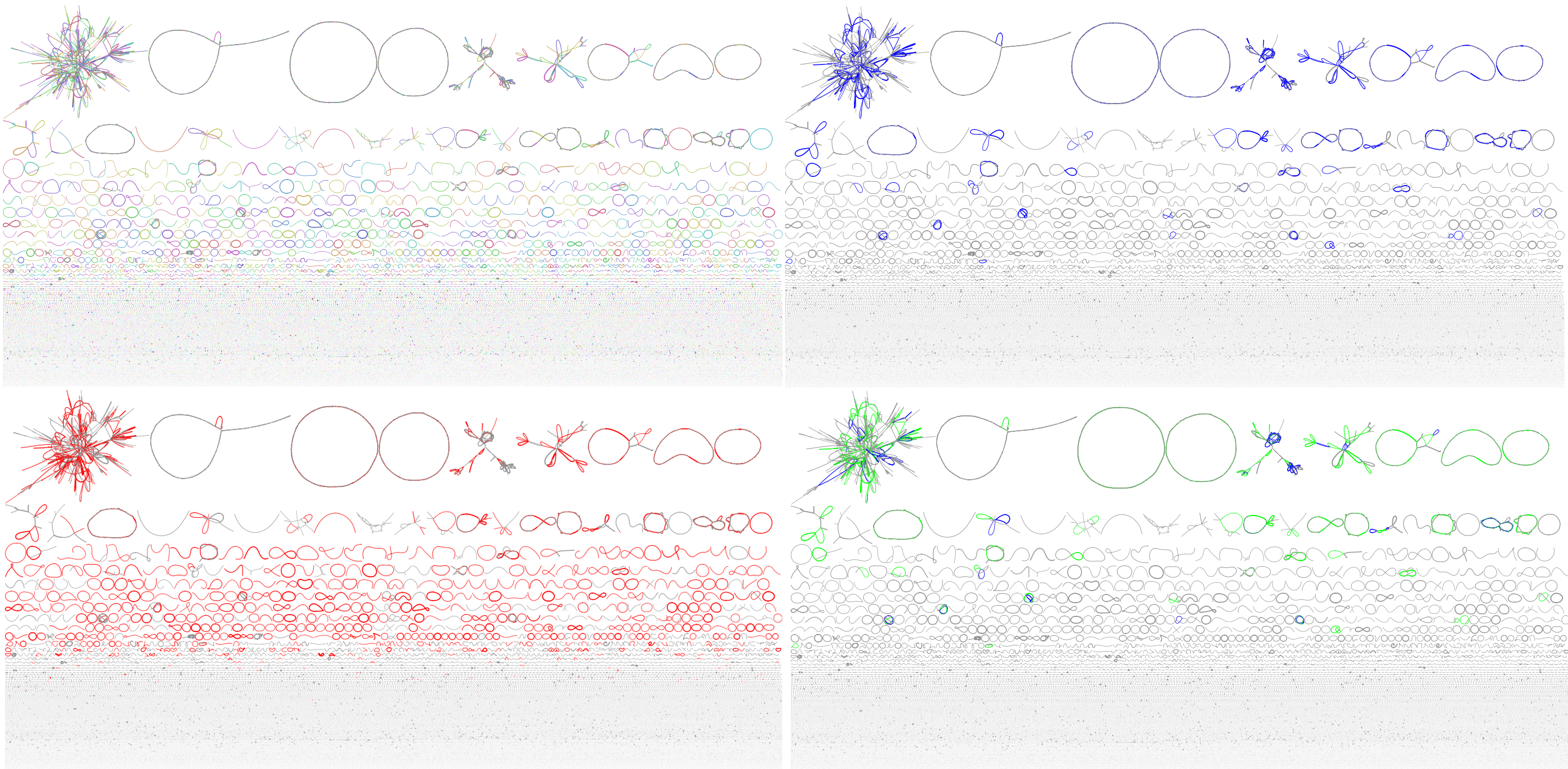}
\caption{(a) Sankey plot showing flow of reads and contigs of sheep-gut-1b. 
Left: reads to contig mappability (``{\tt minimap2 -cxmap-hifi}'').
Middle: contig binning categories.
Right: MAGs quality categories. 
Group heights normalized by counts, except the left-most side which is 
normalized by base pairs.
The heights of ``not aligned'' and the ``dismissed'' have no meaning.
(b) Bandage~\cite{wick2015bandage} plot of sheep-gut-1b primary assembly: 
(top-left) primary assembly graph, 
(top-right) circular path candidates to rescue marked in blue, 
(bottom-right) high-quality-or-above rescued circles marked in green, and 
(bottom-left) best possible high-quality-or-above MAGs marked in red.
Note that the size of circle-shaped subgraph 
that consists of numerous short contigs
was exaggerated by the graph layout method.}
\end{figure}

We use two assumptions:
a bacterial chromosome is expected to be circular, and its length 
to fall into the range of 1Mb-8Mb. 
We work on the primary assembly graph.
We first collect cycles of roughly 1Mb-10Mb long 
on the contig graph via depth-first search (DFS).
To balance redundancy and the risk of missing promising cycles,
the DFS always try to use a contig that has not been a node in any
of the collected cycles. 
The criteria are permissive and only terminates the search
if 100-200 nodes have been used. 
If a search extends too long, we halve the stack
and resume from halfway, in order to avoid
missing cycles due to the arbitrary ordering of contigs.

We then remove duplicated cycles using Mash distance~\cite{ondov2016mash}. 
Mash distance correlates well with ANI in the range of 90-100\% 
and we use bottom 1000 hashes with k-mer size of 21 
per recommendation in the Mash paper.
If the two cycles being compared differ in length for more than 1Mb, 
we do not drop any of them regardless of Mash distance.
We set the species boundary at 95\% whole-genome 
Average Nucleotide Identity (ANI),
which was derived from DNA-DNA hybridization values and later 
reconfirmed by large-scale 
studies~\cite{jain2018high,parks2020complete}. % species boundary; GTDB 
We also tried 97\% as a slightly relaxed threshold,
and 99\% which was the highest possible ANI 
for most pairs of long circular contigs
produced in HiFi assemblies (hifiasm-meta, Hicanu or metaFlye).
97\% threshold produced similar completeness evaluation results
to those of 95\%.
99\% threshold resulted in more duplications, although also slightly improved 
the overall assembly completeness.
Some contigs might be used more than once.
We see little such duplication in the final result 
(i.e. checkM-passed MAGs).
Note that the comparison is restricted within the collection of 
rescued circles candidates. A cycle that is very similar to a long circular contig
would not be discarded because of so.

We run MetaBAT2 on raw contigs independent of circle
rescue and then merge MetaBAT2 bins with rescued circles.
We collect all rescued circles (set1) and remember the contigs
used by them (setb).
Next, MetaBAT2 bins that are at least 500kb, no more than 10Mb
and do not only contain a single $>$1Mb circular contig were examined.
If a bin has more than 1Mb of its contents coming from setb, 
or more than 10 of its contigs coming from setb, 
the bin is rejected. Otherwise we accept it into set2 and 
update setb with its contigs.
We then collect linear contigs that are 
longer than 1Mb and not in setb into set3.
All circular contigs longer than 1Mb form set4.
The binning outcome (``merged MAGs'' or ``merged bins'')
is the union of set1, 2, 3 and 4.

\subsection*{MAG quality evaluations}

We first compared MAG quality brackets from rescued circles, 
MetaBAT2 bins and vamb bins (Table 1).
There are 16 HiFi libraries available, all except one are 
gut materials (Table S1, S4).
We follow the minimum information criteria about a single amplified genome (MISAG)
convention~\cite{bowers2017minimum} using checkM:
``Near-complete'' means $ \geq 90\% $ completeness and $<$5\% contamination.
``High-quality'' means $ \geq 70\% $ completeness, $<$10\% contamination but 
does not qualify for near-complete.
``Medium-qualify'' means $ \geq 50\% $ completeness, 
$ \geq 50 $ quality score but
does not qualify for the above two.
Quality score of a MAG is defined as 
``${\rm completeness}-5\times{\rm contamination}$''.
In terms of the yield in the near-complete quality bracket, 
the proposed topology-based rescue performed occasionally
comparable to vamb, but was consistently worse than MetaBAT2 
except for in env-digester-1.
This was because binners were able to pull together
long contigs that are not connected on the assembly graph 
to form a checkM-passing bin,
while the rescue heuristic relies on graph traversal.
Below we demonstrate the rescuing and the merge with regular MetaBAT2 bins
using sheep-gut-1b, a diverse sheep fecal material sequenced 
with extraordinary depth and diversity~\cite{bickhart2022generating};
see Table 2 for results of all samples available
and Table S2 for binning details.
We tried the heurstices on rust-mdbg~\cite{ekim2021minimizer}
assembly graphs. The rescued circles 
can recover up to 1/3$\sim$1/4 of HiFi MAGs in the corresponding samples,
although with more indels.

\begin{table}[bp!]
  \centering
  \caption{CheckM evaluation of rescued circles, MetaBAT2 bins and vamb bins. 
\*Vamb requires at least 4096 contigs to run.}
    \begin{tabular}{ccccccccccc}
      \hline
      sample & binning & near-complete & high-quality & near-complete & high-quality\\ 
      & method & bins & bins & orphaned $>$1Mb & orphaned $>$1Mb\\
      &  &  &  & linear contig & linear contig\\ \hline
      sheep-gut-1a & rescue & 45 & 5 & 18 & 17\\
          & metabat2 & 51 & 35 & 8 & 6\\
          & vamb & 31 & 34 & 22 & 15\\
      sheep-gut-1b & rescue & 99 & 15 & 85 & 59\\
          & metabat2 & 128 & 103 & 20 & 21\\
          & vamb & 64 & 71 & 79 & 62\\
      chicken-gut-1 & rescue & 10 & 0 & 8 & 7\\
          & metabat2 & 26 & 18 & 0 & 3\\
          & vamb & 18 & 14 & 4 & 6\\
      env-digester-1 & rescue & 12 & 3 & 3 & 7\\
          & metabat2 & 9 & 18 & 1 & 1\\
          & vamb & 4 & 14 & 4 & 7\\
      human-gut-1 & rescue & 20 & 3 & 13 & 15\\
          & metabat2 & 28 & 32 & 8 & 10\\
          & vamb & 18 & 33 & 11 & 16\\
      human-gut-2 & rescue & 21 & 2 & 9 & 24\\
          & metabat2 & 33 & 49 & 6 & 14\\
          & vamb & 23 & 41 & 8 & 21\\
      human-gut-3 & rescue & 8 & 1 & 4 & 3\\
          & metabat2 & 15 & 4 & 0 & 1\\
          & vamb & 8 & 6 & 6 & 1\\
      human-gut-4 & rescue & 7 & 4 & 2 & 7\\
          & metabat2 & 24 & 30 & 0 & 1\\
          & vamb & 18 & 28 & 2 & 6\\
      human-gut-5 & rescue & 8 & 3 & 1 & 3\\
          & metabat2 & 20 & 13 & 1 & 0\\
          & vamb & N/A\\
      human-gut-6 & rescue & 5 & 3 & 2 & 2\\
          & metabat2 & 6 & 10 & 1 & 2\\
          & vamb & 8 & 10 & 2 & 2\\
      human-gut-7 & rescue & 5 & 7 & 8 & 6\\
          & metabat2 & 22 & 17 & 1 & 1\\
          & vamb & 17 & 14 & 5 & 6\\
      huamn-gut-8 & rescue & 3 & 3 & 1 & 4\\
          & metabat2 & 13 & 12 & 0 & 0\\
          & vamb & 8 & 8 & 1 & 5\\
      human-gut-9 & rescue & 4 & 1 & 0 & 3\\
          & metabat2 & 10 & 14 & 0 & 0\\
          & vamb & 22 & 11 & 0 & 0\\
      human-gut-10 & rescue & 12 & 3 & 4 & 6\\
          & metabat2 & 23 & 21 & 0 & 1\\
          & vamb & 19 & 20 & 2 & 3\\ \hline
      
  \end{tabular}
\end{table}

\begin{landscape}
\begin{table}[bp!]
  \centering
  \caption{Merging the rescued circles and MetaBAT2 bins.
``90-5'': near-complete. 
CheckM evaluation columns delimited by commas
contain bin counts of in the order of 
near-complete, high-quality and medium-quality.
Total MetaBAT2 bins did not double count the long circular contigs.}
\begin{tabular}{ccccccccccc}
  \hline
  sample & \#circular & \#rescue & \#metabat2 & checkM & checkM & checkM & checkM & \#orphan & \#present/miss\\ 
& contig &  & accept/all & circular & rescue & metabat2 & merged & 90-5/ & 90-5\\
&  &  &  & contig &  &  &  & all & metabat2\\ \hline
sheep-gut-1a & 145 & 61 & 611/668 & 132,10,2 & 45,5,0 & 51,35,37 & 69,32,36 & 1/4 & 27/27\\
sheep-gut-1b & 279 & 161 & 1269/1431 & 249,22,2 & 99,15,2 & 130,104,111 & 188,102,96 & 4/25 & 37/41\\
chicken-gut-1 & 73 & 12 & 332/379 & 62,10,1 & 10,0,0 & 26,18,14 & 28,18,14 & 0/0 & 8/8\\
env-digester-1 & 21 & 41 & 292/316 & 18,2,0 & 12,3,1 & 9,19,11 & 20,19,12 & 0/2 & 1/1\\
human-gut-1 & 42 & 52 & 470/511 & 37,5,0 & 20,3,2 & 28,32,29 & 37,28,29 & 2/22 & 8/11\\
human-gut-2 & 46 & 39 & 545/581 & 37,7,1 & 21,2,1 & 34,49,40 & 44,44,37 & 2/11 & 9/11\\
human-gut-3 & 37 & 12 & 86/128 & 36,0,1 & 8,1,0 & 15,4,8 & 17,4,8 & 0/4 & 5/6\\
human-gut-4 & 18 & 37 & 234/267 & 16,2,0 & 7,4,3 & 24,30,40 & 17,28,39 & 0/9 & 6/14\\
human-gut-5 & 10 & 19 & 107/123 & 7,0,0 & 8,3,0 & 20,13,8 & 18,15,7 & 0/4 & 8/10\\
human-gut-6 & 7 & 15 & 76/93 & 4,1,2 & 5,3,1 & 6,10,11 & 8,10,12 & 0/1 & 2/3\\
human-gut-7 & 17 & 22 & 166/188 & 16,1,0 & 5,7,1 & 22,17,18 & 21,18,17 & 0/2 & 5/6\\
huamn-gut-8 & 4 & 15 & 112/124 & 3,0,0 & 3,3,0 & 13,12,15 & 10,11,15 & 0/0 & 3/6\\
human-gut-9 & 13 & 18 & 107/125 & 7,1,1 & 4,1,3 & 10,14,19 & 7,9,21 & 0/1 & 4/7\\
human-gut-10 & 23 & 30 & 180/206 & 19,3,0 & 12,3,4 & 23,21,24 & 26,21,26 & 0/6 & 5/9\\ \hline
\end{tabular}

\end{table}
\end{landscape}

For sheep-gut-1b (Figure 2), the circle-rescue heuristic first considered 3340 cycles. 
After deduplication, 161 were reported, 99 of which were near-complete, 
15 were high-quality and 1 was medium-quality.
Additionally, 85 $>$1Mb linear contigs that were not used by 
medium-quality-and-above cycles were near complete. 
Only 25 of them could align more than 60\% of their length
to a long circular contig or a high-quality-or-above rescued circle (-c -xasm20).
The proportion of candidates failing the checkM quality check was comparable
to traditional binning. For example, MetaBAT2 reported 438 bins that are
at least 1Mb and less than 8Mb,
121 of them were near-complete, 80 were high-quality 
and 75 were medium-quality.
We did not find obvious difference between 
the rescued circles that failed checkM, and the rescued circles in the better quality bracket,
in terms of sequence homology to the long circular contigs.
The true positive rate from rescued circles could be improved by 
discarding obviously wrong cycles (e.g. repeating similar subsequences),
but it is not useful in practice, as 
MAGs will always be quality-controlled by checkM and other 
reference-based tools.

%%%%%%figure 2
\begin{figure}[bp!]
\includegraphics[width=0.95\linewidth]{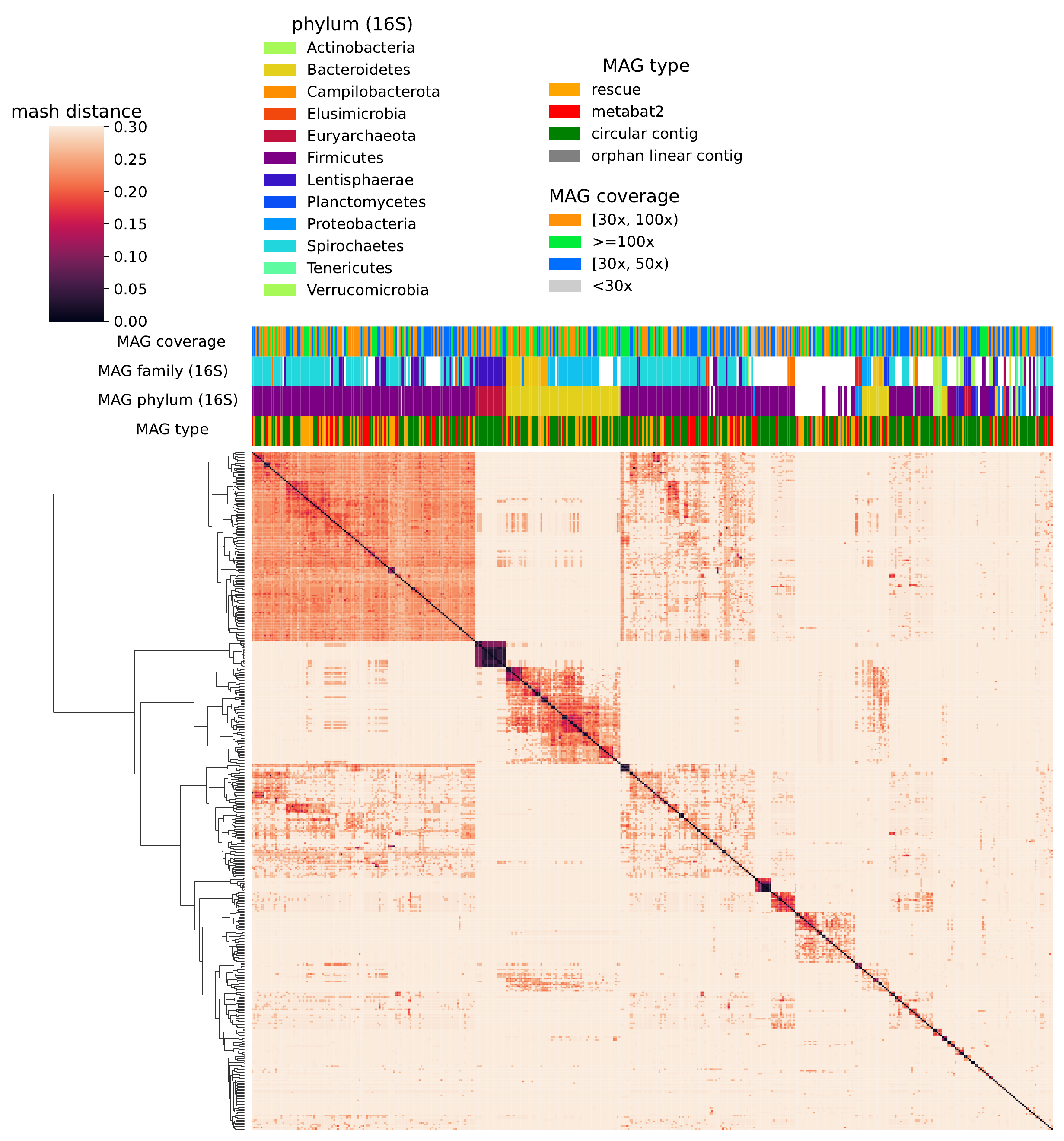}
  \caption{Clustering of near-complete MAGs of sheep-gut-1b. 
The middle-left dense purple square represents Archaea.
Mash distance more than 30\% are shown as 30\% for a better visual.}
\end{figure}

When merging the rescued circles with the regular MetaBAT2 binning (Table 2), 
most MetaBAT2 bins were accepted (1269/1431), which yielded 
89 near-complete bins and 87 high-quality bins.
25 $>$1Mb linear contigs were accepted due to orphanage 
and all of them evaluated to be near-complete.
Along with 279 $>$1Mb circular contigs (249 near-complete and 22 high-quality),
this summed to 462 near-complete MAGs.
41 near-complete MetaBAT2 bins were rejected 
due to overlapping with rescued circles.
We found 37 of them have counterparts in 
the collection of the rescued circles with less than 1\% mash distance
(30 had less than 0.1\% mash distance match), 
effectively being replaced by them.
Out of these 37 pairs,
11 rescued circles matches were better than the rejected MetaBAT2 bins in terms of 
better checkM completeness and/or contamination,
22 had identical checkM evaluations, 
2 were slightly worse but comparable, 
and 2 were worse (rescued circle was wrong).
Among the four MetaBAT2 bins without representation,
3 contained a single long linear contig (wrongly used by rescued circles), 
1 contained two long linear contigs.
Overall, we were able to recover more MAGs than MetaBAT2 alone,
and replace many MetaBAT2 bins with pseudo-circles 
while having acceptable lose.
We note that the lose would be larger in worse assemblies
where the rescued circle is prone to make more errors,
such as human-gut-4 shown in Table 2.

Since both circle-rescuing and the merging allow some contigs to be shared between bins,
there could be duplications.
To check this, we calculate pairwise mash distance 
between near-complete MAGs from the above.
MetaBAT2 bin with more than 1 contig was 
concatenated with 31 N-base paddings for convenience.
Mash sketch does hash k-mer with N-base instead of ignoring them, 
but this has little influence on the distance estimation here.
There were 441 MAGs, therefore 97020 unique pairs.
2/97020 had less than 1\% mash distance.
One pair was between a linear long contig and a rescued circle's path
belonging to the same not fully resolved subgraph 
(0.7\%; both 91\% complete and 0.6\% contaminated).
The other pair was between two circular contigs 
(0.7\%; both 100\% complete and 1.1\% contaminated).
Overall, the MAG collection had little redundancy.

The MAG merging procedure described above does not look
at existing gene annotations. It
is possible to achieve better binning if we use checkM to 
guide binning
(e.g. DAS Tools~\cite{sieber2018recovery}).
We showcase a simple checkM-guided bin merging to 
demonstrate the gap between merged MAGs and the almost-best-possible outcome 
in the near-complete quality bracket (Table 3).
We say ``almost'' because we do not try to create new bins 
or swap bin contents, 
but just to avoid shadowing valid MetaBAT2 bins with wrong rescued circles.
First, the circle-rescuing heuristics and MetaBAT2 binning were executed 
as described above. 
We then ran checkM on the rescued circles, 
MetaBAT2 bins and all 1Mb linear contigs separately.
For each quality bracket, 
we first accept all rescued circles that qualify,
then accept MetaBAT2 bins that qualify and have less than 100kb of its content
used by rescued circles, 
followed by accepting $>$1Mb linear contigs that qualify and 
have not been used by the above two categories,
and finally all $>$1Mb circular contigs that qualify.

\begin{table}[bp!]
  \caption{CheckM-assisted best-possible MAG recovery for each sample. 
  First four numbers of each field give the count of MAGs 
  from $>$1Mb circular contigs, rescued circles, metabat2 bins and 
  $>$1Mb orphaned linear contigs of the 
  corresponding quality bracket, respectively. 
  The fifth number gives the sum.}
    \begin{tabular}{cccc}
      \hline
      sample & \#near-complete & \#high-quality & \#medium-quality\\ \hline
      sheep-gut-1a & 132+45+26+6=209 & 10+5+33+8=56 & 2+0+37+19=58\\
      sheep-gut-1b & 249+99+94+19=461 & 22+15+101+30=168 & 2+2+110+49=163\\
      chicken-gut-1 & 62+10+18+0=90 & 10+0+18+7=35 & 1+0+14+13=28\\
      env-digester-1 & 18+12+8+0=38 & 2+3+19+5=29 & 0+1+11+5=17\\
      human-gut-1 & 37+20+21+7=85 & 5+3+31+15=54 & 0+2+29+30=61\\
      human-gut-2 & 37+21+25+5=88 & 7+2+48+19=76 & 1+1+39+34=75\\
      human-gut-3 & 36+8+10+0=54 & 0+1+4+2=7 & 1+0+8+6=15\\
      human-gut-4 & 16+7+18+0=41 & 2+4+30+4=40 & 0+3+39+10=52\\
      human-gut-5 & 7+8+15+0=30 & 0+3+13+3=19 & 0+0+8+17=25\\
      human-gut-6 & 4+5+4+1=14 & 1+3+9+3=16 & 2+1+11+2=16\\
      human-gut-7 & 16+5+19+1=41 & 1+7+14+6=28 & 0+1+18+3=22\\
      huamn-gut-8 & 3+3+11+0=17 & 0+3+11+3=17 & 0+0+15+5=20\\
      human-gut-9 & 7+4+6+0=17 & 1+1+14+1=17 & 1+3+19+4=27\\
      human-gut-10 & 19+12+18+0=49 & 3+3+20+6=32 & 0+4+23+7=34\\ \hline
    \end{tabular}
  \end{table}

\subsection*{Sample representation completeness assessed using k-mer spectrum}

We borrow the k-mer spectrum plot presented such as in 
KAT~\cite{mapleson2017kat} and mercury~\cite{rhie2020merqury}.
In single genome assemblies, the main interest is to grasp 
assembly redundancies, incompleteness, correction errors and phasing
by looking at the plot.
For metagenome assembly, we focus on illustrating the first two aspects.
Metagenome libraries have extremely high counts of low multiplicity k-mers, 
resulted from the combination of sequencing errors and reads 
from low prevalence species.
We defined ratios based on k-mer counts in read occurrence ranges (rather than 
exact read occurrence values) to 
render the spectrum plot visually easier to interpret (Figure 3).
In the modified k-mer spectrum plot,
if haplotypes are not closely related, contain neglectable amount of 
repeats and the assembly is perfect, 
we expect to the see the band representing $1\times$ assembly multiplicity 
(saturated orange in such plots throughout this manuscript; 
``$1\times$ band'' from now on)
to dominate the plot within genome-wide read coverage range, and 
fall sharply outside of it.
There will be very few k-mers with read multiplicities 
higher than genome-wide coverage due to shared sequences that are longer
than the k-mers (see below).
We also hope to see few k-mers that exist in the reads but not in contigs, 
i.e. $0\times$ assembly multiplicity plotted saturated blue at the bottom of each plot,
as they imply unassembled contents.
If haplotypes are somewhat related,
we expect to see several peaks formed by the $1\times$ band
due to some k-mers being shared by more than one haplotypes.
As a real near-ideal empirical example, 
we compared the human HiFi datasets with HRGM~\cite{kim2021human} (Figure S3), 
which is a diverse yet non-redundant collection of 
5414 species from human gut microbiome.
This plot marks what the plot would look like for 
a reasonably good sample representation.
If we can achieve a smaller $0\times$ band than Figure S3 
through {\it de novo assembly}, 
then the assembly offers no less information than comparing the sample to 
a comprehensive reference set.

\begin{figure}[bp!]
\includegraphics[width=0.95\linewidth]{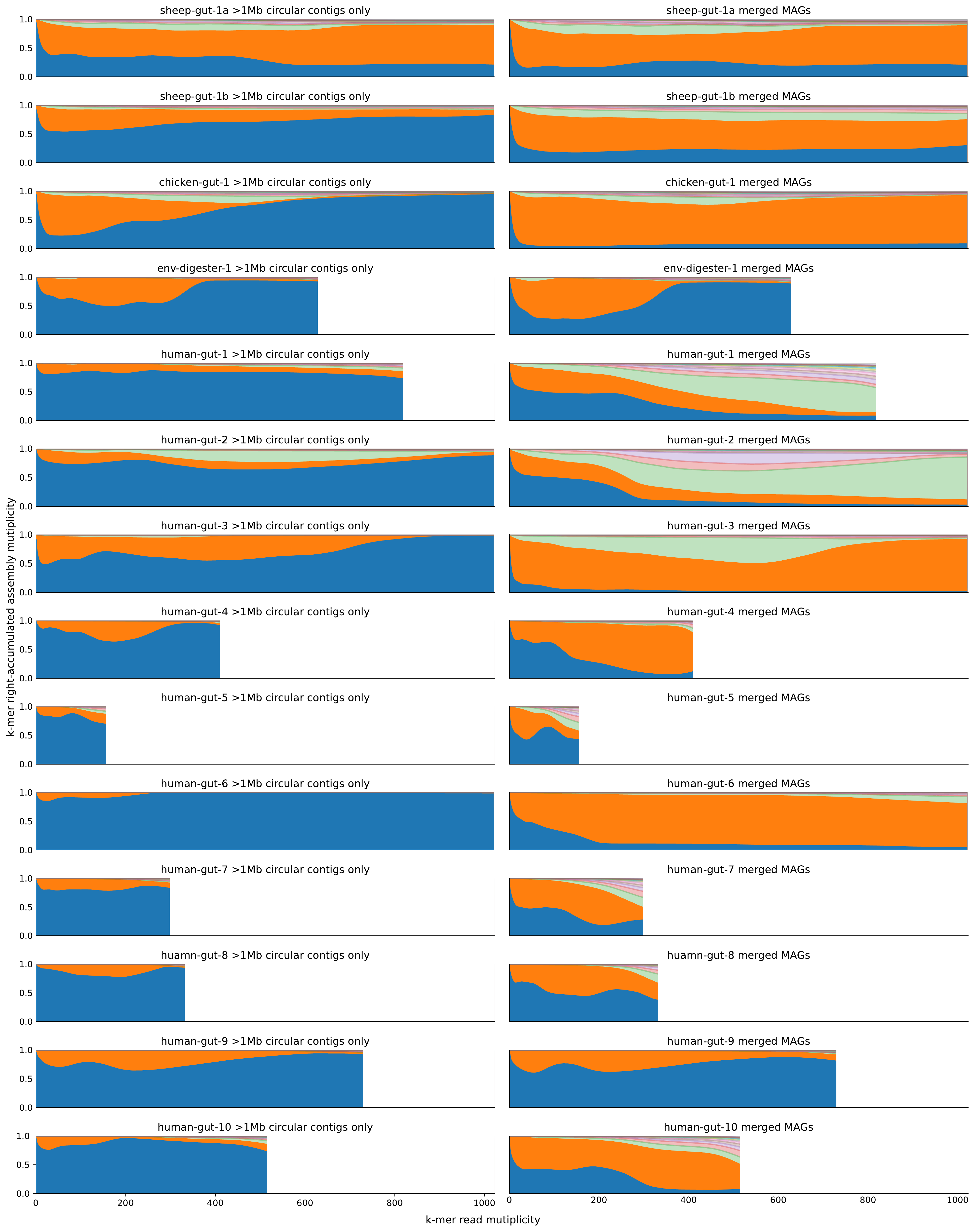}
  \caption{K-mer spectrum plots of all samples. 
In each subplot: 
let $N^{(c)}_x$ be the number of k-mers occurring $\ge x$ times in reads and 
exactly $c$ times in contigs. 
Then $N^{}_x=\sum_c N^{(c)}_x$ is the number of k-mers occurring $\ge x$ times 
in reads. 
The height of the blue area intersecting at $x$ equals $N^{(0)}_x/N^{}_x$; 
the height of the orange area intersecting at $x$ equals $N^{(1)}_x/N^{}_x$. 
Similarly we have green for $c=2$, red for $c=3$, etc. 
Intuitively speaking, a large orange region suggests a good assembly with 
most reads k-mers assembled into a single contig; a large blue region 
suggests an incomplete assembly that misses many high-abundance k-mers in reads; 
a large region of other colors suggests a potentially duplicated assembly with 
k-mers assembled into multiple contigs. 
We truncate at x if there are less than 1 million 
k-mers occurring x or more times in reads.
}
\end{figure}

Using the k-mer spectrum plots, 
we found that although hifiasm-meta could 
generate more near-complete circular contigs than previous studies, 
the circular contigs alone do not provide a complete view of 
the corresponding libraries (Figure 3, left column).
Unresolved subgraphs in the primary assembly
accounted for quite a few dominant haplotypes (Figure S4).
All samples did not receive good representation from just the circular contigs 
except for sheep-gut-1a, but this could be greatly improved
if the merged MAGs are used instead (Figure 3, right column),
except for human-gut-9 and env-digester-1.
In four libraries, merged MAG were close to the near-ideal situation
demonstrated above:
chicken-gut-1, human-gut-3, sheep-gut-1a and sheep-gut-1b.
We extracted k-mers (n=31) of the $0\times$ band and 
aligned them to MAGs in question
with bwa aln~\cite{li2009fast} to see if they can be found by allowing 
a few mismatches or indels. 
This had very insignificant impact to the plot, however.
We also examined the $2\times$ or higher bands in the high 
read-multiplicity range by dumping all kmers from human-gut-10
that were from $2\times$ to $15\times$ bands, and had read multiplicities
higher than 800.
They were likely from ubiquitous genes (e.g. tRNAs) 
or horizontally transferable sequences (see Methods).

\subsection*{Cross validation with full length 16S rRNA compositional estimation}

Complement to the the k-mer spectrum-based 
completeness evaluation is the 16S rRNA-based methods.
Most HiFi reads are 
a few folds longer than the full length rRNA genes.
This provides composition estimation for free.
Predicting the rRNA genes is
well-studied~\cite{seemann2018barrnap,nawrocki2013infernal}.
The performance was reasonable in HiFi reads
(e.g. for barrnap, around 13\% 16S genes were marked as partial).
16S-based taxonomy annotation has been similarly
extensively explored~\cite{wang2007naive,seemann2018barrnap,edgar2018accuracy},
but it suffers from reference scope bias.
For example, 
the human-gut-9 library has 1.8M reads.
24445 16S sequences were identified (1.4\%),
21954 of them (90.0\%) were assigned with confident genus level annotation.
In contrast, 
the env-digester-1 library has 1.0M reads. 
Similarly, 1.3\% of them contain 16S genes, 
but only 21.2\% which would have confident genus level annotation.
This was not limited to 16S genes.
The SRA minhash-based read taxonomy analysis only identified 41.78\% reads
as of cellular organism origin. 
Non-human gut samples were in the middleground:
sheep-gut-1b has 1.2\% reads containing 16S genes, 
with 47.3\% of them confidently annotated to genus level.

Based on these observations, 
although reference-based methods are widely used in practice and 
could resolve to species level~\cite{edgar2018updating,johnson2019evaluation},
we use greedy incremental clustering to define OTUs 
with boundary set to mismatch sequence identity 99\%~\cite{edgar2018updating}
(see Methods).
No assembly could recover all abundant OTUs, but those evaluated 
to be better in k-mer spectrum approach missed less (Figure 4; 
Figure S5 provides plots of all samples and other OTU boundaries).
Some MAGs had more than one OTU assignment
due to possible erroneous OTU clustering, 
16S diversity within genome, 
or collapsing of very similar haplotypes during assembly.
Missing high-prevalence OTUs are usually partially assembled, 
i.e. linear contigs or subgraphs that failed to 
form near-complete MAGs (see Methods).

%%%%%%figure 4
\begin{figure}[bp!]
\includegraphics[width=0.95\linewidth]{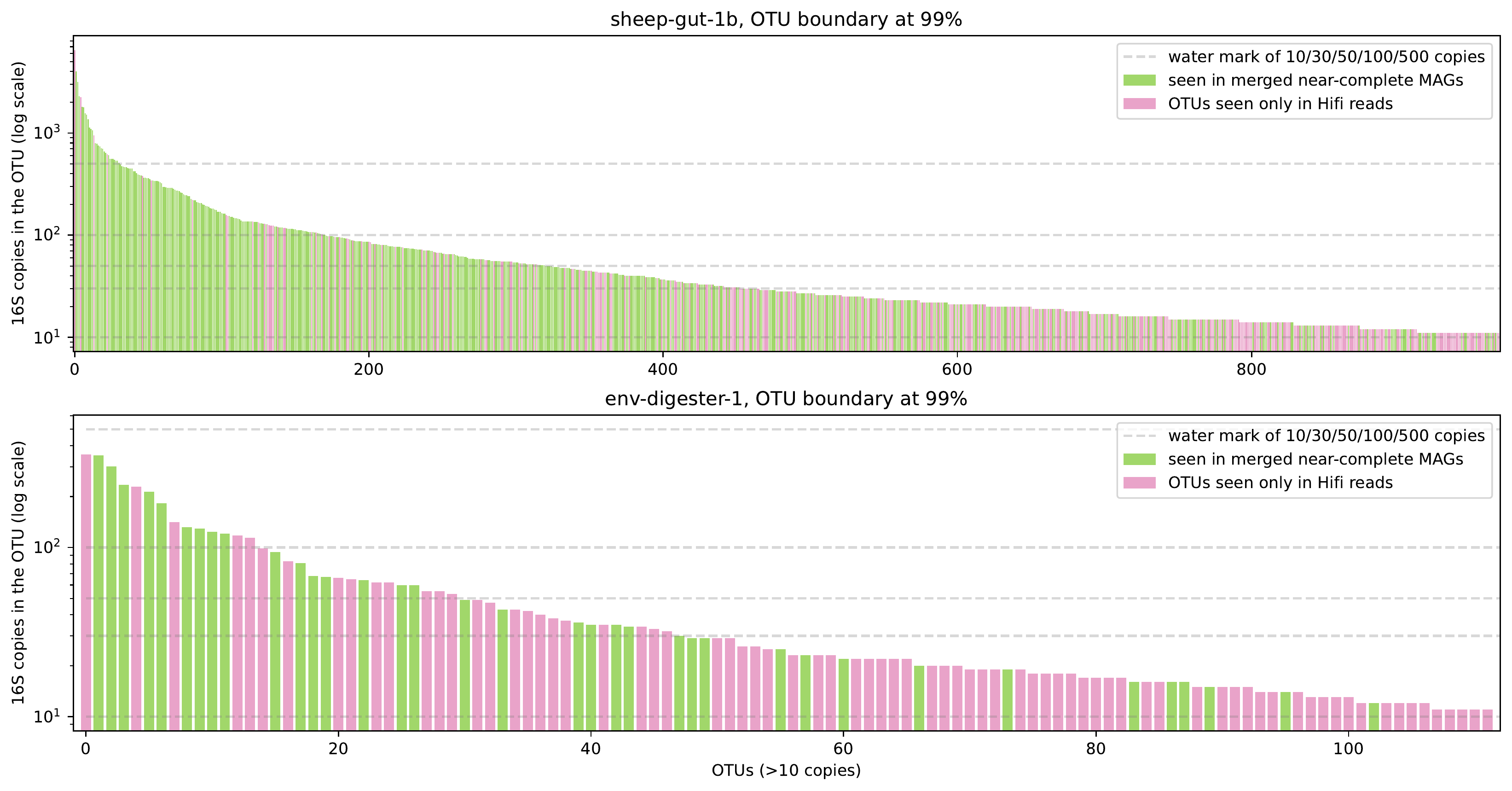}
  \caption{OTU recovery of merged bins, showing 
sheep-gut-1b and env-digester-1, which evaluated poorly and good in k-mer spectrum plot, 
respectively.
}
\end{figure}

\section*{Discussion}

\subsection*{Drawbacks of short read metagenome assemblies and the justification 
of the cost of using accurate long reads.}

With binning, the near-complete MAG yield
from HiFi assemblies, usually 1$\sim$2 such MAGs per Gb per sample,
is similar to or less than the yields of past short read studies (Figure S1)
and currently at a substantially higher cost.
It might seem that the only advantage of HiFi assembly
is the contig continuity, 
which is important but not so much for gene-focused approaches.
However, SR assemblies often miss the major species at MAG level,
and what have been missed is not easy to guess.
While this could be overcame if using a comprehensive catalog as reference
and approach the sequencing library from alignment-based or
pangenome point of view,
actual studies tend to analyze or compare MAGs without 
accounting for possible missing species.
To illustrate the issue, 
we use Almeida et al.'s collection of human gut MAGs~\cite{almeida2019new}.
The authors assembled 13133 human gut metagenomic datasets from 75 
studies with a unified pipeline
(metaSPAdes and MetaBAT2, quality controlled by checkM),
ruling out impact of any pipeline difference.
We collected k-mer spectrum plots for 3490 assemblies (Table S3) 
that had at least 10 MAGs.
We define a good assembly to have more than half of its 0x band
below the 50\% watermark, a nice assembly if using the 30\% watermark,
and a failed assembly otherwise.
Only 0.46\% assemblies were nice. 2.1\% were good.
The completeness of the final catalog might be at the mercy of 
assemblies of unrelated samples luckily recovering each others' most
abundant species.
In contrast, except for env-digester-1 and human-gut-9, 
all other HiFi assemblies presented by the merged MAGs were nice.
Moreover, combined with the representation-completeness evaluations
from two different aspects, we hope that it will be possible to 
transform the ideal sequencing depth from simply common practices
(e.g. $\le$10 Gb for human gut samples)
to be educated guesses.
This would be especially important when faced with novel 
environmental samples.

Other than incomplete sample representation, 
due to short contig length and binning errors, 
SR MAGs are known to be less complete than they appear to be.
Meziti et al. noticed that on average, 23\% population core genes and
50\% variable genes were missing from 
near 95\% complete MAGs~\cite{meziti2021reliability}.
We found most SR MAGs from short read studies to be smaller than 
their counterparts found in the HiFi MAGs (Figure 5);
mash distance $<$5\%).
12 pairs between Almeida et al. MAGs and HiFi contigs
had checkM completeness $>$99\% and contamination $<$1\%,
yet the SR MAGs were smaller than the HiFi contigs for more than 100Kb.
When compared to HRGM, this trend held. 
However, some HiFi circular contigs appeared smaller.
HRGM is a non-redundant species-level collection.
It is not clear whether we actually had smaller genomes 
(or if HRGM favored larger genomes when deduplicating), 
or the smaller size was caused by assembly errors.
Anyway, this is outside the scope of this paper and 
we will investigate in future studies.

%%%%%%figure 5
\begin{figure}[bp!]
\includegraphics[width=0.95\linewidth]{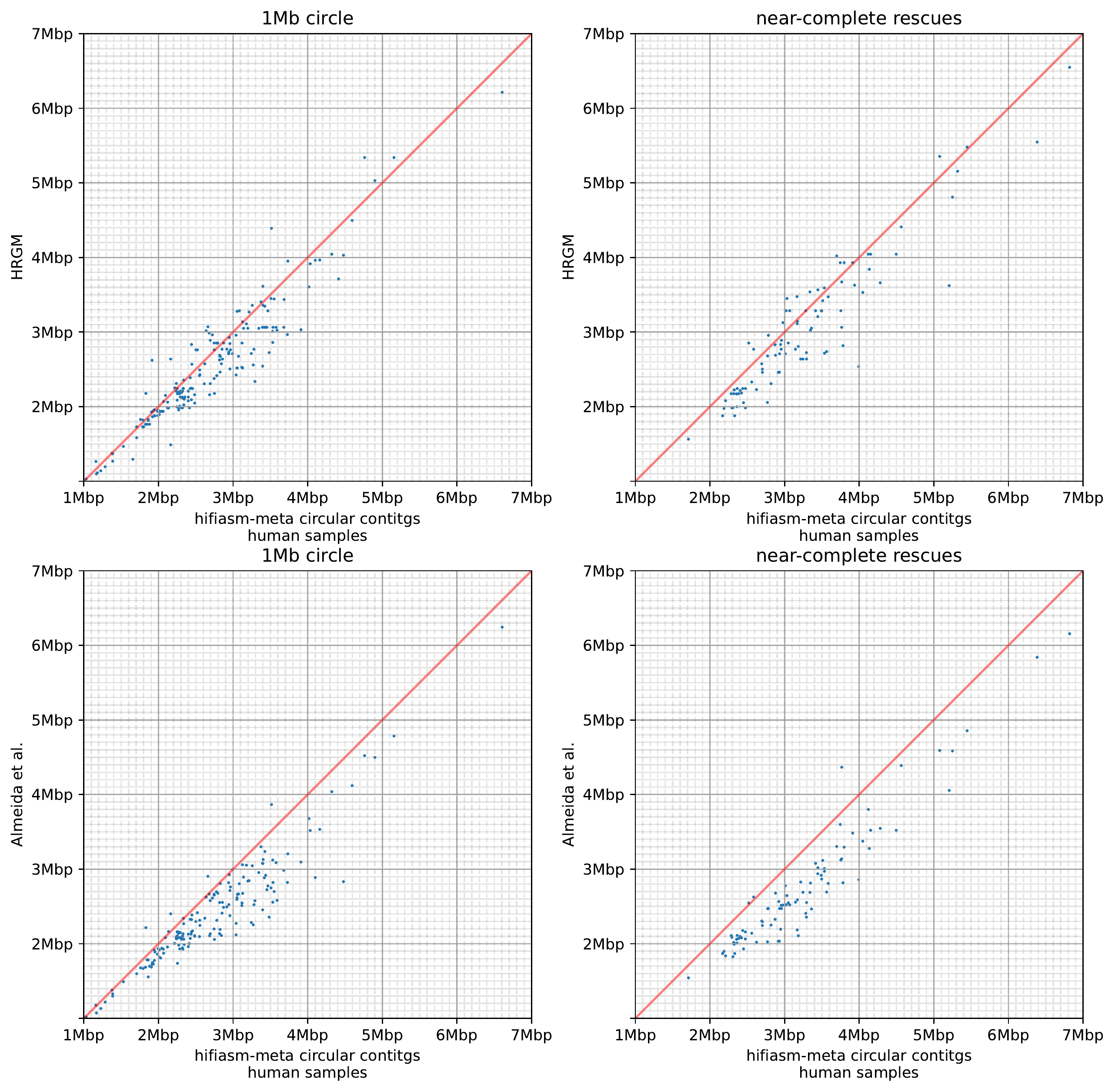}
  \caption{Most near-complete circular contigs and rescued cycles
of human gut samples are longer than assemblies in 
human gut short read MAG collection (Almeida et al.) 
or reference catalog (HRGM, the UHGG extended).
We pair our MAGs with theirs if the pair's mash distance
is less than 5\%. 
Reference MAGs are allowed to be used more than one time.
If there is a tie between reference choices, 
we arbitrarily pick one.
We have no requirements for MAG length when pairing. 
The Jaccard index (which is the basis of mash distance) 
already penalizes set size difference, 
i.e. comparing a set with its strict superset gives imperfect score.
}
\end{figure}

Haplotypes represented by the rescued circles do not appear to be 
particularly harder to recover than those assembled into circular contigs 
in the general sample pool (Figure 2).
At species level (mash distance $<$5\%), 
most of our near-complete circular contigs 
(180/182; 163/180 if comparing to Almeida et al.)
or near-complete rescued circles (93/93; 90/93 if Almeida et al.)
from human samples had matches in HRGM.
At strain level (mash distance $<$1\%), 
51/182 (28\%) near-complete circular contigs and 
16/93 (17\%) near-complete rescued circles were found in Almeida et al.'s
(difference not significant between the two: fisher's exact, p-value=0.07).

\subsection*{Undocumented or unseen MAGs.}

For our near-complete MAGs in human samples (8 individuals plus 2 
4-pooled libraries), 
we noticed some species or genera were absent from public repositories 
or MAG collections from large scale studies, using a cutoff at mash dist 
0.05 (roughly $\ge 95\%$ ANI).
47 (11\%) unseen if compared to only near-complete MAGs from Almeida et al,
17 (4\%) if compared to all reported MAGs from them.
217 (51\%) were not in refseq.
For $\ge$ 1Mb circular contigs, that would be 33 (15\%), 11 (5\%) and 122 (56\%), respectively.
These ratios are consistent with previous study. 
Jin et al. reported~\cite{jin2022hybrid} 475 
high quality (checkM completeness $\ge 80\%$, 
contamination $\le 5\%$, quality score $\ge 60$)
from pacbio-illumina hybrid assembly and binning. 
4 MAGs were circular.
The MAGs were compared to the UHGG dataset~\cite{almeida2021unified} with ANI cutoff at 95\%
and reported 24 MAGs as novel. 
In our comparison settings, 
the number of unseen MAGs are 43 (9\%), 21 (4\%) and 193 (41\%), respectively. 
It is hard to verify whether these observations are due to 
true compositional diversity between individual samples~\cite{visconti2019interplay}
or representation-incomplete assembly of past projects.
We did not find significant coverage difference 
between our MAGs that were unseen in previous works and the rest.

MAG catalogs other than the human gut microbiome's are far from complete. 
We compared the chicken-gut-1 and the sheep samples 
to their relevant MAG catalogs, 
ICRGGC (chicken-gut-1) and PRJNA657473(ruminants), respectively.
ICRGGC~\cite{feng2021metagenome} contained 12339 MAGs derived from 799 samples.
PRJNA657473~\cite{xie2021integrated} contained 10371 MAGs derived from
370 samples (7 sample species, 10 gastrointestinal tract regions).
Near-complete circular contigs and rescued circles from chicken-gut-1 
found 55/62 (contig) and 8/10 (rescued) species-level matches,
7/62 (contig) and 0/10 (rescued) strain-level matches.
sheep-gut-1b found no matches for 249 contigs and 99 rescued circles in both category,
probably due to insufficient relevant reference.

\section*{Conclusions}

In this study, we implemented a topology-based binning heuristic
on the contig graph for hifiasm-meta and described two approaches,
namely k-mer spectrum and species-evel OTUs based on full length 16S rRNAs,
to evaluate an metagenome assembly's completeness in terms of 
sample representation.
We further showed that {\it de novo} HiFi assemblies plus binning 
would have the potential 
to be both genome-complete and representation-complete,
bringing MAGs closer to their original goal
which is to delegate their biosample with minimum bias.
Note that inferring whether a sequencing library has sampled the 
microbiome of interest exhaustively is hard and out of the scope of this work.
There are other drawbacks, too. 
For example, the binning heuristic and the traditional binners 
still assume some reference bias due to being dependent 
on checkM as the final genome quality validation. 

Nonetheless, we anticipate that high quality metagenome assemblies
and further method improvements 
could transform previously inaccessible approaches,
such as analyzing horizontal gene transfers,
{\it de novo} variant calling in unusual samples
and direct comparison between microbial communities. 

\section*{Methods}

\subsection*{Assembly, evaluation and simulation}
We generated HiFi assemblies using hifiasm-meta r63 with default settings.
See Table S5 for versions of tools used.
Short read assemblies were downloaded from their releases.
We used MetaBAT2~\cite{kang2019metabat} with contig coverage 
Contig coverage for MetaBAT2 and vamb binning 
was estimated with minimap2 alignment and
MetaBAT2's jgi module.
Coverage: we ran minimap2~\cite{li2018minimap2} with 
``{\tt minimap2 -ak19 -w10 -I10G -g5k -r2k -{}-lj-min-ratio 0.5 -A2 -B5 
-O5,56 -E4,1 -z400,50 -{}-sam-hit-only contigs.fa reads.fa}''.
BAM file handling used SAMtools~\cite{li2009sequence}.

Coverage was estimated by 
``{\tt jgi\_summa\_rsize\_bam\_contig\_depths 
-{}-outputDepth depth.txt input.bam}''.
Binning: we ran MetaBAT2 using
``{\tt metabat2 -{}-seed 1 -i contigs.fa -a depth.txt}'',
and vamb using
``{\tt vamb -p 48 -{}-outdir ./ -{}-fasta contigs.fa -{}-jgi 
jgi\_depth -{}-minfasta 500000}''.
MetaBAT2's random seed has little influence. 
We separate circular contigs of 1Mb or longer into a separate MAG 
if it is binned together with other contigs.

We used checkM module ``{\tt lineage\_wf}'' to evaluate MAG quality. 
Its outputs were formatted by ``{\tt checkm qa -o 2}'' before parsing. 
We did not try DAS tools's evaluation in this work, 
but it should give consistent but more generous results. 

We did HiFi read simulation using PBSIM2~\cite{ono2021pbsim2}:
``{\tt pbsim2 -{}-depth INT}'' ``{\tt -{}-sample-fastq sample.fp}''.
We generated ``{\tt sample.fq}'', the empirical error profile, 
by randomly sampling 100k reads from sheep-gut-1a. 
We used seqtk (\url{https://github.com/lh3/seqtk}):
``{\tt seqtk sample sheep-gut-1a.fq 100000 > out.fq}''.

We forked yak (\url{https://github.com/lh3/yak}) for k-mer spectrum plot.
The fork used k=31 rather than the k=27 default of KAT.
This did not have significant impact on the plots; 
k=21 generated overall similar outputs as well.

We ran rust-mdBG with ``{\tt -k 21 -l 14 --density 0.003 -p asm}'',
then ``{\tt magic\_simplify\_meta asm}'' to generate
the final assembly graph and the sequences
per developer's recommendation.
A freestanding implementation of the circle-finding heuristics
was used, and we used mash distance (cutoff: 90\%) to compare between 
the reported circular paths to merged MAGs of hifiasm-meta's.
In sludge, humanO3 and sheepB, this reported 
22, 22 and 120 rescued circles, respectively.
We did not try to do metaBAT2 and bin merging because
checkM was sensitive to indels.

\subsection*{Examining high read- and assembly-multiplicity kmers}

We use human-gut-10 as an example. 
We identified kmers with at least 2x and up to 15x
assembly multiplicity, 
and at least 800x read multiplicity,
i.e. the right-most part of k-mer spectrum plot above 1x band.
Their location on the contigs were collected.
We merged overlapping intervals and dumped the sequences.
There were 5469 unique sequences (max length 18.6kb, 
N50 2.0kb).
We randomly select 20 from these 
(``{\tt seqtk subseq in.fa 20}'') and 
did BLAST (blastn web cgi, defaults) against nr/nt.
All queries had full length BLAST hits with low sequence divergence
and frequently overlap with genes encoding
DNA-related enzymes, transposase and tRNA or rRNA (Table S6).

\subsection*{16S rRNA methods}

We identified and annotated 16S rRNA genes from HiFi libraries
with the following steps.
First, HiFi reads that could aligned to SILVA reference 
were extracted, with base qualities stripped:
``{\tt seqtk subseq hifi.fq <(minimap2 SILVA.fa 
hifi.fq | cut -f1 | uniq) | seqtk seq -A > SSUreads.fa}''
.
We ran barrnap to identify rRNA genes:
``{\tt barrnap -{}-kingdom bac -{}-outseq rRNA.fa SSUreads.fa}''.
INFERNAL cmsearch might identify a few more rRNAs than barrnap.
We believe this would not have major influence on the conclusion
based on previous observations.
We then annotated rRNA genes with RDP classifier:
``{\tt java -Xmx16g -jar RDPTools/classifier.jar classify 
-o RDP.tsv -h RDP.hier rRNA.fa}''
.
We accept annotations of 16S rRNAs with genus scores of at least 0.9.

To define OTUs from HiFi reads,
we first selected 16S genes not marked with ``partial'' from the barrnap.
We used greedy incremental clustering: 
we initialize an empty collection $S$ to collect seed sequences.
For each 16S gene $q$, if it could align to 
any sequence $s$ in $S$ with alignment block longer than 1000bp and at least
99\% mismatch identity, it is assigned the same OTU label as $s$.
(If multiple seeds are available, the one with highest identity 
will be chosen. If a tie, the seed is arbitrarily chosen from the bests.)
Otherwise, $q$ is added to $S$ and assigned a new OTU label.
Alignment is done with minimap2's python binding, mappy,
with ``{\tt preset=map-hifi}''.
Alignment block length is given by ``{\tt mappy.Alignment.blen}''.
Mismatch identity is calculated as 
``{\tt mappy.Alignment.mlen/mappy.Alignment.blen}''.
Assigning OTU label for an unseen 16S copy is done similarly.
If a sequence can not align to any seed sequences, its OTU label is undefined.

There are two ways to assign OTU labels to MAGs:
1) collected reads belonging to contigs of a MAG and their OTU labels, 
or 2)identify 16S copies from contigs then assign labels.
We did both and found them to be mostly consistent. 
We only considered near-complete MAGs.

When evaluating MAGs as in Figure 4 and Figure S5, 
we drop OTUs with less than 10 16S copies to rule out 
artifacts from sequencing errors
and to ignore species with very low coverage.
This was done after the OTU label assignment.
A MAG could have more than one OTU assignment.
It was difficult to distinguish wrong cases (i.e. suboptimal clustering result)
from true cases, i.e. a genome having multiple distinct 16S copies, 
therefore we simply accept all OTU labels of a MAG.
For example, if the read set yields 3 OTUs ($a$,$b$ and $c$),
the assembly has a single MAG from which we identify three 16S copies
that are labeled $a$, $a$, $b$.
Then in the plot, both $a$ and $b$ would be colored as ``seen in MAG''.

\begin{backmatter}
% https://genomebiology.biomedcentral.com/submission-guidelines/preparing-your-manuscript/research

\section*{Supplementary Information}

%\subsection*{Acknowledgements}%% if any
%TBD

%\subsection*{Abbreviations}%% if any
%TBD - is this required?

\section*{Declarations}
\subsection*{Ethics approval and consent to participate}%% if any
Not applicable.

\subsection*{Consent for publication}%% if any
Not applicable.

\subsection*{Availability of data and materials}%% if any
Hifiasm-meta is open source at \url{https://github.com/xfengnefx/hifiasm-meta/}.
The yak fork used is at \url{https://github.com/xfengnefx/yam}.
Scripts and assemblies are at \url{https://github.com/xfengnefx/snpt_mtgncomp} 
and \url{ftp://ftp.dfci.harvard.edu/pub/hli/hifiasm-meta/hifiasm-binning/asm/}, 
respectively.
Hifi libraries~\cite{Kolmogorov2020-gu,bickhart2022generating,sereika2022oxford,kim2022hifi} 
are all publicly available and 
can be found under their accession IDs (table S4).
For convenience, in Table S3 we supply urls to the short read libraries 
we used when evaluating short read MAGs from Almeida et al 
(the accession IDs were originally available in their supplementary tables).  % the almeida fastqs

\subsection*{Competing interests}
The authors declare that they have no competing interests.

\subsection*{Funding}
This work was funded by National Human Genome Research Institute (NHGRI) 
R01HG010040 and U01HG010961, the Chan Zuckerberg Initiative and 
the Sloan foundation.

\subsection*{Authors' contributions}
HL and XF conceived the projected, carried out analysis and wrote the manuscript.
All authors read and approved the final manuscript.

\subsection*{Acknowledgements}
Not applicable.

\bibliographystyle{bmc-mathphys} % Style BST file (bmc-mathphys, vancouver, spbasic).
\bibliography{main}      % Bibliography file (usually '*.bib' )


\begin{thebibliography}{59}
\ifx \bisbn   \undefined \def \bisbn  #1{ISBN #1}\fi
\ifx \binits  \undefined \def \binits#1{#1}\fi
\ifx \bauthor  \undefined \def \bauthor#1{#1}\fi
\ifx \batitle  \undefined \def \batitle#1{#1}\fi
\ifx \bjtitle  \undefined \def \bjtitle#1{#1}\fi
\ifx \bvolume  \undefined \def \bvolume#1{\textbf{#1}}\fi
\ifx \byear  \undefined \def \byear#1{#1}\fi
\ifx \bissue  \undefined \def \bissue#1{#1}\fi
\ifx \bfpage  \undefined \def \bfpage#1{#1}\fi
\ifx \blpage  \undefined \def \blpage #1{#1}\fi
\ifx \burl  \undefined \def \burl#1{\textsf{#1}}\fi
\ifx \doiurl  \undefined \def \doiurl#1{\textsf{#1}}\fi
\ifx \betal  \undefined \def \betal{\textit{et al.}}\fi
\ifx \binstitute  \undefined \def \binstitute#1{#1}\fi
\ifx \binstitutionaled  \undefined \def \binstitutionaled#1{#1}\fi
\ifx \bctitle  \undefined \def \bctitle#1{#1}\fi
\ifx \beditor  \undefined \def \beditor#1{#1}\fi
\ifx \bpublisher  \undefined \def \bpublisher#1{#1}\fi
\ifx \bbtitle  \undefined \def \bbtitle#1{#1}\fi
\ifx \bedition  \undefined \def \bedition#1{#1}\fi
\ifx \bseriesno  \undefined \def \bseriesno#1{#1}\fi
\ifx \blocation  \undefined \def \blocation#1{#1}\fi
\ifx \bsertitle  \undefined \def \bsertitle#1{#1}\fi
\ifx \bsnm \undefined \def \bsnm#1{#1}\fi
\ifx \bsuffix \undefined \def \bsuffix#1{#1}\fi
\ifx \bparticle \undefined \def \bparticle#1{#1}\fi
\ifx \barticle \undefined \def \barticle#1{#1}\fi
\ifx \bconfdate \undefined \def \bconfdate #1{#1}\fi
\ifx \botherref \undefined \def \botherref #1{#1}\fi
\ifx \url \undefined \def \url#1{\textsf{#1}}\fi
\ifx \bchapter \undefined \def \bchapter#1{#1}\fi
\ifx \bbook \undefined \def \bbook#1{#1}\fi
\ifx \bcomment \undefined \def \bcomment#1{#1}\fi
\ifx \oauthor \undefined \def \oauthor#1{#1}\fi
\ifx \citeauthoryear \undefined \def \citeauthoryear#1{#1}\fi
\ifx \endbibitem  \undefined \def \endbibitem {}\fi
\ifx \bconflocation  \undefined \def \bconflocation#1{#1}\fi
\ifx \arxivurl  \undefined \def \arxivurl#1{\textsf{#1}}\fi
\csname PreBibitemsHook\endcsname

%%% 1
\bibitem{tully2018reconstruction}
\begin{barticle}
\bauthor{\bsnm{Tully}, \binits{B.J.}},
\bauthor{\bsnm{Graham}, \binits{E.D.}},
\bauthor{\bsnm{Heidelberg}, \binits{J.F.}}:
\batitle{The reconstruction of 2,631 draft metagenome-assembled genomes from
  the global oceans}.
\bjtitle{Scientific data}
\bvolume{5}(\bissue{1}),
\bfpage{1}--\blpage{8}
(\byear{2018})
\end{barticle}
\endbibitem

%%% 2
\bibitem{howe2014tackling}
\begin{barticle}
\bauthor{\bsnm{Howe}, \binits{A.C.}},
\bauthor{\bsnm{Jansson}, \binits{J.K.}},
\bauthor{\bsnm{Malfatti}, \binits{S.A.}},
\bauthor{\bsnm{Tringe}, \binits{S.G.}},
\bauthor{\bsnm{Tiedje}, \binits{J.M.}},
\bauthor{\bsnm{Brown}, \binits{C.T.}}:
\batitle{Tackling soil diversity with the assembly of large, complex
  metagenomes}.
\bjtitle{Proceedings of the National Academy of Sciences}
\bvolume{111}(\bissue{13}),
\bfpage{4904}--\blpage{4909}
(\byear{2014})
\end{barticle}
\endbibitem

%%% 3
\bibitem{kroeger2018new}
\begin{barticle}
\bauthor{\bsnm{Kroeger}, \binits{M.E.}},
\bauthor{\bsnm{Delmont}, \binits{T.O.}},
\bauthor{\bsnm{Eren}, \binits{A.M.}},
\bauthor{\bsnm{Meyer}, \binits{K.M.}},
\bauthor{\bsnm{Guo}, \binits{J.}},
\bauthor{\bsnm{Khan}, \binits{K.}},
\bauthor{\bsnm{Rodrigues}, \binits{J.L.}},
\bauthor{\bsnm{Bohannan}, \binits{B.J.}},
\bauthor{\bsnm{Tringe}, \binits{S.G.}},
\bauthor{\bsnm{Borges}, \binits{C.D.}}, \betal:
\batitle{New biological insights into how deforestation in amazonia affects
  soil microbial communities using metagenomics and metagenome-assembled
  genomes}.
\bjtitle{Frontiers in microbiology}
\bvolume{9},
\bfpage{1635}
(\byear{2018})
\end{barticle}
\endbibitem

%%% 4
\bibitem{Parks2015ua}
\begin{barticle}
\bauthor{\bsnm{Parks}, \binits{D.H.}},
\bauthor{\bsnm{Imelfort}, \binits{M.}},
\bauthor{\bsnm{Skennerton}, \binits{C.T.}},
\bauthor{\bsnm{Hugenholtz}, \binits{P.}},
\bauthor{\bsnm{Tyson}, \binits{G.W.}}:
\batitle{{CheckM}: assessing the quality of microbial genomes recovered from
  isolates, single cells, and metagenomes}.
\bjtitle{Genome Res.}
\bvolume{25}(\bissue{7}),
\bfpage{1043}--\blpage{1055}
(\byear{2015})
\end{barticle}
\endbibitem

%%% 5
\bibitem{bowers2017minimum}
\begin{barticle}
\bauthor{\bsnm{Bowers}, \binits{R.M.}},
\bauthor{\bsnm{Kyrpides}, \binits{N.C.}},
\bauthor{\bsnm{Stepanauskas}, \binits{R.}},
\bauthor{\bsnm{Harmon-Smith}, \binits{M.}},
\bauthor{\bsnm{Doud}, \binits{D.}},
\bauthor{\bsnm{Reddy}, \binits{T.}},
\bauthor{\bsnm{Schulz}, \binits{F.}},
\bauthor{\bsnm{Jarett}, \binits{J.}},
\bauthor{\bsnm{Rivers}, \binits{A.R.}},
\bauthor{\bsnm{Eloe-Fadrosh}, \binits{E.A.}}, \betal:
\batitle{Minimum information about a single amplified genome (misag) and a
  metagenome-assembled genome (mimag) of bacteria and archaea}.
\bjtitle{Nature biotechnology}
\bvolume{35}(\bissue{8}),
\bfpage{725}--\blpage{731}
(\byear{2017})
\end{barticle}
\endbibitem

%%% 6
\bibitem{albertsen2013genome}
\begin{barticle}
\bauthor{\bsnm{Albertsen}, \binits{M.}},
\bauthor{\bsnm{Hugenholtz}, \binits{P.}},
\bauthor{\bsnm{Skarshewski}, \binits{A.}},
\bauthor{\bsnm{Nielsen}, \binits{K.L.}},
\bauthor{\bsnm{Tyson}, \binits{G.W.}},
\bauthor{\bsnm{Nielsen}, \binits{P.H.}}:
\batitle{Genome sequences of rare, uncultured bacteria obtained by differential
  coverage binning of multiple metagenomes}.
\bjtitle{Nature biotechnology}
\bvolume{31}(\bissue{6}),
\bfpage{533}--\blpage{538}
(\byear{2013})
\end{barticle}
\endbibitem

%%% 7
\bibitem{vicedomini2021strainberry}
\begin{barticle}
\bauthor{\bsnm{Vicedomini}, \binits{R.}},
\bauthor{\bsnm{Quince}, \binits{C.}},
\bauthor{\bsnm{Darling}, \binits{A.E.}},
\bauthor{\bsnm{Chikhi}, \binits{R.}}:
\batitle{Strainberry: automated strain separation in low-complexity metagenomes
  using long reads}.
\bjtitle{Nature Communications}
\bvolume{12}(\bissue{1}),
\bfpage{1}--\blpage{14}
(\byear{2021})
\end{barticle}
\endbibitem

%%% 8
\bibitem{nayfach2019new}
\begin{barticle}
\bauthor{\bsnm{Nayfach}, \binits{S.}},
\bauthor{\bsnm{Shi}, \binits{Z.J.}},
\bauthor{\bsnm{Seshadri}, \binits{R.}},
\bauthor{\bsnm{Pollard}, \binits{K.S.}},
\bauthor{\bsnm{Kyrpides}, \binits{N.C.}}:
\batitle{New insights from uncultivated genomes of the global human gut
  microbiome}.
\bjtitle{Nature}
\bvolume{568}(\bissue{7753}),
\bfpage{505}--\blpage{510}
(\byear{2019})
\end{barticle}
\endbibitem

%%% 9
\bibitem{luo2012individual}
\begin{barticle}
\bauthor{\bsnm{Luo}, \binits{C.}},
\bauthor{\bsnm{Tsementzi}, \binits{D.}},
\bauthor{\bsnm{Kyrpides}, \binits{N.C.}},
\bauthor{\bsnm{Konstantinidis}, \binits{K.T.}}:
\batitle{Individual genome assembly from complex community short-read
  metagenomic datasets}.
\bjtitle{The ISME journal}
\bvolume{6}(\bissue{4}),
\bfpage{898}--\blpage{901}
(\byear{2012})
\end{barticle}
\endbibitem

%%% 10
\bibitem{stackebrandt1994taxonomic}
\begin{barticle}
\bauthor{\bsnm{Stackebrandt}, \binits{E.}},
\bauthor{\bsnm{Goebel}, \binits{B.M.}}:
\batitle{Taxonomic note: a place for dna-dna reassociation and 16s rrna
  sequence analysis in the present species definition in bacteriology}.
\bjtitle{International journal of systematic and evolutionary microbiology}
\bvolume{44}(\bissue{4}),
\bfpage{846}--\blpage{849}
(\byear{1994})
\end{barticle}
\endbibitem

%%% 11
\bibitem{brumfield2020microbial}
\begin{barticle}
\bauthor{\bsnm{Brumfield}, \binits{K.D.}},
\bauthor{\bsnm{Huq}, \binits{A.}},
\bauthor{\bsnm{Colwell}, \binits{R.R.}},
\bauthor{\bsnm{Olds}, \binits{J.L.}},
\bauthor{\bsnm{Leddy}, \binits{M.B.}}:
\batitle{Microbial resolution of whole genome shotgun and 16s amplicon
  metagenomic sequencing using publicly available neon data}.
\bjtitle{PLoS One}
\bvolume{15}(\bissue{2}),
\bfpage{0228899}
(\byear{2020})
\end{barticle}
\endbibitem

%%% 12
\bibitem{edgar2018updating}
\begin{barticle}
\bauthor{\bsnm{Edgar}, \binits{R.C.}}:
\batitle{Updating the 97\% identity threshold for 16s ribosomal rna otus}.
\bjtitle{Bioinformatics}
\bvolume{34}(\bissue{14}),
\bfpage{2371}--\blpage{2375}
(\byear{2018})
\end{barticle}
\endbibitem

%%% 13
\bibitem{yuan2015reconstructing}
\begin{barticle}
\bauthor{\bsnm{Yuan}, \binits{C.}},
\bauthor{\bsnm{Lei}, \binits{J.}},
\bauthor{\bsnm{Cole}, \binits{J.}},
\bauthor{\bsnm{Sun}, \binits{Y.}}:
\batitle{Reconstructing 16s rrna genes in metagenomic data}.
\bjtitle{Bioinformatics}
\bvolume{31}(\bissue{12}),
\bfpage{35}--\blpage{43}
(\byear{2015})
\end{barticle}
\endbibitem

%%% 14
\bibitem{poretsky2014strengths}
\begin{barticle}
\bauthor{\bsnm{Poretsky}, \binits{R.}},
\bauthor{\bsnm{Rodriguez-R}, \binits{L.M.}},
\bauthor{\bsnm{Luo}, \binits{C.}},
\bauthor{\bsnm{Tsementzi}, \binits{D.}},
\bauthor{\bsnm{Konstantinidis}, \binits{K.T.}}:
\batitle{Strengths and limitations of 16s rrna gene amplicon sequencing in
  revealing temporal microbial community dynamics}.
\bjtitle{PloS one}
\bvolume{9}(\bissue{4}),
\bfpage{93827}
(\byear{2014})
\end{barticle}
\endbibitem

%%% 15
\bibitem{kai2019rapid}
\begin{barticle}
\bauthor{\bsnm{Kai}, \binits{S.}},
\bauthor{\bsnm{Matsuo}, \binits{Y.}},
\bauthor{\bsnm{Nakagawa}, \binits{S.}},
\bauthor{\bsnm{Kryukov}, \binits{K.}},
\bauthor{\bsnm{Matsukawa}, \binits{S.}},
\bauthor{\bsnm{Tanaka}, \binits{H.}},
\bauthor{\bsnm{Iwai}, \binits{T.}},
\bauthor{\bsnm{Imanishi}, \binits{T.}},
\bauthor{\bsnm{Hirota}, \binits{K.}}:
\batitle{Rapid bacterial identification by direct pcr amplification of 16s rrna
  genes using the minion™ nanopore sequencer}.
\bjtitle{FEBS open bio}
\bvolume{9}(\bissue{3}),
\bfpage{548}--\blpage{557}
(\byear{2019})
\end{barticle}
\endbibitem

%%% 16
\bibitem{johnson2019evaluation}
\begin{barticle}
\bauthor{\bsnm{Johnson}, \binits{J.S.}},
\bauthor{\bsnm{Spakowicz}, \binits{D.J.}},
\bauthor{\bsnm{Hong}, \binits{B.-Y.}},
\bauthor{\bsnm{Petersen}, \binits{L.M.}},
\bauthor{\bsnm{Demkowicz}, \binits{P.}},
\bauthor{\bsnm{Chen}, \binits{L.}},
\bauthor{\bsnm{Leopold}, \binits{S.R.}},
\bauthor{\bsnm{Hanson}, \binits{B.M.}},
\bauthor{\bsnm{Agresta}, \binits{H.O.}},
\bauthor{\bsnm{Gerstein}, \binits{M.}}, \betal:
\batitle{Evaluation of 16s rrna gene sequencing for species and strain-level
  microbiome analysis}.
\bjtitle{Nature communications}
\bvolume{10}(\bissue{1}),
\bfpage{1}--\blpage{11}
(\byear{2019})
\end{barticle}
\endbibitem

%%% 17
\bibitem{wang2019metagenomic}
\begin{barticle}
\bauthor{\bsnm{Wang}, \binits{W.}},
\bauthor{\bsnm{Hu}, \binits{H.}},
\bauthor{\bsnm{Zijlstra}, \binits{R.T.}},
\bauthor{\bsnm{Zheng}, \binits{J.}},
\bauthor{\bsnm{G{\"a}nzle}, \binits{M.G.}}:
\batitle{Metagenomic reconstructions of gut microbial metabolism in weanling
  pigs}.
\bjtitle{Microbiome}
\bvolume{7}(\bissue{1}),
\bfpage{1}--\blpage{11}
(\byear{2019})
\end{barticle}
\endbibitem

%%% 18
\bibitem{almeida2019new}
\begin{barticle}
\bauthor{\bsnm{Almeida}, \binits{A.}},
\bauthor{\bsnm{Mitchell}, \binits{A.L.}},
\bauthor{\bsnm{Boland}, \binits{M.}},
\bauthor{\bsnm{Forster}, \binits{S.C.}},
\bauthor{\bsnm{Gloor}, \binits{G.B.}},
\bauthor{\bsnm{Tarkowska}, \binits{A.}},
\bauthor{\bsnm{Lawley}, \binits{T.D.}},
\bauthor{\bsnm{Finn}, \binits{R.D.}}:
\batitle{A new genomic blueprint of the human gut microbiota}.
\bjtitle{Nature}
\bvolume{568}(\bissue{7753}),
\bfpage{499}--\blpage{504}
(\byear{2019})
\end{barticle}
\endbibitem

%%% 19
\bibitem{kim2021human}
\begin{barticle}
\bauthor{\bsnm{Kim}, \binits{C.Y.}},
\bauthor{\bsnm{Lee}, \binits{M.}},
\bauthor{\bsnm{Yang}, \binits{S.}},
\bauthor{\bsnm{Kim}, \binits{K.}},
\bauthor{\bsnm{Yong}, \binits{D.}},
\bauthor{\bsnm{Kim}, \binits{H.R.}},
\bauthor{\bsnm{Lee}, \binits{I.}}:
\batitle{Human reference gut microbiome catalog including newly assembled
  genomes from under-represented asian metagenomes}.
\bjtitle{Genome Medicine}
\bvolume{13}(\bissue{1}),
\bfpage{1}--\blpage{20}
(\byear{2021})
\end{barticle}
\endbibitem

%%% 20
\bibitem{pasolli2019extensive}
\begin{barticle}
\bauthor{\bsnm{Pasolli}, \binits{E.}},
\bauthor{\bsnm{Asnicar}, \binits{F.}},
\bauthor{\bsnm{Manara}, \binits{S.}},
\bauthor{\bsnm{Zolfo}, \binits{M.}},
\bauthor{\bsnm{Karcher}, \binits{N.}},
\bauthor{\bsnm{Armanini}, \binits{F.}},
\bauthor{\bsnm{Beghini}, \binits{F.}},
\bauthor{\bsnm{Manghi}, \binits{P.}},
\bauthor{\bsnm{Tett}, \binits{A.}},
\bauthor{\bsnm{Ghensi}, \binits{P.}}, \betal:
\batitle{Extensive unexplored human microbiome diversity revealed by over
  150,000 genomes from metagenomes spanning age, geography, and lifestyle}.
\bjtitle{Cell}
\bvolume{176}(\bissue{3}),
\bfpage{649}--\blpage{662}
(\byear{2019})
\end{barticle}
\endbibitem

%%% 21
\bibitem{chen2021expanded}
\begin{barticle}
\bauthor{\bsnm{Chen}, \binits{C.}},
\bauthor{\bsnm{Zhou}, \binits{Y.}},
\bauthor{\bsnm{Fu}, \binits{H.}},
\bauthor{\bsnm{Xiong}, \binits{X.}},
\bauthor{\bsnm{Fang}, \binits{S.}},
\bauthor{\bsnm{Jiang}, \binits{H.}},
\bauthor{\bsnm{Wu}, \binits{J.}},
\bauthor{\bsnm{Yang}, \binits{H.}},
\bauthor{\bsnm{Gao}, \binits{J.}},
\bauthor{\bsnm{Huang}, \binits{L.}}:
\batitle{Expanded catalog of microbial genes and metagenome-assembled genomes
  from the pig gut microbiome}.
\bjtitle{Nature communications}
\bvolume{12}(\bissue{1}),
\bfpage{1}--\blpage{13}
(\byear{2021})
\end{barticle}
\endbibitem

%%% 22
\bibitem{visconti2019interplay}
\begin{barticle}
\bauthor{\bsnm{Visconti}, \binits{A.}},
\bauthor{\bsnm{Le~Roy}, \binits{C.I.}},
\bauthor{\bsnm{Rosa}, \binits{F.}},
\bauthor{\bsnm{Rossi}, \binits{N.}},
\bauthor{\bsnm{Martin}, \binits{T.C.}},
\bauthor{\bsnm{Mohney}, \binits{R.P.}},
\bauthor{\bsnm{Li}, \binits{W.}},
\bauthor{\bparticle{de} \bsnm{Rinaldis}, \binits{E.}},
\bauthor{\bsnm{Bell}, \binits{J.T.}},
\bauthor{\bsnm{Venter}, \binits{J.C.}}, \betal:
\batitle{Interplay between the human gut microbiome and host metabolism}.
\bjtitle{Nature communications}
\bvolume{10}(\bissue{1}),
\bfpage{1}--\blpage{10}
(\byear{2019})
\end{barticle}
\endbibitem

%%% 23
\bibitem{liu2022towards}
\begin{botherref}
\oauthor{\bsnm{Liu}, \binits{P.}},
\oauthor{\bsnm{Hu}, \binits{S.}},
\oauthor{\bsnm{He}, \binits{Z.}},
\oauthor{\bsnm{Feng}, \binits{C.}},
\oauthor{\bsnm{Dong}, \binits{G.}},
\oauthor{\bsnm{An}, \binits{S.}},
\oauthor{\bsnm{Liu}, \binits{R.}},
\oauthor{\bsnm{Xu}, \binits{F.}},
\oauthor{\bsnm{Chen}, \binits{Y.}},
\oauthor{\bsnm{Ying}, \binits{X.}}:
Towards strain-level complexity: Sequencing depth required for comprehensive
  single-nucleotide polymorphism analysis of the human gut microbiome.
Frontiers in Microbiology
\textbf{13}
(2022)
\end{botherref}
\endbibitem

%%% 24
\bibitem{hillmann2018evaluating}
\begin{barticle}
\bauthor{\bsnm{Hillmann}, \binits{B.}},
\bauthor{\bsnm{Al-Ghalith}, \binits{G.A.}},
\bauthor{\bsnm{Shields-Cutler}, \binits{R.R.}},
\bauthor{\bsnm{Zhu}, \binits{Q.}},
\bauthor{\bsnm{Gohl}, \binits{D.M.}},
\bauthor{\bsnm{Beckman}, \binits{K.B.}},
\bauthor{\bsnm{Knight}, \binits{R.}},
\bauthor{\bsnm{Knights}, \binits{D.}}:
\batitle{Evaluating the information content of shallow shotgun metagenomics}.
\bjtitle{Msystems}
\bvolume{3}(\bissue{6}),
\bfpage{00069}--\blpage{18}
(\byear{2018})
\end{barticle}
\endbibitem

%%% 25
\bibitem{SRwetland}
\begin{botherref}
\oauthor{\bsnm{Institute}, \binits{D.J.G.}}
\url{https://www.ncbi.nlm.nih.gov/bioproject/?term=Wetland+microbial+communities+from+Twitchell+Island}
(2011)
\end{botherref}
\endbibitem

%%% 26
\bibitem{kanehisa2016blastkoala}
\begin{barticle}
\bauthor{\bsnm{Kanehisa}, \binits{M.}},
\bauthor{\bsnm{Sato}, \binits{Y.}},
\bauthor{\bsnm{Morishima}, \binits{K.}}:
\batitle{Blastkoala and ghostkoala: Kegg tools for functional characterization
  of genome and metagenome sequences}.
\bjtitle{Journal of molecular biology}
\bvolume{428}(\bissue{4}),
\bfpage{726}--\blpage{731}
(\byear{2016})
\end{barticle}
\endbibitem

%%% 27
\bibitem{frioux2020bag}
\begin{barticle}
\bauthor{\bsnm{Frioux}, \binits{C.}},
\bauthor{\bsnm{Singh}, \binits{D.}},
\bauthor{\bsnm{Korcsmaros}, \binits{T.}},
\bauthor{\bsnm{Hildebrand}, \binits{F.}}:
\batitle{From bag-of-genes to bag-of-genomes: metabolic modelling of
  communities in the era of metagenome-assembled genomes}.
\bjtitle{Computational and Structural Biotechnology Journal}
\bvolume{18},
\bfpage{1722}--\blpage{1734}
(\byear{2020})
\end{barticle}
\endbibitem

%%% 28
\bibitem{laudadio2018quantitative}
\begin{barticle}
\bauthor{\bsnm{Laudadio}, \binits{I.}},
\bauthor{\bsnm{Fulci}, \binits{V.}},
\bauthor{\bsnm{Palone}, \binits{F.}},
\bauthor{\bsnm{Stronati}, \binits{L.}},
\bauthor{\bsnm{Cucchiara}, \binits{S.}},
\bauthor{\bsnm{Carissimi}, \binits{C.}}:
\batitle{Quantitative assessment of shotgun metagenomics and 16s rdna amplicon
  sequencing in the study of human gut microbiome}.
\bjtitle{OMICS: A Journal of Integrative Biology}
\bvolume{22}(\bissue{4}),
\bfpage{248}--\blpage{254}
(\byear{2018})
\end{barticle}
\endbibitem

%%% 29
\bibitem{Kolmogorov2020-gu}
\begin{barticle}
\bauthor{\bsnm{Kolmogorov}, \binits{M.}},
\bauthor{\bsnm{Bickhart}, \binits{D.M.}},
\bauthor{\bsnm{Behsaz}, \binits{B.}},
\bauthor{\bsnm{Gurevich}, \binits{A.}},
\bauthor{\bsnm{Rayko}, \binits{M.}},
\bauthor{\bsnm{Shin}, \binits{S.B.}},
\bauthor{\bsnm{Kuhn}, \binits{K.}},
\bauthor{\bsnm{Yuan}, \binits{J.}},
\bauthor{\bsnm{Polevikov}, \binits{E.}},
\bauthor{\bsnm{Smith}, \binits{T.P.L.}},
\bauthor{\bsnm{Pevzner}, \binits{P.A.}}:
\batitle{{metaFlye}: scalable long-read metagenome assembly using repeat
  graphs}.
\bjtitle{Nat. Methods}
\bvolume{17}(\bissue{11}),
\bfpage{1103}--\blpage{1110}
(\byear{2020})
\end{barticle}
\endbibitem

%%% 30
\bibitem{Nurk2020-zh}
\begin{barticle}
\bauthor{\bsnm{Nurk}, \binits{S.}},
\bauthor{\bsnm{Walenz}, \binits{B.P.}},
\bauthor{\bsnm{Rhie}, \binits{A.}},
\bauthor{\bsnm{Vollger}, \binits{M.R.}},
\bauthor{\bsnm{Logsdon}, \binits{G.A.}},
\bauthor{\bsnm{Grothe}, \binits{R.}},
\bauthor{\bsnm{Miga}, \binits{K.H.}},
\bauthor{\bsnm{Eichler}, \binits{E.E.}},
\bauthor{\bsnm{Phillippy}, \binits{A.M.}},
\bauthor{\bsnm{Koren}, \binits{S.}}:
\batitle{{HiCanu}: accurate assembly of segmental duplications, satellites, and
  allelic variants from high-fidelity long reads}.
\bjtitle{Genome Res.}
\bvolume{30}(\bissue{9}),
\bfpage{1291}--\blpage{1305}
(\byear{2020})
\end{barticle}
\endbibitem

%%% 31
\bibitem{feng2022metagenome}
\begin{botherref}
\oauthor{\bsnm{Feng}, \binits{X.}},
\oauthor{\bsnm{Cheng}, \binits{H.}},
\oauthor{\bsnm{Portik}, \binits{D.}},
\oauthor{\bsnm{Li}, \binits{H.}}:
Metagenome assembly of high-fidelity long reads with hifiasm-meta.
Nature Methods,
1--4
(2022)
\end{botherref}
\endbibitem

%%% 32
\bibitem{kang2019metabat}
\begin{barticle}
\bauthor{\bsnm{Kang}, \binits{D.D.}},
\bauthor{\bsnm{Li}, \binits{F.}},
\bauthor{\bsnm{Kirton}, \binits{E.}},
\bauthor{\bsnm{Thomas}, \binits{A.}},
\bauthor{\bsnm{Egan}, \binits{R.}},
\bauthor{\bsnm{An}, \binits{H.}},
\bauthor{\bsnm{Wang}, \binits{Z.}}:
\batitle{{MetaBAT} 2: an adaptive binning algorithm for robust and efficient
  genome reconstruction from metagenome assemblies}.
\bjtitle{PeerJ}
\bvolume{7},
\bfpage{7359}
(\byear{2019})
\end{barticle}
\endbibitem

%%% 33
\bibitem{nissen2021improved}
\begin{barticle}
\bauthor{\bsnm{Nissen}, \binits{J.N.}},
\bauthor{\bsnm{Johansen}, \binits{J.}},
\bauthor{\bsnm{Alles{\o}e}, \binits{R.L.}},
\bauthor{\bsnm{S{\o}nderby}, \binits{C.K.}},
\bauthor{\bsnm{Armenteros}, \binits{J.J.A.}},
\bauthor{\bsnm{Gr{\o}nbech}, \binits{C.H.}},
\bauthor{\bsnm{Jensen}, \binits{L.J.}},
\bauthor{\bsnm{Nielsen}, \binits{H.B.}},
\bauthor{\bsnm{Petersen}, \binits{T.N.}},
\bauthor{\bsnm{Winther}, \binits{O.}}, \betal:
\batitle{Improved metagenome binning and assembly using deep variational
  autoencoders}.
\bjtitle{Nature biotechnology}
\bvolume{39}(\bissue{5}),
\bfpage{555}--\blpage{560}
(\byear{2021})
\end{barticle}
\endbibitem

%%% 34
\bibitem{mallawaarachchi2020graphbin}
\begin{barticle}
\bauthor{\bsnm{Mallawaarachchi}, \binits{V.}},
\bauthor{\bsnm{Wickramarachchi}, \binits{A.}},
\bauthor{\bsnm{Lin}, \binits{Y.}}:
\batitle{Graphbin: refined binning of metagenomic contigs using assembly
  graphs}.
\bjtitle{Bioinformatics}
\bvolume{36}(\bissue{11}),
\bfpage{3307}--\blpage{3313}
(\byear{2020})
\end{barticle}
\endbibitem

%%% 35
\bibitem{wu2014maxbin}
\begin{barticle}
\bauthor{\bsnm{Wu}, \binits{Y.-W.}},
\bauthor{\bsnm{Tang}, \binits{Y.-H.}},
\bauthor{\bsnm{Tringe}, \binits{S.G.}},
\bauthor{\bsnm{Simmons}, \binits{B.A.}},
\bauthor{\bsnm{Singer}, \binits{S.W.}}:
\batitle{Maxbin: an automated binning method to recover individual genomes from
  metagenomes using an expectation-maximization algorithm}.
\bjtitle{Microbiome}
\bvolume{2}(\bissue{1}),
\bfpage{1}--\blpage{18}
(\byear{2014})
\end{barticle}
\endbibitem

%%% 36
\bibitem{sieber2018recovery}
\begin{barticle}
\bauthor{\bsnm{Sieber}, \binits{C.M.}},
\bauthor{\bsnm{Probst}, \binits{A.J.}},
\bauthor{\bsnm{Sharrar}, \binits{A.}},
\bauthor{\bsnm{Thomas}, \binits{B.C.}},
\bauthor{\bsnm{Hess}, \binits{M.}},
\bauthor{\bsnm{Tringe}, \binits{S.G.}},
\bauthor{\bsnm{Banfield}, \binits{J.F.}}:
\batitle{Recovery of genomes from metagenomes via a dereplication, aggregation
  and scoring strategy}.
\bjtitle{Nature microbiology}
\bvolume{3}(\bissue{7}),
\bfpage{836}--\blpage{843}
(\byear{2018})
\end{barticle}
\endbibitem

%%% 37
\bibitem{wick2015bandage}
\begin{barticle}
\bauthor{\bsnm{Wick}, \binits{R.R.}},
\bauthor{\bsnm{Schultz}, \binits{M.B.}},
\bauthor{\bsnm{Zobel}, \binits{J.}},
\bauthor{\bsnm{Holt}, \binits{K.E.}}:
\batitle{Bandage: interactive visualization of de novo genome assemblies}.
\bjtitle{Bioinformatics}
\bvolume{31}(\bissue{20}),
\bfpage{3350}--\blpage{3352}
(\byear{2015})
\end{barticle}
\endbibitem

%%% 38
\bibitem{ondov2016mash}
\begin{barticle}
\bauthor{\bsnm{Ondov}, \binits{B.D.}},
\bauthor{\bsnm{Treangen}, \binits{T.J.}},
\bauthor{\bsnm{Melsted}, \binits{P.}},
\bauthor{\bsnm{Mallonee}, \binits{A.B.}},
\bauthor{\bsnm{Bergman}, \binits{N.H.}},
\bauthor{\bsnm{Koren}, \binits{S.}},
\bauthor{\bsnm{Phillippy}, \binits{A.M.}}:
\batitle{Mash: fast genome and metagenome distance estimation using minhash}.
\bjtitle{Genome biology}
\bvolume{17}(\bissue{1}),
\bfpage{1}--\blpage{14}
(\byear{2016})
\end{barticle}
\endbibitem

%%% 39
\bibitem{jain2018high}
\begin{barticle}
\bauthor{\bsnm{Jain}, \binits{C.}},
\bauthor{\bsnm{Rodriguez-R}, \binits{L.M.}},
\bauthor{\bsnm{Phillippy}, \binits{A.M.}},
\bauthor{\bsnm{Konstantinidis}, \binits{K.T.}},
\bauthor{\bsnm{Aluru}, \binits{S.}}:
\batitle{High throughput ani analysis of 90k prokaryotic genomes reveals clear
  species boundaries}.
\bjtitle{Nature communications}
\bvolume{9}(\bissue{1}),
\bfpage{1}--\blpage{8}
(\byear{2018})
\end{barticle}
\endbibitem

%%% 40
\bibitem{parks2020complete}
\begin{barticle}
\bauthor{\bsnm{Parks}, \binits{D.H.}},
\bauthor{\bsnm{Chuvochina}, \binits{M.}},
\bauthor{\bsnm{Chaumeil}, \binits{P.-A.}},
\bauthor{\bsnm{Rinke}, \binits{C.}},
\bauthor{\bsnm{Mussig}, \binits{A.J.}},
\bauthor{\bsnm{Hugenholtz}, \binits{P.}}:
\batitle{A complete domain-to-species taxonomy for bacteria and archaea}.
\bjtitle{Nature biotechnology}
\bvolume{38}(\bissue{9}),
\bfpage{1079}--\blpage{1086}
(\byear{2020})
\end{barticle}
\endbibitem

%%% 41
\bibitem{bickhart2022generating}
\begin{botherref}
\oauthor{\bsnm{Bickhart}, \binits{D.M.}},
\oauthor{\bsnm{Kolmogorov}, \binits{M.}},
\oauthor{\bsnm{Tseng}, \binits{E.}},
\oauthor{\bsnm{Portik}, \binits{D.M.}},
\oauthor{\bsnm{Korobeynikov}, \binits{A.}},
\oauthor{\bsnm{Tolstoganov}, \binits{I.}},
\oauthor{\bsnm{Uritskiy}, \binits{G.}},
\oauthor{\bsnm{Liachko}, \binits{I.}},
\oauthor{\bsnm{Sullivan}, \binits{S.T.}},
\oauthor{\bsnm{Shin}, \binits{S.B.}}, et al.:
Generating lineage-resolved, complete metagenome-assembled genomes from complex
  microbial communities.
Nature biotechnology,
1--9
(2022)
\end{botherref}
\endbibitem

%%% 42
\bibitem{ekim2021minimizer}
\begin{barticle}
\bauthor{\bsnm{Ekim}, \binits{B.}},
\bauthor{\bsnm{Berger}, \binits{B.}},
\bauthor{\bsnm{Chikhi}, \binits{R.}}:
\batitle{Minimizer-space de bruijn graphs: Whole-genome assembly of long reads
  in minutes on a personal computer}.
\bjtitle{Cell systems}
\bvolume{12}(\bissue{10}),
\bfpage{958}--\blpage{968}
(\byear{2021})
\end{barticle}
\endbibitem

%%% 43
\bibitem{mapleson2017kat}
\begin{barticle}
\bauthor{\bsnm{Mapleson}, \binits{D.}},
\bauthor{\bsnm{Garcia~Accinelli}, \binits{G.}},
\bauthor{\bsnm{Kettleborough}, \binits{G.}},
\bauthor{\bsnm{Wright}, \binits{J.}},
\bauthor{\bsnm{Clavijo}, \binits{B.J.}}:
\batitle{Kat: a k-mer analysis toolkit to quality control ngs datasets and
  genome assemblies}.
\bjtitle{Bioinformatics}
\bvolume{33}(\bissue{4}),
\bfpage{574}--\blpage{576}
(\byear{2017})
\end{barticle}
\endbibitem

%%% 44
\bibitem{rhie2020merqury}
\begin{barticle}
\bauthor{\bsnm{Rhie}, \binits{A.}},
\bauthor{\bsnm{Walenz}, \binits{B.P.}},
\bauthor{\bsnm{Koren}, \binits{S.}},
\bauthor{\bsnm{Phillippy}, \binits{A.M.}}:
\batitle{Merqury: reference-free quality, completeness, and phasing assessment
  for genome assemblies}.
\bjtitle{Genome biology}
\bvolume{21}(\bissue{1}),
\bfpage{1}--\blpage{27}
(\byear{2020})
\end{barticle}
\endbibitem

%%% 45
\bibitem{li2009fast}
\begin{barticle}
\bauthor{\bsnm{Li}, \binits{H.}},
\bauthor{\bsnm{Durbin}, \binits{R.}}:
\batitle{Fast and accurate short read alignment with burrows--wheeler
  transform}.
\bjtitle{bioinformatics}
\bvolume{25}(\bissue{14}),
\bfpage{1754}--\blpage{1760}
(\byear{2009})
\end{barticle}
\endbibitem

%%% 46
\bibitem{seemann2018barrnap}
\begin{botherref}
\oauthor{\bsnm{Seemann}, \binits{T.}},
\oauthor{\bsnm{Booth}, \binits{T.}}:
Barrnap: BAsic Rapid Ribosomal RNA Predictor
(2018)
\end{botherref}
\endbibitem

%%% 47
\bibitem{nawrocki2013infernal}
\begin{barticle}
\bauthor{\bsnm{Nawrocki}, \binits{E.P.}},
\bauthor{\bsnm{Eddy}, \binits{S.R.}}:
\batitle{Infernal 1.1: 100-fold faster rna homology searches}.
\bjtitle{Bioinformatics}
\bvolume{29}(\bissue{22}),
\bfpage{2933}--\blpage{2935}
(\byear{2013})
\end{barticle}
\endbibitem

%%% 48
\bibitem{wang2007naive}
\begin{barticle}
\bauthor{\bsnm{Wang}, \binits{Q.}},
\bauthor{\bsnm{Garrity}, \binits{G.M.}},
\bauthor{\bsnm{Tiedje}, \binits{J.M.}},
\bauthor{\bsnm{Cole}, \binits{J.R.}}:
\batitle{Naive bayesian classifier for rapid assignment of rrna sequences into
  the new bacterial taxonomy}.
\bjtitle{Applied and environmental microbiology}
\bvolume{73}(\bissue{16}),
\bfpage{5261}--\blpage{5267}
(\byear{2007})
\end{barticle}
\endbibitem

%%% 49
\bibitem{edgar2018accuracy}
\begin{barticle}
\bauthor{\bsnm{Edgar}, \binits{R.C.}}:
\batitle{Accuracy of taxonomy prediction for 16s rrna and fungal its
  sequences}.
\bjtitle{PeerJ}
\bvolume{6},
\bfpage{4652}
(\byear{2018})
\end{barticle}
\endbibitem

%%% 50
\bibitem{meziti2021reliability}
\begin{barticle}
\bauthor{\bsnm{Meziti}, \binits{A.}},
\bauthor{\bsnm{Rodriguez-R}, \binits{L.M.}},
\bauthor{\bsnm{Hatt}, \binits{J.K.}},
\bauthor{\bsnm{Pe{\~n}a-Gonzalez}, \binits{A.}},
\bauthor{\bsnm{Levy}, \binits{K.}},
\bauthor{\bsnm{Konstantinidis}, \binits{K.T.}}:
\batitle{The reliability of metagenome-assembled genomes (mags) in representing
  natural populations: Insights from comparing mags against isolate genomes
  derived from the same fecal sample}.
\bjtitle{Applied and environmental microbiology}
\bvolume{87}(\bissue{6}),
\bfpage{02593}--\blpage{20}
(\byear{2021})
\end{barticle}
\endbibitem

%%% 51
\bibitem{jin2022hybrid}
\begin{barticle}
\bauthor{\bsnm{Jin}, \binits{H.}},
\bauthor{\bsnm{You}, \binits{L.}},
\bauthor{\bsnm{Zhao}, \binits{F.}},
\bauthor{\bsnm{Li}, \binits{S.}},
\bauthor{\bsnm{Ma}, \binits{T.}},
\bauthor{\bsnm{Kwok}, \binits{L.-Y.}},
\bauthor{\bsnm{Xu}, \binits{H.}},
\bauthor{\bsnm{Sun}, \binits{Z.}}:
\batitle{Hybrid, ultra-deep metagenomic sequencing enables genomic and
  functional characterization of low-abundance species in the human gut
  microbiome}.
\bjtitle{Gut microbes}
\bvolume{14}(\bissue{1}),
\bfpage{2021790}
(\byear{2022})
\end{barticle}
\endbibitem

%%% 52
\bibitem{almeida2021unified}
\begin{barticle}
\bauthor{\bsnm{Almeida}, \binits{A.}},
\bauthor{\bsnm{Nayfach}, \binits{S.}},
\bauthor{\bsnm{Boland}, \binits{M.}},
\bauthor{\bsnm{Strozzi}, \binits{F.}},
\bauthor{\bsnm{Beracochea}, \binits{M.}},
\bauthor{\bsnm{Shi}, \binits{Z.J.}},
\bauthor{\bsnm{Pollard}, \binits{K.S.}},
\bauthor{\bsnm{Sakharova}, \binits{E.}},
\bauthor{\bsnm{Parks}, \binits{D.H.}},
\bauthor{\bsnm{Hugenholtz}, \binits{P.}}, \betal:
\batitle{A unified catalog of 204,938 reference genomes from the human gut
  microbiome}.
\bjtitle{Nature biotechnology}
\bvolume{39}(\bissue{1}),
\bfpage{105}--\blpage{114}
(\byear{2021})
\end{barticle}
\endbibitem

%%% 53
\bibitem{feng2021metagenome}
\begin{barticle}
\bauthor{\bsnm{Feng}, \binits{Y.}},
\bauthor{\bsnm{Wang}, \binits{Y.}},
\bauthor{\bsnm{Zhu}, \binits{B.}},
\bauthor{\bsnm{Gao}, \binits{G.F.}},
\bauthor{\bsnm{Guo}, \binits{Y.}},
\bauthor{\bsnm{Hu}, \binits{Y.}}:
\batitle{Metagenome-assembled genomes and gene catalog from the chicken gut
  microbiome aid in deciphering antibiotic resistomes}.
\bjtitle{Communications biology}
\bvolume{4}(\bissue{1}),
\bfpage{1}--\blpage{9}
(\byear{2021})
\end{barticle}
\endbibitem

%%% 54
\bibitem{xie2021integrated}
\begin{barticle}
\bauthor{\bsnm{Xie}, \binits{F.}},
\bauthor{\bsnm{Jin}, \binits{W.}},
\bauthor{\bsnm{Si}, \binits{H.}},
\bauthor{\bsnm{Yuan}, \binits{Y.}},
\bauthor{\bsnm{Tao}, \binits{Y.}},
\bauthor{\bsnm{Liu}, \binits{J.}},
\bauthor{\bsnm{Wang}, \binits{X.}},
\bauthor{\bsnm{Yang}, \binits{C.}},
\bauthor{\bsnm{Li}, \binits{Q.}},
\bauthor{\bsnm{Yan}, \binits{X.}}, \betal:
\batitle{An integrated gene catalog and over 10,000 metagenome-assembled
  genomes from the gastrointestinal microbiome of ruminants}.
\bjtitle{Microbiome}
\bvolume{9}(\bissue{1}),
\bfpage{1}--\blpage{20}
(\byear{2021})
\end{barticle}
\endbibitem

%%% 55
\bibitem{li2018minimap2}
\begin{barticle}
\bauthor{\bsnm{Li}, \binits{H.}}:
\batitle{Minimap2: pairwise alignment for nucleotide sequences}.
\bjtitle{Bioinformatics}
\bvolume{34}(\bissue{18}),
\bfpage{3094}--\blpage{3100}
(\byear{2018})
\end{barticle}
\endbibitem

%%% 56
\bibitem{li2009sequence}
\begin{barticle}
\bauthor{\bsnm{Li}, \binits{H.}},
\bauthor{\bsnm{Handsaker}, \binits{B.}},
\bauthor{\bsnm{Wysoker}, \binits{A.}},
\bauthor{\bsnm{Fennell}, \binits{T.}},
\bauthor{\bsnm{Ruan}, \binits{J.}},
\bauthor{\bsnm{Homer}, \binits{N.}},
\bauthor{\bsnm{Marth}, \binits{G.}},
\bauthor{\bsnm{Abecasis}, \binits{G.}},
\bauthor{\bsnm{Durbin}, \binits{R.}}:
\batitle{The sequence alignment/map format and samtools}.
\bjtitle{Bioinformatics}
\bvolume{25}(\bissue{16}),
\bfpage{2078}--\blpage{2079}
(\byear{2009})
\end{barticle}
\endbibitem

%%% 57
\bibitem{ono2021pbsim2}
\begin{barticle}
\bauthor{\bsnm{Ono}, \binits{Y.}},
\bauthor{\bsnm{Asai}, \binits{K.}},
\bauthor{\bsnm{Hamada}, \binits{M.}}:
\batitle{Pbsim2: a simulator for long-read sequencers with a novel generative
  model of quality scores}.
\bjtitle{Bioinformatics}
\bvolume{37}(\bissue{5}),
\bfpage{589}--\blpage{595}
(\byear{2021})
\end{barticle}
\endbibitem

%%% 58
\bibitem{sereika2022oxford}
\begin{barticle}
\bauthor{\bsnm{Sereika}, \binits{M.}},
\bauthor{\bsnm{Kirkegaard}, \binits{R.H.}},
\bauthor{\bsnm{Karst}, \binits{S.M.}},
\bauthor{\bsnm{Michaelsen}, \binits{T.Y.}},
\bauthor{\bsnm{S{\o}rensen}, \binits{E.A.}},
\bauthor{\bsnm{Wollenberg}, \binits{R.D.}},
\bauthor{\bsnm{Albertsen}, \binits{M.}}:
\batitle{Oxford nanopore r10. 4 long-read sequencing enables the generation of
  near-finished bacterial genomes from pure cultures and metagenomes without
  short-read or reference polishing}.
\bjtitle{Nature methods}
\bvolume{19}(\bissue{7}),
\bfpage{823}--\blpage{826}
(\byear{2022})
\end{barticle}
\endbibitem

%%% 59
\bibitem{kim2022hifi}
\begin{botherref}
\oauthor{\bsnm{Kim}, \binits{C.Y.}},
\oauthor{\bsnm{Ma}, \binits{J.}},
\oauthor{\bsnm{Lee}, \binits{I.}}:
Hifi metagenomic sequencing enables assembly of accurate and complete genomes
  from human gut microbiota.
bioRxiv
(2022)
\end{botherref}
\endbibitem

\end{thebibliography}

\newcommand{\BMCxmlcomment}[1]{}

\BMCxmlcomment{

<refgrp>

<bibl id="B1">
  <title><p>The reconstruction of 2,631 draft metagenome-assembled genomes from
  the global oceans</p></title>
  <aug>
    <au><snm>Tully</snm><fnm>BJ</fnm></au>
    <au><snm>Graham</snm><fnm>ED</fnm></au>
    <au><snm>Heidelberg</snm><fnm>JF</fnm></au>
  </aug>
  <source>Scientific data</source>
  <publisher>Nature Publishing Group</publisher>
  <pubdate>2018</pubdate>
  <volume>5</volume>
  <issue>1</issue>
  <fpage>1</fpage>
  <lpage>-8</lpage>
</bibl>

<bibl id="B2">
  <title><p>Tackling soil diversity with the assembly of large, complex
  metagenomes</p></title>
  <aug>
    <au><snm>Howe</snm><fnm>AC</fnm></au>
    <au><snm>Jansson</snm><fnm>JK</fnm></au>
    <au><snm>Malfatti</snm><fnm>SA</fnm></au>
    <au><snm>Tringe</snm><fnm>SG</fnm></au>
    <au><snm>Tiedje</snm><fnm>JM</fnm></au>
    <au><snm>Brown</snm><fnm>CT</fnm></au>
  </aug>
  <source>Proceedings of the National Academy of Sciences</source>
  <publisher>National Acad Sciences</publisher>
  <pubdate>2014</pubdate>
  <volume>111</volume>
  <issue>13</issue>
  <fpage>4904</fpage>
  <lpage>-4909</lpage>
</bibl>

<bibl id="B3">
  <title><p>New biological insights into how deforestation in Amazonia affects
  soil microbial communities using metagenomics and metagenome-assembled
  genomes</p></title>
  <aug>
    <au><snm>Kroeger</snm><fnm>ME</fnm></au>
    <au><snm>Delmont</snm><fnm>TO</fnm></au>
    <au><snm>Eren</snm><fnm>AM</fnm></au>
    <au><snm>Meyer</snm><fnm>KM</fnm></au>
    <au><snm>Guo</snm><fnm>J</fnm></au>
    <au><snm>Khan</snm><fnm>K</fnm></au>
    <au><snm>Rodrigues</snm><fnm>JL</fnm></au>
    <au><snm>Bohannan</snm><fnm>BJ</fnm></au>
    <au><snm>Tringe</snm><fnm>SG</fnm></au>
    <au><snm>Borges</snm><fnm>CD</fnm></au>
    <au><cnm>others</cnm></au>
  </aug>
  <source>Frontiers in microbiology</source>
  <publisher>Frontiers Media SA</publisher>
  <pubdate>2018</pubdate>
  <volume>9</volume>
  <fpage>1635</fpage>
</bibl>

<bibl id="B4">
  <title><p>{CheckM}: assessing the quality of microbial genomes recovered from
  isolates, single cells, and metagenomes</p></title>
  <aug>
    <au><snm>Parks</snm><fnm>DH</fnm></au>
    <au><snm>Imelfort</snm><fnm>M</fnm></au>
    <au><snm>Skennerton</snm><fnm>CT</fnm></au>
    <au><snm>Hugenholtz</snm><fnm>P</fnm></au>
    <au><snm>Tyson</snm><fnm>GW</fnm></au>
  </aug>
  <source>Genome Res.</source>
  <pubdate>2015</pubdate>
  <volume>25</volume>
  <issue>7</issue>
  <fpage>1043</fpage>
  <lpage>-1055</lpage>
</bibl>

<bibl id="B5">
  <title><p>Minimum information about a single amplified genome (MISAG) and a
  metagenome-assembled genome (MIMAG) of bacteria and archaea</p></title>
  <aug>
    <au><snm>Bowers</snm><fnm>RM</fnm></au>
    <au><snm>Kyrpides</snm><fnm>NC</fnm></au>
    <au><snm>Stepanauskas</snm><fnm>R</fnm></au>
    <au><snm>Harmon Smith</snm><fnm>M</fnm></au>
    <au><snm>Doud</snm><fnm>D</fnm></au>
    <au><snm>Reddy</snm><fnm>TBK</fnm></au>
    <au><snm>Schulz</snm><fnm>F</fnm></au>
    <au><snm>Jarett</snm><fnm>J</fnm></au>
    <au><snm>Rivers</snm><fnm>AR</fnm></au>
    <au><snm>Eloe Fadrosh</snm><fnm>EA</fnm></au>
    <au><cnm>others</cnm></au>
  </aug>
  <source>Nature biotechnology</source>
  <publisher>Nature Publishing Group</publisher>
  <pubdate>2017</pubdate>
  <volume>35</volume>
  <issue>8</issue>
  <fpage>725</fpage>
  <lpage>-731</lpage>
</bibl>

<bibl id="B6">
  <title><p>Genome sequences of rare, uncultured bacteria obtained by
  differential coverage binning of multiple metagenomes</p></title>
  <aug>
    <au><snm>Albertsen</snm><fnm>M</fnm></au>
    <au><snm>Hugenholtz</snm><fnm>P</fnm></au>
    <au><snm>Skarshewski</snm><fnm>A</fnm></au>
    <au><snm>Nielsen</snm><fnm>KL</fnm></au>
    <au><snm>Tyson</snm><fnm>GW</fnm></au>
    <au><snm>Nielsen</snm><fnm>PH</fnm></au>
  </aug>
  <source>Nature biotechnology</source>
  <publisher>Nature Publishing Group</publisher>
  <pubdate>2013</pubdate>
  <volume>31</volume>
  <issue>6</issue>
  <fpage>533</fpage>
  <lpage>-538</lpage>
</bibl>

<bibl id="B7">
  <title><p>Strainberry: automated strain separation in low-complexity
  metagenomes using long reads</p></title>
  <aug>
    <au><snm>Vicedomini</snm><fnm>R</fnm></au>
    <au><snm>Quince</snm><fnm>C</fnm></au>
    <au><snm>Darling</snm><fnm>AE</fnm></au>
    <au><snm>Chikhi</snm><fnm>R</fnm></au>
  </aug>
  <source>Nature Communications</source>
  <publisher>Nature Publishing Group</publisher>
  <pubdate>2021</pubdate>
  <volume>12</volume>
  <issue>1</issue>
  <fpage>1</fpage>
  <lpage>-14</lpage>
</bibl>

<bibl id="B8">
  <title><p>New insights from uncultivated genomes of the global human gut
  microbiome</p></title>
  <aug>
    <au><snm>Nayfach</snm><fnm>S</fnm></au>
    <au><snm>Shi</snm><fnm>ZJ</fnm></au>
    <au><snm>Seshadri</snm><fnm>R</fnm></au>
    <au><snm>Pollard</snm><fnm>KS</fnm></au>
    <au><snm>Kyrpides</snm><fnm>NC</fnm></au>
  </aug>
  <source>Nature</source>
  <publisher>Nature Publishing Group</publisher>
  <pubdate>2019</pubdate>
  <volume>568</volume>
  <issue>7753</issue>
  <fpage>505</fpage>
  <lpage>-510</lpage>
</bibl>

<bibl id="B9">
  <title><p>Individual genome assembly from complex community short-read
  metagenomic datasets</p></title>
  <aug>
    <au><snm>Luo</snm><fnm>C</fnm></au>
    <au><snm>Tsementzi</snm><fnm>D</fnm></au>
    <au><snm>Kyrpides</snm><fnm>NC</fnm></au>
    <au><snm>Konstantinidis</snm><fnm>KT</fnm></au>
  </aug>
  <source>The ISME journal</source>
  <publisher>Nature Publishing Group</publisher>
  <pubdate>2012</pubdate>
  <volume>6</volume>
  <issue>4</issue>
  <fpage>898</fpage>
  <lpage>-901</lpage>
</bibl>

<bibl id="B10">
  <title><p>Taxonomic note: a place for DNA-DNA reassociation and 16S rRNA
  sequence analysis in the present species definition in
  bacteriology</p></title>
  <aug>
    <au><snm>Stackebrandt</snm><fnm>E</fnm></au>
    <au><snm>Goebel</snm><fnm>BM</fnm></au>
  </aug>
  <source>International journal of systematic and evolutionary
  microbiology</source>
  <publisher>Microbiology Society</publisher>
  <pubdate>1994</pubdate>
  <volume>44</volume>
  <issue>4</issue>
  <fpage>846</fpage>
  <lpage>-849</lpage>
</bibl>

<bibl id="B11">
  <title><p>Microbial resolution of whole genome shotgun and 16S amplicon
  metagenomic sequencing using publicly available NEON data</p></title>
  <aug>
    <au><snm>Brumfield</snm><fnm>KD</fnm></au>
    <au><snm>Huq</snm><fnm>A</fnm></au>
    <au><snm>Colwell</snm><fnm>RR</fnm></au>
    <au><snm>Olds</snm><fnm>JL</fnm></au>
    <au><snm>Leddy</snm><fnm>MB</fnm></au>
  </aug>
  <source>PLoS One</source>
  <publisher>Public Library of Science San Francisco, CA USA</publisher>
  <pubdate>2020</pubdate>
  <volume>15</volume>
  <issue>2</issue>
  <fpage>e0228899</fpage>
</bibl>

<bibl id="B12">
  <title><p>Updating the 97\% identity threshold for 16S ribosomal RNA
  OTUs</p></title>
  <aug>
    <au><snm>Edgar</snm><fnm>RC</fnm></au>
  </aug>
  <source>Bioinformatics</source>
  <publisher>Oxford University Press</publisher>
  <pubdate>2018</pubdate>
  <volume>34</volume>
  <issue>14</issue>
  <fpage>2371</fpage>
  <lpage>-2375</lpage>
</bibl>

<bibl id="B13">
  <title><p>Reconstructing 16S rRNA genes in metagenomic data</p></title>
  <aug>
    <au><snm>Yuan</snm><fnm>C</fnm></au>
    <au><snm>Lei</snm><fnm>J</fnm></au>
    <au><snm>Cole</snm><fnm>J</fnm></au>
    <au><snm>Sun</snm><fnm>Y</fnm></au>
  </aug>
  <source>Bioinformatics</source>
  <publisher>Oxford University Press</publisher>
  <pubdate>2015</pubdate>
  <volume>31</volume>
  <issue>12</issue>
  <fpage>i35</fpage>
  <lpage>-i43</lpage>
</bibl>

<bibl id="B14">
  <title><p>Strengths and limitations of 16S rRNA gene amplicon sequencing in
  revealing temporal microbial community dynamics</p></title>
  <aug>
    <au><snm>Poretsky</snm><fnm>R</fnm></au>
    <au><snm>Rodriguez R</snm><fnm>LM</fnm></au>
    <au><snm>Luo</snm><fnm>C</fnm></au>
    <au><snm>Tsementzi</snm><fnm>D</fnm></au>
    <au><snm>Konstantinidis</snm><fnm>KT</fnm></au>
  </aug>
  <source>PloS one</source>
  <publisher>Public Library of Science San Francisco, USA</publisher>
  <pubdate>2014</pubdate>
  <volume>9</volume>
  <issue>4</issue>
  <fpage>e93827</fpage>
</bibl>

<bibl id="B15">
  <title><p>Rapid bacterial identification by direct PCR amplification of 16S
  rRNA genes using the MinION™ nanopore sequencer</p></title>
  <aug>
    <au><snm>Kai</snm><fnm>S</fnm></au>
    <au><snm>Matsuo</snm><fnm>Y</fnm></au>
    <au><snm>Nakagawa</snm><fnm>S</fnm></au>
    <au><snm>Kryukov</snm><fnm>K</fnm></au>
    <au><snm>Matsukawa</snm><fnm>S</fnm></au>
    <au><snm>Tanaka</snm><fnm>H</fnm></au>
    <au><snm>Iwai</snm><fnm>T</fnm></au>
    <au><snm>Imanishi</snm><fnm>T</fnm></au>
    <au><snm>Hirota</snm><fnm>K</fnm></au>
  </aug>
  <source>FEBS open bio</source>
  <publisher>Wiley Online Library</publisher>
  <pubdate>2019</pubdate>
  <volume>9</volume>
  <issue>3</issue>
  <fpage>548</fpage>
  <lpage>-557</lpage>
</bibl>

<bibl id="B16">
  <title><p>Evaluation of 16S rRNA gene sequencing for species and strain-level
  microbiome analysis</p></title>
  <aug>
    <au><snm>Johnson</snm><fnm>JS</fnm></au>
    <au><snm>Spakowicz</snm><fnm>DJ</fnm></au>
    <au><snm>Hong</snm><fnm>BY</fnm></au>
    <au><snm>Petersen</snm><fnm>LM</fnm></au>
    <au><snm>Demkowicz</snm><fnm>P</fnm></au>
    <au><snm>Chen</snm><fnm>L</fnm></au>
    <au><snm>Leopold</snm><fnm>SR</fnm></au>
    <au><snm>Hanson</snm><fnm>BM</fnm></au>
    <au><snm>Agresta</snm><fnm>HO</fnm></au>
    <au><snm>Gerstein</snm><fnm>M</fnm></au>
    <au><cnm>others</cnm></au>
  </aug>
  <source>Nature communications</source>
  <publisher>Nature Publishing Group</publisher>
  <pubdate>2019</pubdate>
  <volume>10</volume>
  <issue>1</issue>
  <fpage>1</fpage>
  <lpage>-11</lpage>
</bibl>

<bibl id="B17">
  <title><p>Metagenomic reconstructions of gut microbial metabolism in weanling
  pigs</p></title>
  <aug>
    <au><snm>Wang</snm><fnm>W</fnm></au>
    <au><snm>Hu</snm><fnm>H</fnm></au>
    <au><snm>Zijlstra</snm><fnm>RT</fnm></au>
    <au><snm>Zheng</snm><fnm>J</fnm></au>
    <au><snm>G{\"a}nzle</snm><fnm>MG</fnm></au>
  </aug>
  <source>Microbiome</source>
  <publisher>Springer</publisher>
  <pubdate>2019</pubdate>
  <volume>7</volume>
  <issue>1</issue>
  <fpage>1</fpage>
  <lpage>-11</lpage>
</bibl>

<bibl id="B18">
  <title><p>A new genomic blueprint of the human gut microbiota</p></title>
  <aug>
    <au><snm>Almeida</snm><fnm>A</fnm></au>
    <au><snm>Mitchell</snm><fnm>AL</fnm></au>
    <au><snm>Boland</snm><fnm>M</fnm></au>
    <au><snm>Forster</snm><fnm>SC</fnm></au>
    <au><snm>Gloor</snm><fnm>GB</fnm></au>
    <au><snm>Tarkowska</snm><fnm>A</fnm></au>
    <au><snm>Lawley</snm><fnm>TD</fnm></au>
    <au><snm>Finn</snm><fnm>RD</fnm></au>
  </aug>
  <source>Nature</source>
  <publisher>Nature Publishing Group</publisher>
  <pubdate>2019</pubdate>
  <volume>568</volume>
  <issue>7753</issue>
  <fpage>499</fpage>
  <lpage>-504</lpage>
</bibl>

<bibl id="B19">
  <title><p>Human reference gut microbiome catalog including newly assembled
  genomes from under-represented Asian metagenomes</p></title>
  <aug>
    <au><snm>Kim</snm><fnm>CY</fnm></au>
    <au><snm>Lee</snm><fnm>M</fnm></au>
    <au><snm>Yang</snm><fnm>S</fnm></au>
    <au><snm>Kim</snm><fnm>K</fnm></au>
    <au><snm>Yong</snm><fnm>D</fnm></au>
    <au><snm>Kim</snm><fnm>HR</fnm></au>
    <au><snm>Lee</snm><fnm>I</fnm></au>
  </aug>
  <source>Genome Medicine</source>
  <publisher>BioMed Central</publisher>
  <pubdate>2021</pubdate>
  <volume>13</volume>
  <issue>1</issue>
  <fpage>1</fpage>
  <lpage>-20</lpage>
</bibl>

<bibl id="B20">
  <title><p>Extensive unexplored human microbiome diversity revealed by over
  150,000 genomes from metagenomes spanning age, geography, and
  lifestyle</p></title>
  <aug>
    <au><snm>Pasolli</snm><fnm>E</fnm></au>
    <au><snm>Asnicar</snm><fnm>F</fnm></au>
    <au><snm>Manara</snm><fnm>S</fnm></au>
    <au><snm>Zolfo</snm><fnm>M</fnm></au>
    <au><snm>Karcher</snm><fnm>N</fnm></au>
    <au><snm>Armanini</snm><fnm>F</fnm></au>
    <au><snm>Beghini</snm><fnm>F</fnm></au>
    <au><snm>Manghi</snm><fnm>P</fnm></au>
    <au><snm>Tett</snm><fnm>A</fnm></au>
    <au><snm>Ghensi</snm><fnm>P</fnm></au>
    <au><cnm>others</cnm></au>
  </aug>
  <source>Cell</source>
  <publisher>Elsevier</publisher>
  <pubdate>2019</pubdate>
  <volume>176</volume>
  <issue>3</issue>
  <fpage>649</fpage>
  <lpage>-662</lpage>
</bibl>

<bibl id="B21">
  <title><p>Expanded catalog of microbial genes and metagenome-assembled
  genomes from the pig gut microbiome</p></title>
  <aug>
    <au><snm>Chen</snm><fnm>C</fnm></au>
    <au><snm>Zhou</snm><fnm>Y</fnm></au>
    <au><snm>Fu</snm><fnm>H</fnm></au>
    <au><snm>Xiong</snm><fnm>X</fnm></au>
    <au><snm>Fang</snm><fnm>S</fnm></au>
    <au><snm>Jiang</snm><fnm>H</fnm></au>
    <au><snm>Wu</snm><fnm>J</fnm></au>
    <au><snm>Yang</snm><fnm>H</fnm></au>
    <au><snm>Gao</snm><fnm>J</fnm></au>
    <au><snm>Huang</snm><fnm>L</fnm></au>
  </aug>
  <source>Nature communications</source>
  <publisher>Nature Publishing Group</publisher>
  <pubdate>2021</pubdate>
  <volume>12</volume>
  <issue>1</issue>
  <fpage>1</fpage>
  <lpage>-13</lpage>
</bibl>

<bibl id="B22">
  <title><p>Interplay between the human gut microbiome and host
  metabolism</p></title>
  <aug>
    <au><snm>Visconti</snm><fnm>A</fnm></au>
    <au><snm>Le Roy</snm><fnm>CI</fnm></au>
    <au><snm>Rosa</snm><fnm>F</fnm></au>
    <au><snm>Rossi</snm><fnm>N</fnm></au>
    <au><snm>Martin</snm><fnm>TC</fnm></au>
    <au><snm>Mohney</snm><fnm>RP</fnm></au>
    <au><snm>Li</snm><fnm>W</fnm></au>
    <au><snm>Rinaldis</snm><fnm>E</fnm></au>
    <au><snm>Bell</snm><fnm>JT</fnm></au>
    <au><snm>Venter</snm><fnm>JC</fnm></au>
    <au><cnm>others</cnm></au>
  </aug>
  <source>Nature communications</source>
  <publisher>Nature Publishing Group</publisher>
  <pubdate>2019</pubdate>
  <volume>10</volume>
  <issue>1</issue>
  <fpage>1</fpage>
  <lpage>-10</lpage>
</bibl>

<bibl id="B23">
  <title><p>Towards Strain-Level Complexity: Sequencing Depth Required for
  Comprehensive Single-Nucleotide Polymorphism Analysis of the Human Gut
  Microbiome</p></title>
  <aug>
    <au><snm>Liu</snm><fnm>P</fnm></au>
    <au><snm>Hu</snm><fnm>S</fnm></au>
    <au><snm>He</snm><fnm>Z</fnm></au>
    <au><snm>Feng</snm><fnm>C</fnm></au>
    <au><snm>Dong</snm><fnm>G</fnm></au>
    <au><snm>An</snm><fnm>S</fnm></au>
    <au><snm>Liu</snm><fnm>R</fnm></au>
    <au><snm>Xu</snm><fnm>F</fnm></au>
    <au><snm>Chen</snm><fnm>Y</fnm></au>
    <au><snm>Ying</snm><fnm>X</fnm></au>
  </aug>
  <source>Frontiers in Microbiology</source>
  <publisher>Frontiers Media SA</publisher>
  <pubdate>2022</pubdate>
  <volume>13</volume>
</bibl>

<bibl id="B24">
  <title><p>Evaluating the information content of shallow shotgun
  metagenomics</p></title>
  <aug>
    <au><snm>Hillmann</snm><fnm>B</fnm></au>
    <au><snm>Al Ghalith</snm><fnm>GA</fnm></au>
    <au><snm>Shields Cutler</snm><fnm>RR</fnm></au>
    <au><snm>Zhu</snm><fnm>Q</fnm></au>
    <au><snm>Gohl</snm><fnm>DM</fnm></au>
    <au><snm>Beckman</snm><fnm>KB</fnm></au>
    <au><snm>Knight</snm><fnm>R</fnm></au>
    <au><snm>Knights</snm><fnm>D</fnm></au>
  </aug>
  <source>Msystems</source>
  <publisher>Am Soc Microbiol</publisher>
  <pubdate>2018</pubdate>
  <volume>3</volume>
  <issue>6</issue>
  <fpage>e00069</fpage>
  <lpage>-18</lpage>
</bibl>

<bibl id="B25">
  <aug>
    <au><snm>Institute</snm><fnm>DJG</fnm></au>
  </aug>
  <source>\url{https://www.ncbi.nlm.nih.gov/bioproject/?term=Wetland+microbial+communities+from+Twitchell+Island}</source>
  <pubdate>2011</pubdate>
</bibl>

<bibl id="B26">
  <title><p>BlastKOALA and GhostKOALA: KEGG tools for functional
  characterization of genome and metagenome sequences</p></title>
  <aug>
    <au><snm>Kanehisa</snm><fnm>M</fnm></au>
    <au><snm>Sato</snm><fnm>Y</fnm></au>
    <au><snm>Morishima</snm><fnm>K</fnm></au>
  </aug>
  <source>Journal of molecular biology</source>
  <publisher>Elsevier</publisher>
  <pubdate>2016</pubdate>
  <volume>428</volume>
  <issue>4</issue>
  <fpage>726</fpage>
  <lpage>-731</lpage>
</bibl>

<bibl id="B27">
  <title><p>From bag-of-genes to bag-of-genomes: metabolic modelling of
  communities in the era of metagenome-assembled genomes</p></title>
  <aug>
    <au><snm>Frioux</snm><fnm>C</fnm></au>
    <au><snm>Singh</snm><fnm>D</fnm></au>
    <au><snm>Korcsmaros</snm><fnm>T</fnm></au>
    <au><snm>Hildebrand</snm><fnm>F</fnm></au>
  </aug>
  <source>Computational and Structural Biotechnology Journal</source>
  <publisher>Elsevier</publisher>
  <pubdate>2020</pubdate>
  <volume>18</volume>
  <fpage>1722</fpage>
  <lpage>-1734</lpage>
</bibl>

<bibl id="B28">
  <title><p>Quantitative assessment of shotgun metagenomics and 16S rDNA
  amplicon sequencing in the study of human gut microbiome</p></title>
  <aug>
    <au><snm>Laudadio</snm><fnm>I</fnm></au>
    <au><snm>Fulci</snm><fnm>V</fnm></au>
    <au><snm>Palone</snm><fnm>F</fnm></au>
    <au><snm>Stronati</snm><fnm>L</fnm></au>
    <au><snm>Cucchiara</snm><fnm>S</fnm></au>
    <au><snm>Carissimi</snm><fnm>C</fnm></au>
  </aug>
  <source>OMICS: A Journal of Integrative Biology</source>
  <publisher>Mary Ann Liebert, Inc. 140 Huguenot Street, 3rd Floor New
  Rochelle, NY 10801 USA</publisher>
  <pubdate>2018</pubdate>
  <volume>22</volume>
  <issue>4</issue>
  <fpage>248</fpage>
  <lpage>-254</lpage>
</bibl>

<bibl id="B29">
  <title><p>{metaFlye}: scalable long-read metagenome assembly using repeat
  graphs</p></title>
  <aug>
    <au><snm>Kolmogorov</snm><fnm>M</fnm></au>
    <au><snm>Bickhart</snm><fnm>DM</fnm></au>
    <au><snm>Behsaz</snm><fnm>B</fnm></au>
    <au><snm>Gurevich</snm><fnm>A</fnm></au>
    <au><snm>Rayko</snm><fnm>M</fnm></au>
    <au><snm>Shin</snm><fnm>SB</fnm></au>
    <au><snm>Kuhn</snm><fnm>K</fnm></au>
    <au><snm>Yuan</snm><fnm>J</fnm></au>
    <au><snm>Polevikov</snm><fnm>E</fnm></au>
    <au><snm>Smith</snm><fnm>TPL</fnm></au>
    <au><snm>Pevzner</snm><fnm>PA</fnm></au>
  </aug>
  <source>Nat. Methods</source>
  <pubdate>2020</pubdate>
  <volume>17</volume>
  <issue>11</issue>
  <fpage>1103</fpage>
  <lpage>-1110</lpage>
</bibl>

<bibl id="B30">
  <title><p>{HiCanu}: accurate assembly of segmental duplications, satellites,
  and allelic variants from high-fidelity long reads</p></title>
  <aug>
    <au><snm>Nurk</snm><fnm>S</fnm></au>
    <au><snm>Walenz</snm><fnm>BP</fnm></au>
    <au><snm>Rhie</snm><fnm>A</fnm></au>
    <au><snm>Vollger</snm><fnm>MR</fnm></au>
    <au><snm>Logsdon</snm><fnm>GA</fnm></au>
    <au><snm>Grothe</snm><fnm>R</fnm></au>
    <au><snm>Miga</snm><fnm>KH</fnm></au>
    <au><snm>Eichler</snm><fnm>EE</fnm></au>
    <au><snm>Phillippy</snm><fnm>AM</fnm></au>
    <au><snm>Koren</snm><fnm>S</fnm></au>
  </aug>
  <source>Genome Res.</source>
  <pubdate>2020</pubdate>
  <volume>30</volume>
  <issue>9</issue>
  <fpage>1291</fpage>
  <lpage>-1305</lpage>
</bibl>

<bibl id="B31">
  <title><p>Metagenome assembly of high-fidelity long reads with
  hifiasm-meta</p></title>
  <aug>
    <au><snm>Feng</snm><fnm>X</fnm></au>
    <au><snm>Cheng</snm><fnm>H</fnm></au>
    <au><snm>Portik</snm><fnm>D</fnm></au>
    <au><snm>Li</snm><fnm>H</fnm></au>
  </aug>
  <source>Nature Methods</source>
  <publisher>Nature Publishing Group</publisher>
  <pubdate>2022</pubdate>
  <fpage>1</fpage>
  <lpage>-4</lpage>
</bibl>

<bibl id="B32">
  <title><p>{MetaBAT} 2: an adaptive binning algorithm for robust and efficient
  genome reconstruction from metagenome assemblies</p></title>
  <aug>
    <au><snm>Kang</snm><fnm>DD</fnm></au>
    <au><snm>Li</snm><fnm>F</fnm></au>
    <au><snm>Kirton</snm><fnm>E</fnm></au>
    <au><snm>Thomas</snm><fnm>A</fnm></au>
    <au><snm>Egan</snm><fnm>R</fnm></au>
    <au><snm>An</snm><fnm>H</fnm></au>
    <au><snm>Wang</snm><fnm>Z</fnm></au>
  </aug>
  <source>PeerJ</source>
  <publisher>PeerJ Inc.</publisher>
  <pubdate>2019</pubdate>
  <volume>7</volume>
  <fpage>e7359</fpage>
</bibl>

<bibl id="B33">
  <title><p>Improved metagenome binning and assembly using deep variational
  autoencoders</p></title>
  <aug>
    <au><snm>Nissen</snm><fnm>JN</fnm></au>
    <au><snm>Johansen</snm><fnm>J</fnm></au>
    <au><snm>Alles{\o}e</snm><fnm>RL</fnm></au>
    <au><snm>S{\o}nderby</snm><fnm>CK</fnm></au>
    <au><snm>Armenteros</snm><fnm>JJA</fnm></au>
    <au><snm>Gr{\o}nbech</snm><fnm>CH</fnm></au>
    <au><snm>Jensen</snm><fnm>LJ</fnm></au>
    <au><snm>Nielsen</snm><fnm>HB</fnm></au>
    <au><snm>Petersen</snm><fnm>TN</fnm></au>
    <au><snm>Winther</snm><fnm>O</fnm></au>
    <au><cnm>others</cnm></au>
  </aug>
  <source>Nature biotechnology</source>
  <publisher>Nature Publishing Group</publisher>
  <pubdate>2021</pubdate>
  <volume>39</volume>
  <issue>5</issue>
  <fpage>555</fpage>
  <lpage>-560</lpage>
</bibl>

<bibl id="B34">
  <title><p>GraphBin: refined binning of metagenomic contigs using assembly
  graphs</p></title>
  <aug>
    <au><snm>Mallawaarachchi</snm><fnm>V</fnm></au>
    <au><snm>Wickramarachchi</snm><fnm>A</fnm></au>
    <au><snm>Lin</snm><fnm>Y</fnm></au>
  </aug>
  <source>Bioinformatics</source>
  <publisher>Oxford University Press</publisher>
  <pubdate>2020</pubdate>
  <volume>36</volume>
  <issue>11</issue>
  <fpage>3307</fpage>
  <lpage>-3313</lpage>
</bibl>

<bibl id="B35">
  <title><p>MaxBin: an automated binning method to recover individual genomes
  from metagenomes using an expectation-maximization algorithm</p></title>
  <aug>
    <au><snm>Wu</snm><fnm>YW</fnm></au>
    <au><snm>Tang</snm><fnm>YH</fnm></au>
    <au><snm>Tringe</snm><fnm>SG</fnm></au>
    <au><snm>Simmons</snm><fnm>BA</fnm></au>
    <au><snm>Singer</snm><fnm>SW</fnm></au>
  </aug>
  <source>Microbiome</source>
  <publisher>Springer</publisher>
  <pubdate>2014</pubdate>
  <volume>2</volume>
  <issue>1</issue>
  <fpage>1</fpage>
  <lpage>-18</lpage>
</bibl>

<bibl id="B36">
  <title><p>Recovery of genomes from metagenomes via a dereplication,
  aggregation and scoring strategy</p></title>
  <aug>
    <au><snm>Sieber</snm><fnm>CM</fnm></au>
    <au><snm>Probst</snm><fnm>AJ</fnm></au>
    <au><snm>Sharrar</snm><fnm>A</fnm></au>
    <au><snm>Thomas</snm><fnm>BC</fnm></au>
    <au><snm>Hess</snm><fnm>M</fnm></au>
    <au><snm>Tringe</snm><fnm>SG</fnm></au>
    <au><snm>Banfield</snm><fnm>JF</fnm></au>
  </aug>
  <source>Nature microbiology</source>
  <publisher>Nature Publishing Group</publisher>
  <pubdate>2018</pubdate>
  <volume>3</volume>
  <issue>7</issue>
  <fpage>836</fpage>
  <lpage>-843</lpage>
</bibl>

<bibl id="B37">
  <title><p>Bandage: interactive visualization of de novo genome
  assemblies</p></title>
  <aug>
    <au><snm>Wick</snm><fnm>RR</fnm></au>
    <au><snm>Schultz</snm><fnm>MB</fnm></au>
    <au><snm>Zobel</snm><fnm>J</fnm></au>
    <au><snm>Holt</snm><fnm>KE</fnm></au>
  </aug>
  <source>Bioinformatics</source>
  <publisher>Oxford University Press</publisher>
  <pubdate>2015</pubdate>
  <volume>31</volume>
  <issue>20</issue>
  <fpage>3350</fpage>
  <lpage>-3352</lpage>
</bibl>

<bibl id="B38">
  <title><p>Mash: fast genome and metagenome distance estimation using
  MinHash</p></title>
  <aug>
    <au><snm>Ondov</snm><fnm>BD</fnm></au>
    <au><snm>Treangen</snm><fnm>TJ</fnm></au>
    <au><snm>Melsted</snm><fnm>P</fnm></au>
    <au><snm>Mallonee</snm><fnm>AB</fnm></au>
    <au><snm>Bergman</snm><fnm>NH</fnm></au>
    <au><snm>Koren</snm><fnm>S</fnm></au>
    <au><snm>Phillippy</snm><fnm>AM</fnm></au>
  </aug>
  <source>Genome biology</source>
  <publisher>BioMed Central</publisher>
  <pubdate>2016</pubdate>
  <volume>17</volume>
  <issue>1</issue>
  <fpage>1</fpage>
  <lpage>-14</lpage>
</bibl>

<bibl id="B39">
  <title><p>High throughput ANI analysis of 90K prokaryotic genomes reveals
  clear species boundaries</p></title>
  <aug>
    <au><snm>Jain</snm><fnm>C</fnm></au>
    <au><snm>Rodriguez R</snm><fnm>LM</fnm></au>
    <au><snm>Phillippy</snm><fnm>AM</fnm></au>
    <au><snm>Konstantinidis</snm><fnm>KT</fnm></au>
    <au><snm>Aluru</snm><fnm>S</fnm></au>
  </aug>
  <source>Nature communications</source>
  <publisher>Nature Publishing Group</publisher>
  <pubdate>2018</pubdate>
  <volume>9</volume>
  <issue>1</issue>
  <fpage>1</fpage>
  <lpage>-8</lpage>
</bibl>

<bibl id="B40">
  <title><p>A complete domain-to-species taxonomy for Bacteria and
  Archaea</p></title>
  <aug>
    <au><snm>Parks</snm><fnm>DH</fnm></au>
    <au><snm>Chuvochina</snm><fnm>M</fnm></au>
    <au><snm>Chaumeil</snm><fnm>PA</fnm></au>
    <au><snm>Rinke</snm><fnm>C</fnm></au>
    <au><snm>Mussig</snm><fnm>AJ</fnm></au>
    <au><snm>Hugenholtz</snm><fnm>P</fnm></au>
  </aug>
  <source>Nature biotechnology</source>
  <publisher>Nature Publishing Group</publisher>
  <pubdate>2020</pubdate>
  <volume>38</volume>
  <issue>9</issue>
  <fpage>1079</fpage>
  <lpage>-1086</lpage>
</bibl>

<bibl id="B41">
  <title><p>Generating lineage-resolved, complete metagenome-assembled genomes
  from complex microbial communities</p></title>
  <aug>
    <au><snm>Bickhart</snm><fnm>DM</fnm></au>
    <au><snm>Kolmogorov</snm><fnm>M</fnm></au>
    <au><snm>Tseng</snm><fnm>E</fnm></au>
    <au><snm>Portik</snm><fnm>DM</fnm></au>
    <au><snm>Korobeynikov</snm><fnm>A</fnm></au>
    <au><snm>Tolstoganov</snm><fnm>I</fnm></au>
    <au><snm>Uritskiy</snm><fnm>G</fnm></au>
    <au><snm>Liachko</snm><fnm>I</fnm></au>
    <au><snm>Sullivan</snm><fnm>ST</fnm></au>
    <au><snm>Shin</snm><fnm>SB</fnm></au>
    <au><cnm>others</cnm></au>
  </aug>
  <source>Nature biotechnology</source>
  <publisher>Nature Publishing Group</publisher>
  <pubdate>2022</pubdate>
  <fpage>1</fpage>
  <lpage>-9</lpage>
</bibl>

<bibl id="B42">
  <title><p>Minimizer-space de Bruijn graphs: Whole-genome assembly of long
  reads in minutes on a personal computer</p></title>
  <aug>
    <au><snm>Ekim</snm><fnm>B</fnm></au>
    <au><snm>Berger</snm><fnm>B</fnm></au>
    <au><snm>Chikhi</snm><fnm>R</fnm></au>
  </aug>
  <source>Cell systems</source>
  <publisher>Elsevier</publisher>
  <pubdate>2021</pubdate>
  <volume>12</volume>
  <issue>10</issue>
  <fpage>958</fpage>
  <lpage>-968</lpage>
</bibl>

<bibl id="B43">
  <title><p>KAT: a K-mer analysis toolkit to quality control NGS datasets and
  genome assemblies</p></title>
  <aug>
    <au><snm>Mapleson</snm><fnm>D</fnm></au>
    <au><snm>Garcia Accinelli</snm><fnm>G</fnm></au>
    <au><snm>Kettleborough</snm><fnm>G</fnm></au>
    <au><snm>Wright</snm><fnm>J</fnm></au>
    <au><snm>Clavijo</snm><fnm>BJ</fnm></au>
  </aug>
  <source>Bioinformatics</source>
  <publisher>Oxford University Press</publisher>
  <pubdate>2017</pubdate>
  <volume>33</volume>
  <issue>4</issue>
  <fpage>574</fpage>
  <lpage>-576</lpage>
</bibl>

<bibl id="B44">
  <title><p>Merqury: reference-free quality, completeness, and phasing
  assessment for genome assemblies</p></title>
  <aug>
    <au><snm>Rhie</snm><fnm>A</fnm></au>
    <au><snm>Walenz</snm><fnm>BP</fnm></au>
    <au><snm>Koren</snm><fnm>S</fnm></au>
    <au><snm>Phillippy</snm><fnm>AM</fnm></au>
  </aug>
  <source>Genome biology</source>
  <publisher>BioMed Central</publisher>
  <pubdate>2020</pubdate>
  <volume>21</volume>
  <issue>1</issue>
  <fpage>1</fpage>
  <lpage>-27</lpage>
</bibl>

<bibl id="B45">
  <title><p>Fast and accurate short read alignment with Burrows--Wheeler
  transform</p></title>
  <aug>
    <au><snm>Li</snm><fnm>H</fnm></au>
    <au><snm>Durbin</snm><fnm>R</fnm></au>
  </aug>
  <source>bioinformatics</source>
  <publisher>Oxford University Press</publisher>
  <pubdate>2009</pubdate>
  <volume>25</volume>
  <issue>14</issue>
  <fpage>1754</fpage>
  <lpage>-1760</lpage>
</bibl>

<bibl id="B46">
  <title><p>Barrnap: BAsic Rapid Ribosomal RNA Predictor</p></title>
  <aug>
    <au><snm>Seemann</snm><fnm>T</fnm></au>
    <au><snm>Booth</snm><fnm>T</fnm></au>
  </aug>
  <pubdate>2018</pubdate>
</bibl>

<bibl id="B47">
  <title><p>Infernal 1.1: 100-fold faster RNA homology searches</p></title>
  <aug>
    <au><snm>Nawrocki</snm><fnm>EP</fnm></au>
    <au><snm>Eddy</snm><fnm>SR</fnm></au>
  </aug>
  <source>Bioinformatics</source>
  <publisher>Oxford University Press</publisher>
  <pubdate>2013</pubdate>
  <volume>29</volume>
  <issue>22</issue>
  <fpage>2933</fpage>
  <lpage>-2935</lpage>
</bibl>

<bibl id="B48">
  <title><p>Naive Bayesian classifier for rapid assignment of rRNA sequences
  into the new bacterial taxonomy</p></title>
  <aug>
    <au><snm>Wang</snm><fnm>Q</fnm></au>
    <au><snm>Garrity</snm><fnm>GM</fnm></au>
    <au><snm>Tiedje</snm><fnm>JM</fnm></au>
    <au><snm>Cole</snm><fnm>JR</fnm></au>
  </aug>
  <source>Applied and environmental microbiology</source>
  <publisher>Am Soc Microbiol</publisher>
  <pubdate>2007</pubdate>
  <volume>73</volume>
  <issue>16</issue>
  <fpage>5261</fpage>
  <lpage>-5267</lpage>
</bibl>

<bibl id="B49">
  <title><p>Accuracy of taxonomy prediction for 16S rRNA and fungal ITS
  sequences</p></title>
  <aug>
    <au><snm>Edgar</snm><fnm>RC</fnm></au>
  </aug>
  <source>PeerJ</source>
  <publisher>PeerJ Inc.</publisher>
  <pubdate>2018</pubdate>
  <volume>6</volume>
  <fpage>e4652</fpage>
</bibl>

<bibl id="B50">
  <title><p>The reliability of metagenome-assembled genomes (MAGs) in
  representing natural populations: Insights from comparing MAGs against
  isolate genomes derived from the same fecal sample</p></title>
  <aug>
    <au><snm>Meziti</snm><fnm>A</fnm></au>
    <au><snm>Rodriguez R</snm><fnm>LM</fnm></au>
    <au><snm>Hatt</snm><fnm>JK</fnm></au>
    <au><snm>Pe{\~n}a Gonzalez</snm><fnm>A</fnm></au>
    <au><snm>Levy</snm><fnm>K</fnm></au>
    <au><snm>Konstantinidis</snm><fnm>KT</fnm></au>
  </aug>
  <source>Applied and environmental microbiology</source>
  <publisher>Am Soc Microbiol</publisher>
  <pubdate>2021</pubdate>
  <volume>87</volume>
  <issue>6</issue>
  <fpage>e02593</fpage>
  <lpage>-20</lpage>
</bibl>

<bibl id="B51">
  <title><p>Hybrid, ultra-deep metagenomic sequencing enables genomic and
  functional characterization of low-abundance species in the human gut
  microbiome</p></title>
  <aug>
    <au><snm>Jin</snm><fnm>H</fnm></au>
    <au><snm>You</snm><fnm>L</fnm></au>
    <au><snm>Zhao</snm><fnm>F</fnm></au>
    <au><snm>Li</snm><fnm>S</fnm></au>
    <au><snm>Ma</snm><fnm>T</fnm></au>
    <au><snm>Kwok</snm><fnm>LY</fnm></au>
    <au><snm>Xu</snm><fnm>H</fnm></au>
    <au><snm>Sun</snm><fnm>Z</fnm></au>
  </aug>
  <source>Gut microbes</source>
  <publisher>Taylor \& Francis</publisher>
  <pubdate>2022</pubdate>
  <volume>14</volume>
  <issue>1</issue>
  <fpage>2021790</fpage>
</bibl>

<bibl id="B52">
  <title><p>A unified catalog of 204,938 reference genomes from the human gut
  microbiome</p></title>
  <aug>
    <au><snm>Almeida</snm><fnm>A</fnm></au>
    <au><snm>Nayfach</snm><fnm>S</fnm></au>
    <au><snm>Boland</snm><fnm>M</fnm></au>
    <au><snm>Strozzi</snm><fnm>F</fnm></au>
    <au><snm>Beracochea</snm><fnm>M</fnm></au>
    <au><snm>Shi</snm><fnm>ZJ</fnm></au>
    <au><snm>Pollard</snm><fnm>KS</fnm></au>
    <au><snm>Sakharova</snm><fnm>E</fnm></au>
    <au><snm>Parks</snm><fnm>DH</fnm></au>
    <au><snm>Hugenholtz</snm><fnm>P</fnm></au>
    <au><cnm>others</cnm></au>
  </aug>
  <source>Nature biotechnology</source>
  <publisher>Nature Publishing Group</publisher>
  <pubdate>2021</pubdate>
  <volume>39</volume>
  <issue>1</issue>
  <fpage>105</fpage>
  <lpage>-114</lpage>
</bibl>

<bibl id="B53">
  <title><p>Metagenome-assembled genomes and gene catalog from the chicken gut
  microbiome aid in deciphering antibiotic resistomes</p></title>
  <aug>
    <au><snm>Feng</snm><fnm>Y</fnm></au>
    <au><snm>Wang</snm><fnm>Y</fnm></au>
    <au><snm>Zhu</snm><fnm>B</fnm></au>
    <au><snm>Gao</snm><fnm>GF</fnm></au>
    <au><snm>Guo</snm><fnm>Y</fnm></au>
    <au><snm>Hu</snm><fnm>Y</fnm></au>
  </aug>
  <source>Communications biology</source>
  <publisher>Nature Publishing Group</publisher>
  <pubdate>2021</pubdate>
  <volume>4</volume>
  <issue>1</issue>
  <fpage>1</fpage>
  <lpage>-9</lpage>
</bibl>

<bibl id="B54">
  <title><p>An integrated gene catalog and over 10,000 metagenome-assembled
  genomes from the gastrointestinal microbiome of ruminants</p></title>
  <aug>
    <au><snm>Xie</snm><fnm>F</fnm></au>
    <au><snm>Jin</snm><fnm>W</fnm></au>
    <au><snm>Si</snm><fnm>H</fnm></au>
    <au><snm>Yuan</snm><fnm>Y</fnm></au>
    <au><snm>Tao</snm><fnm>Y</fnm></au>
    <au><snm>Liu</snm><fnm>J</fnm></au>
    <au><snm>Wang</snm><fnm>X</fnm></au>
    <au><snm>Yang</snm><fnm>C</fnm></au>
    <au><snm>Li</snm><fnm>Q</fnm></au>
    <au><snm>Yan</snm><fnm>X</fnm></au>
    <au><cnm>others</cnm></au>
  </aug>
  <source>Microbiome</source>
  <publisher>Springer</publisher>
  <pubdate>2021</pubdate>
  <volume>9</volume>
  <issue>1</issue>
  <fpage>1</fpage>
  <lpage>-20</lpage>
</bibl>

<bibl id="B55">
  <title><p>Minimap2: pairwise alignment for nucleotide sequences</p></title>
  <aug>
    <au><snm>Li</snm><fnm>H</fnm></au>
  </aug>
  <source>Bioinformatics</source>
  <publisher>Oxford University Press</publisher>
  <pubdate>2018</pubdate>
  <volume>34</volume>
  <issue>18</issue>
  <fpage>3094</fpage>
  <lpage>-3100</lpage>
</bibl>

<bibl id="B56">
  <title><p>The sequence alignment/map format and SAMtools</p></title>
  <aug>
    <au><snm>Li</snm><fnm>H</fnm></au>
    <au><snm>Handsaker</snm><fnm>B</fnm></au>
    <au><snm>Wysoker</snm><fnm>A</fnm></au>
    <au><snm>Fennell</snm><fnm>T</fnm></au>
    <au><snm>Ruan</snm><fnm>J</fnm></au>
    <au><snm>Homer</snm><fnm>N</fnm></au>
    <au><snm>Marth</snm><fnm>G</fnm></au>
    <au><snm>Abecasis</snm><fnm>G</fnm></au>
    <au><snm>Durbin</snm><fnm>R</fnm></au>
  </aug>
  <source>Bioinformatics</source>
  <publisher>Oxford University Press</publisher>
  <pubdate>2009</pubdate>
  <volume>25</volume>
  <issue>16</issue>
  <fpage>2078</fpage>
  <lpage>-2079</lpage>
</bibl>

<bibl id="B57">
  <title><p>PBSIM2: a simulator for long-read sequencers with a novel
  generative model of quality scores</p></title>
  <aug>
    <au><snm>Ono</snm><fnm>Y</fnm></au>
    <au><snm>Asai</snm><fnm>K</fnm></au>
    <au><snm>Hamada</snm><fnm>M</fnm></au>
  </aug>
  <source>Bioinformatics</source>
  <publisher>Oxford University Press</publisher>
  <pubdate>2021</pubdate>
  <volume>37</volume>
  <issue>5</issue>
  <fpage>589</fpage>
  <lpage>-595</lpage>
</bibl>

<bibl id="B58">
  <title><p>Oxford Nanopore R10. 4 long-read sequencing enables the generation
  of near-finished bacterial genomes from pure cultures and metagenomes without
  short-read or reference polishing</p></title>
  <aug>
    <au><snm>Sereika</snm><fnm>M</fnm></au>
    <au><snm>Kirkegaard</snm><fnm>RH</fnm></au>
    <au><snm>Karst</snm><fnm>SM</fnm></au>
    <au><snm>Michaelsen</snm><fnm>TY</fnm></au>
    <au><snm>S{\o}rensen</snm><fnm>EA</fnm></au>
    <au><snm>Wollenberg</snm><fnm>RD</fnm></au>
    <au><snm>Albertsen</snm><fnm>M</fnm></au>
  </aug>
  <source>Nature methods</source>
  <publisher>Nature Publishing Group</publisher>
  <pubdate>2022</pubdate>
  <volume>19</volume>
  <issue>7</issue>
  <fpage>823</fpage>
  <lpage>-826</lpage>
</bibl>

<bibl id="B59">
  <title><p>HiFi Metagenomic Sequencing Enables Assembly of Accurate and
  Complete Genomes from Human Gut Microbiota</p></title>
  <aug>
    <au><snm>Kim</snm><fnm>CY</fnm></au>
    <au><snm>Ma</snm><fnm>J</fnm></au>
    <au><snm>Lee</snm><fnm>I</fnm></au>
  </aug>
  <source>bioRxiv</source>
  <publisher>Cold Spring Harbor Laboratory</publisher>
  <pubdate>2022</pubdate>
</bibl>

</refgrp>
} % end of \BMCxmlcomment

%%%%%%%%%%%%%%%%%%%%%%%%%%%%%%%%%%%
%%                               %%
%% Figures                       %%
%%                               %%
%% NB: this is for captions and  %%
%% Titles. All graphics must be  %%
%% submitted separately and NOT  %%
%% included in the Tex document  %%
%%                               %%
%%%%%%%%%%%%%%%%%%%%%%%%%%%%%%%%%%%

%%
%% Do not use \listoffigures as most will included as separate files

\section*{Figures}

\section*{Supplementary Figures}

\setcounter{figure}{0}
\renewcommand{\figurename}{Supplementary Figure}

%%%%%%figure S1
\begin{figure}[bp!]
\includegraphics[width=0.95\linewidth]{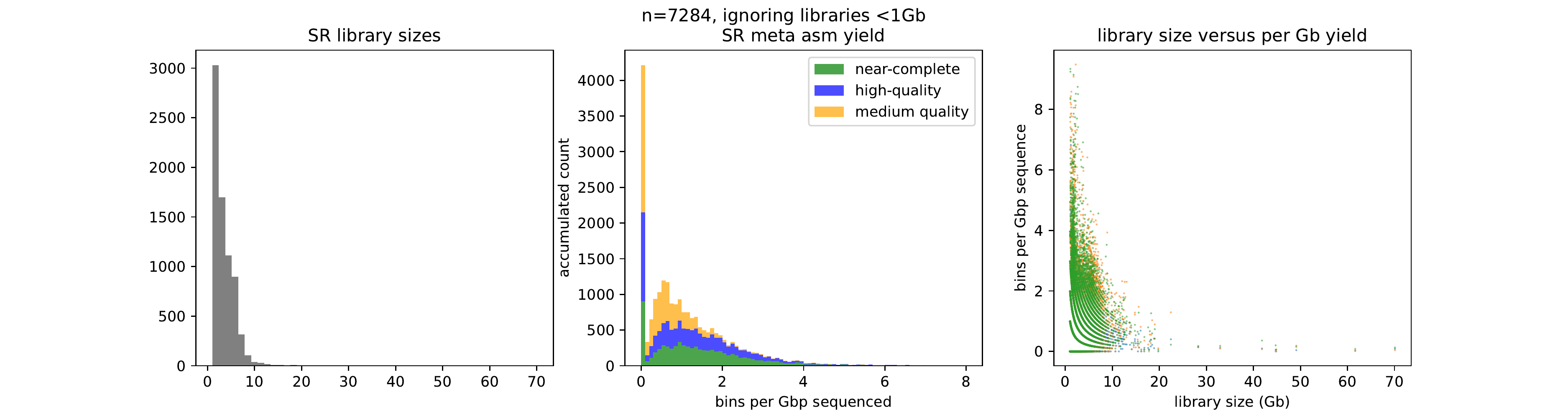}
  \caption{Almeida et al. unified assembly and MAG quality evaluations for 
13,133 human gut metagenomic datasets from 75 short-read-based studies. 
Most libraries are smaller than 10Gbp (99.2\%), and 
per Gbp MAG recovery visually negatively correlated with library size. 
For an ``average'' library, we expect 1-2 near-complete MAGs. 
HiFi assemblies provide closed contigs and much lower intra-MAG heterozygosity 
(0.06\% near-complete circular contigs have >25\% checkM heterogenous score, 
versus 23.8\% for SR) at similar per Gbp yields.
}
\end{figure}

%%%%%%figure S2
\begin{figure}[bp!]
\includegraphics[width=0.95\linewidth]{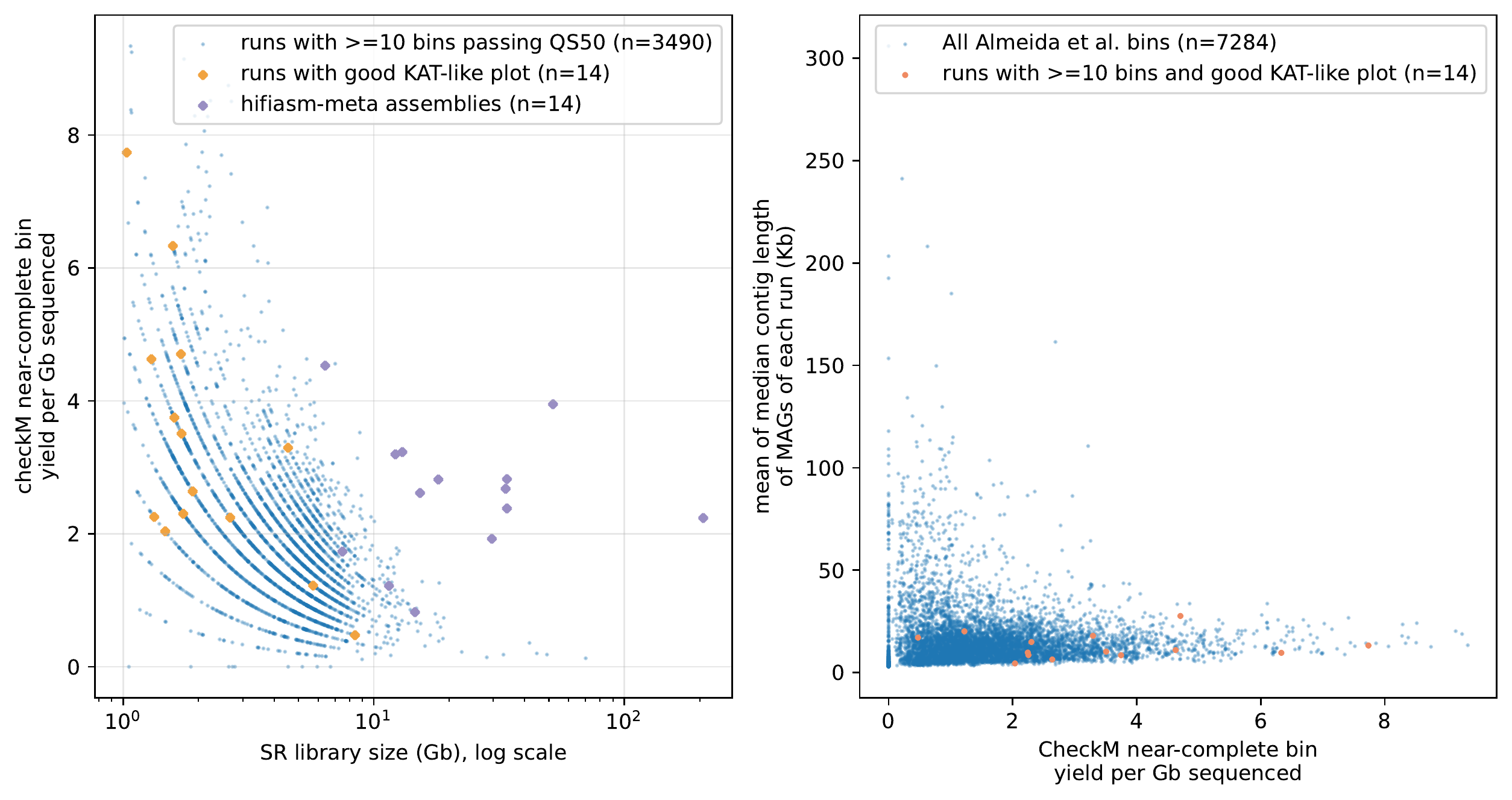}
  \caption{A few runs from short read metagenome assembly studies 
achieved reasonable sample representation. 
However, they were fragmented and larger library size 
did not imply a better outcome.
}
\end{figure}

%%%%%%figure S3: HRGM
\begin{figure}[bp!]
  \includegraphics[width=0.95\linewidth]{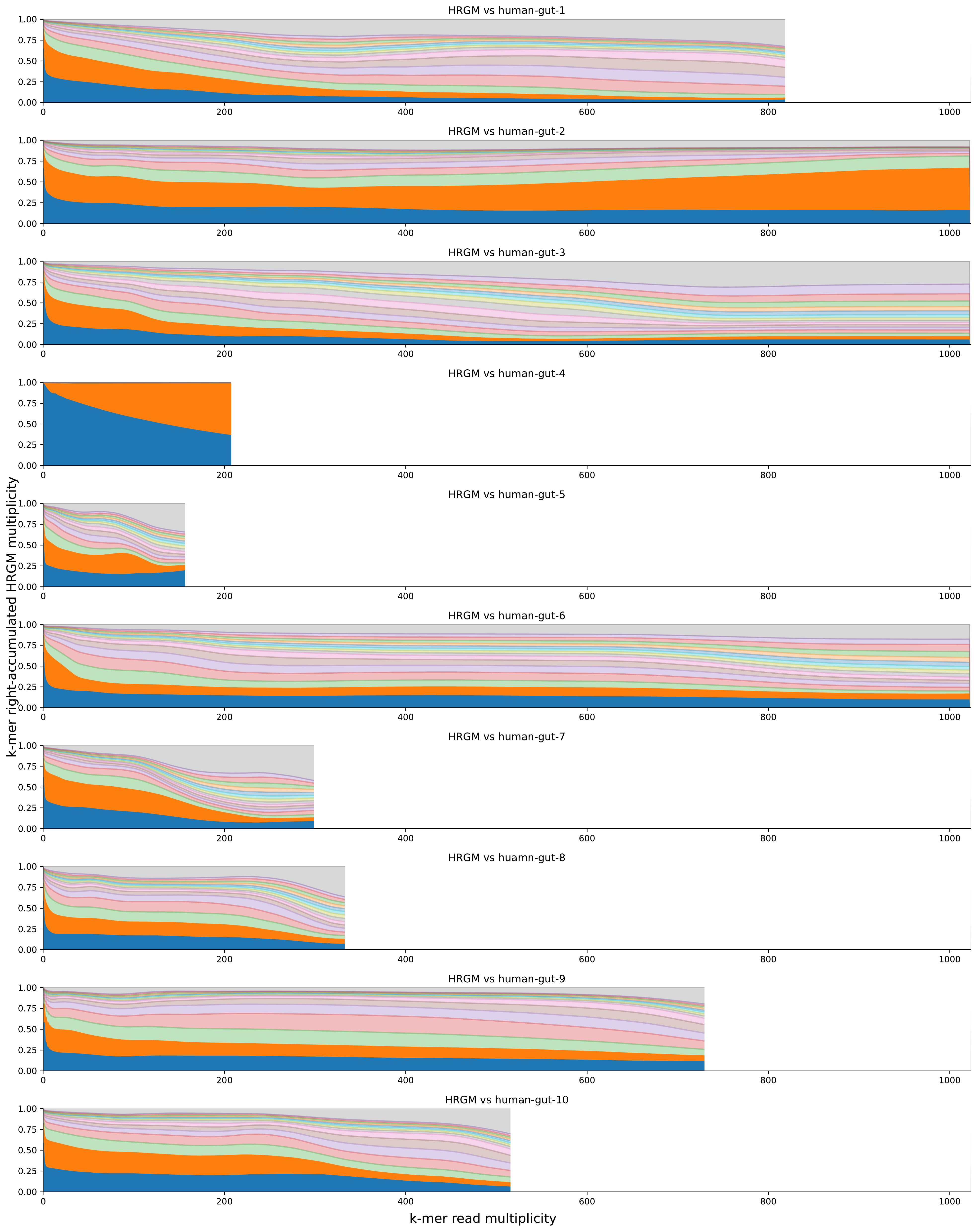}
    \caption{K-mer spectrum plots using HRGM assemblies as
the MAGs, and HiFi reads as the library. 
Intend to show how good would a relatively complete library
represent a new sample.
  }
  \end{figure}

%%%%%%figure S4
\begin{figure}[bp!]
\includegraphics[width=0.95\linewidth]{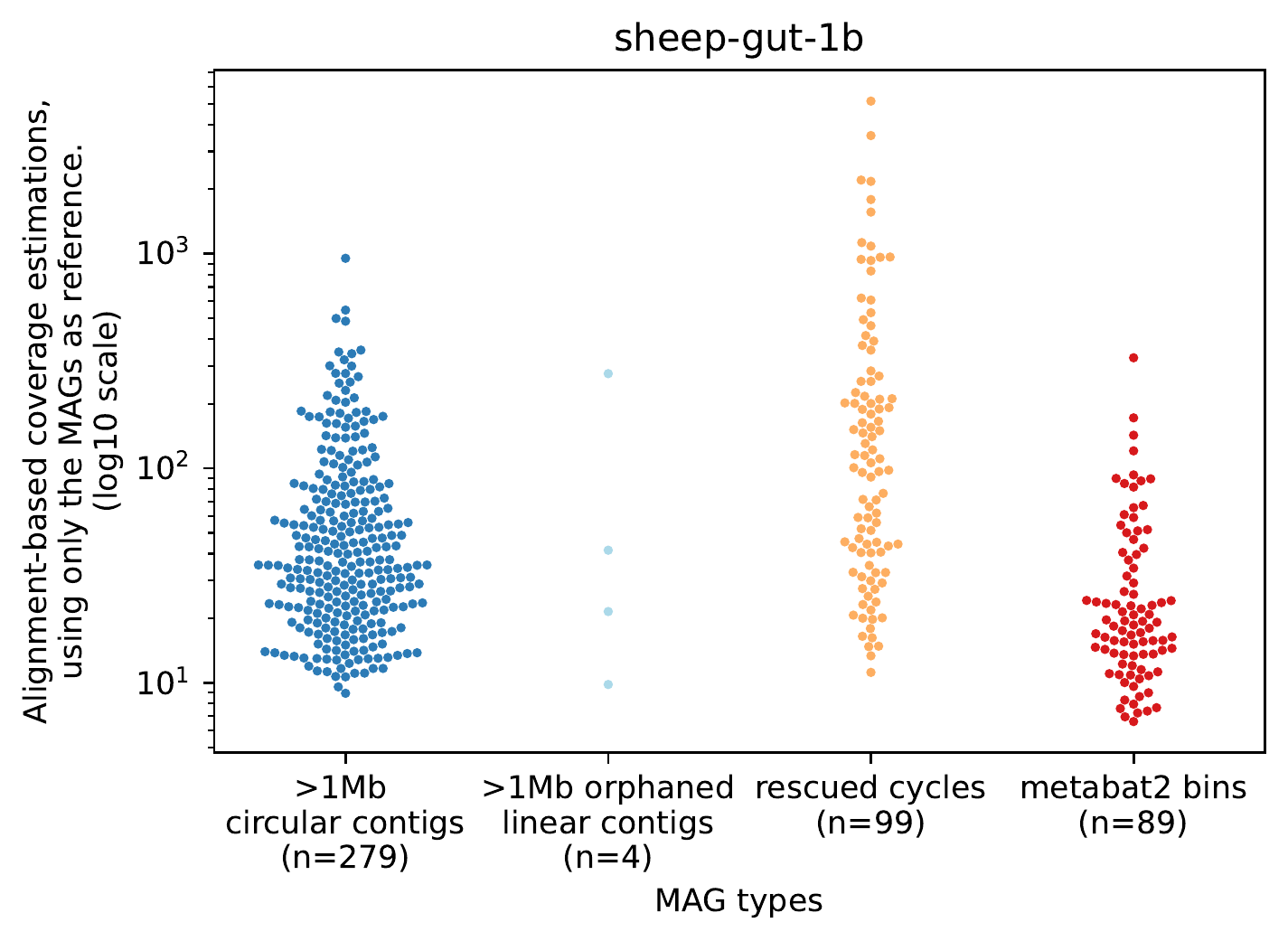}
  \caption{Coverage of sheep-gut-1b merged MAGs by category.
}
\end{figure}

%%%%%%figure S5
\begin{figure}[bp!]
  \includegraphics[width=0.95\linewidth]{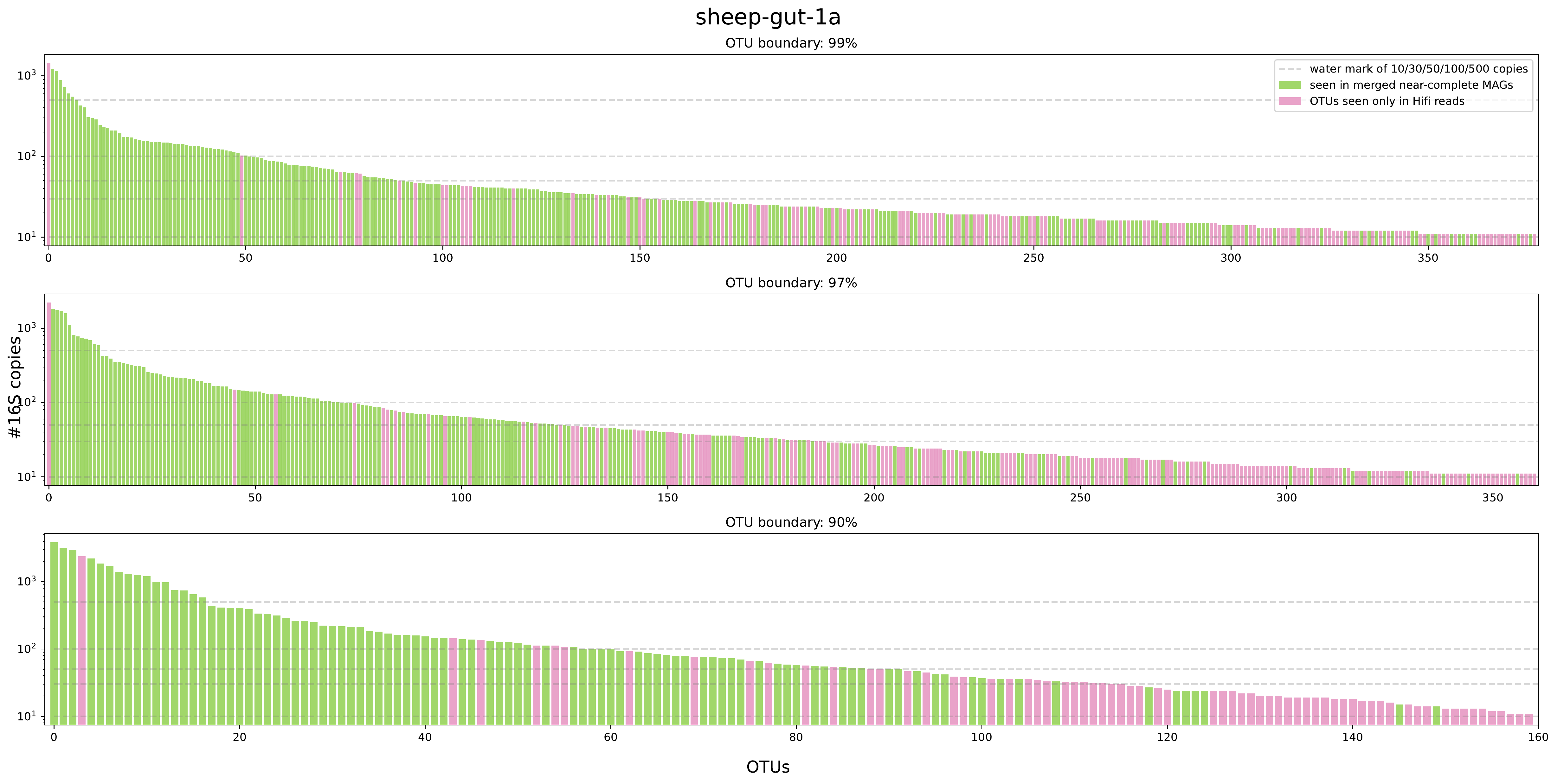}
  \includegraphics[width=0.95\linewidth]{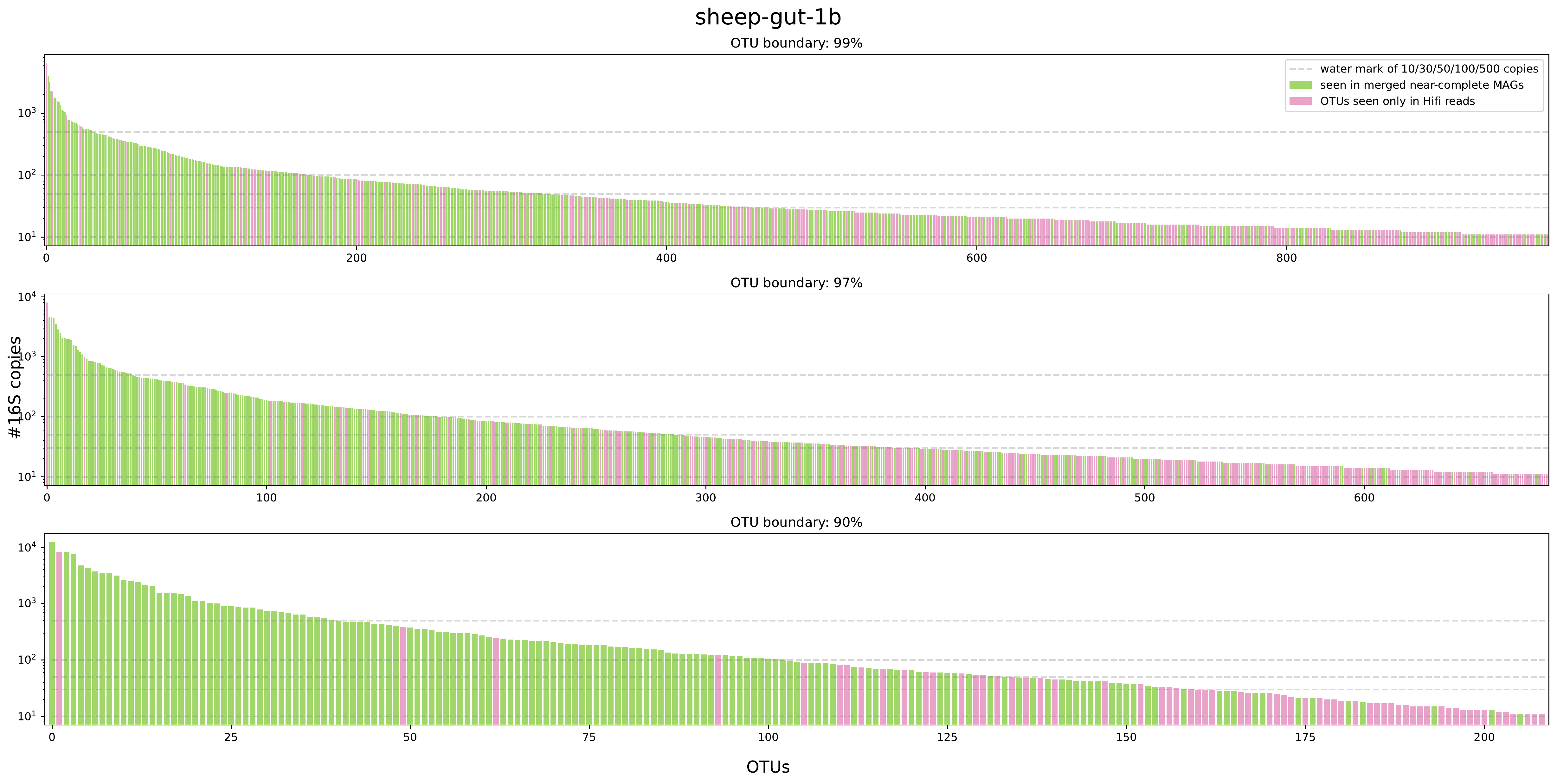}
  \includegraphics[width=0.95\linewidth]{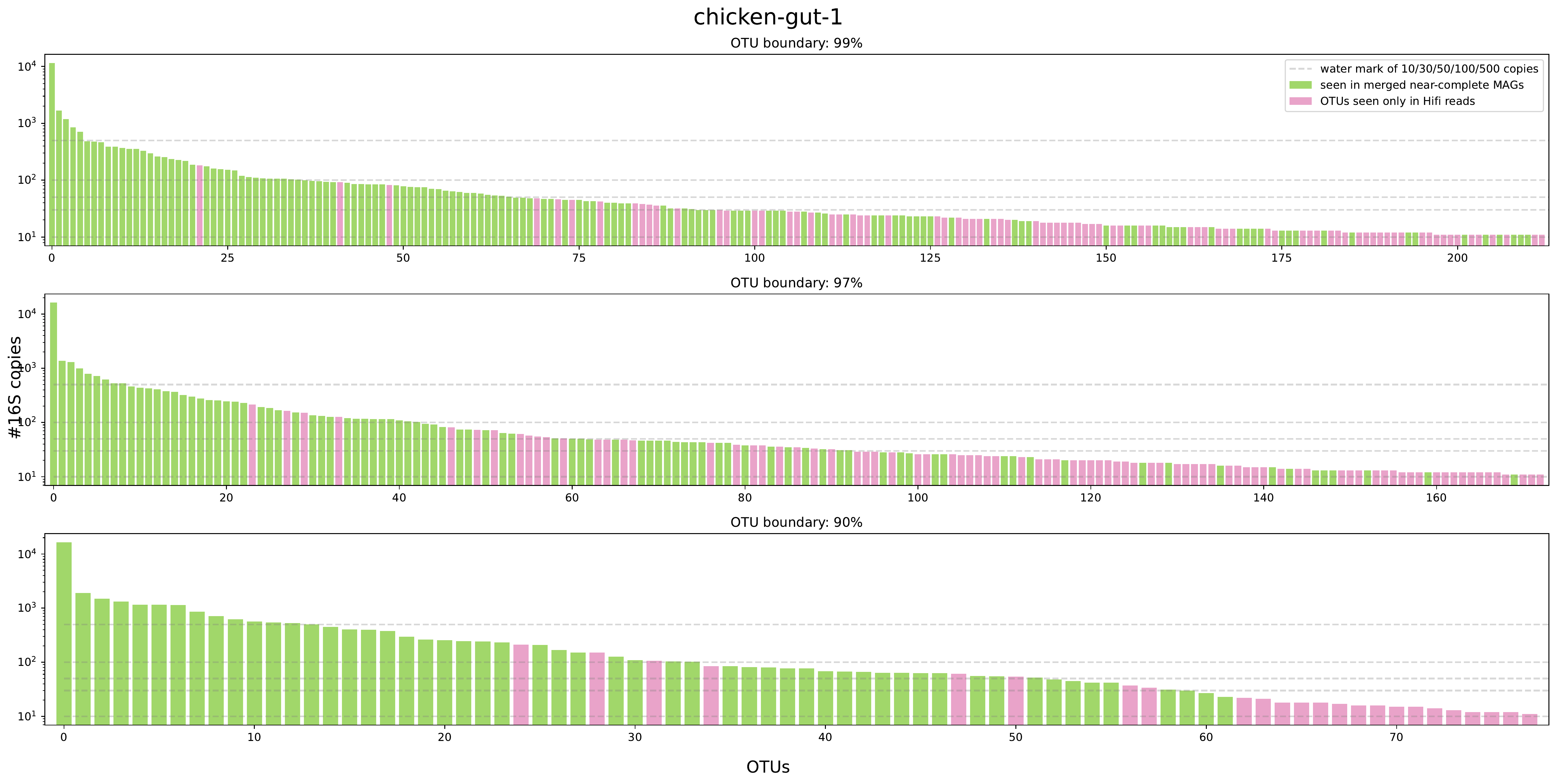}

\end{figure}
\begin{figure}[bp!]
  \includegraphics[width=0.95\linewidth]{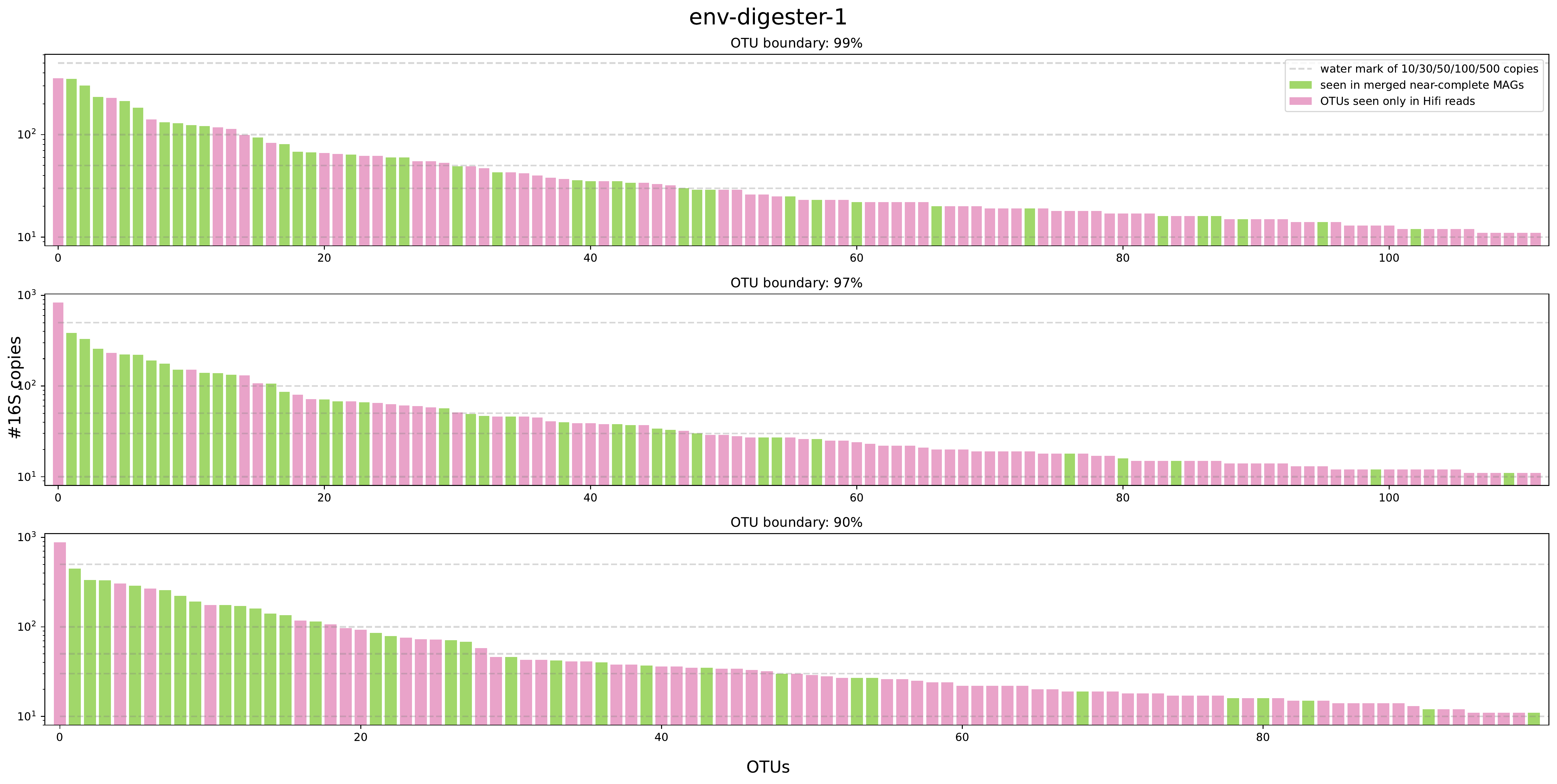}  
  \includegraphics[width=0.95\linewidth]{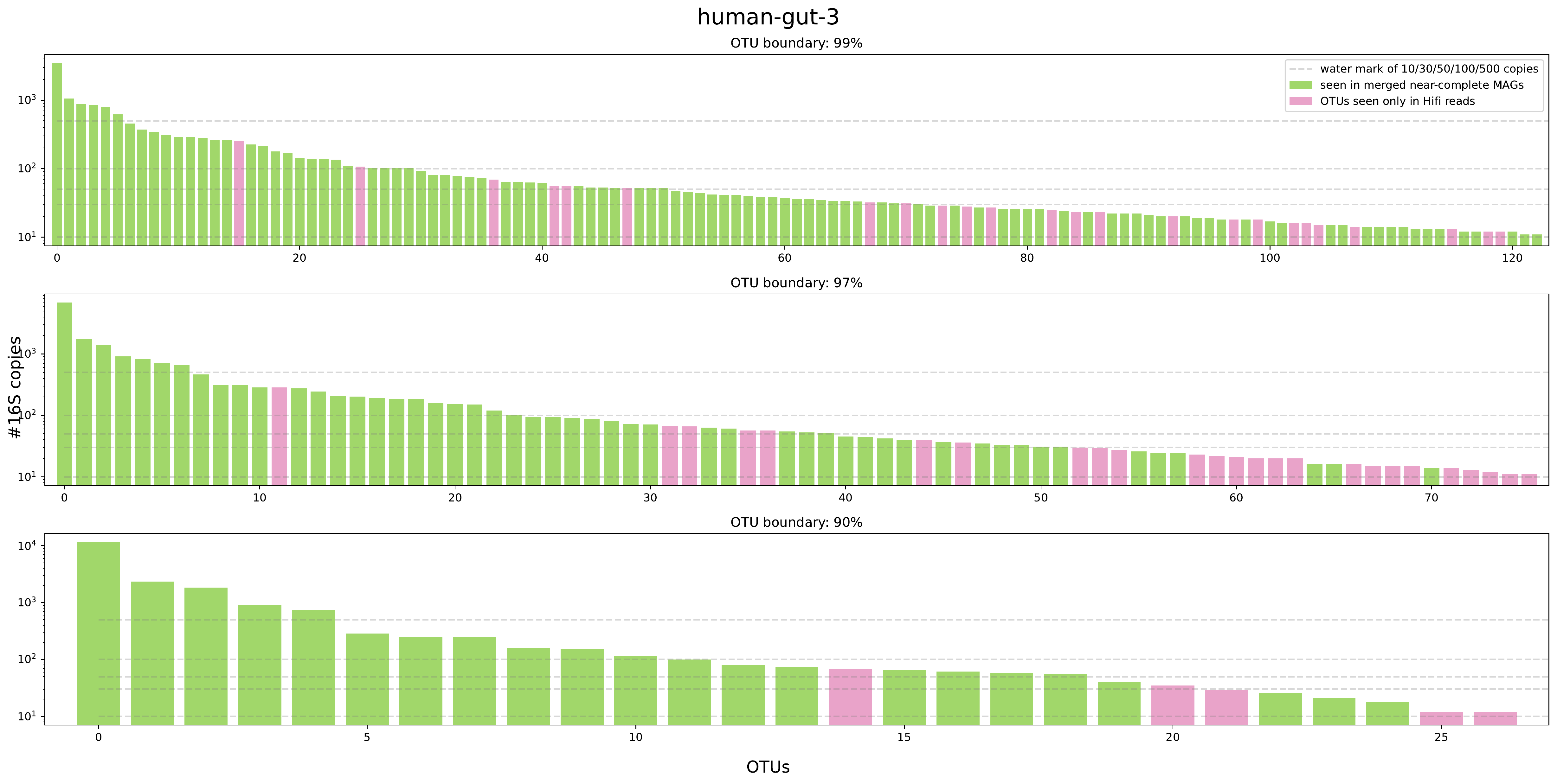}
  \includegraphics[width=0.95\linewidth]{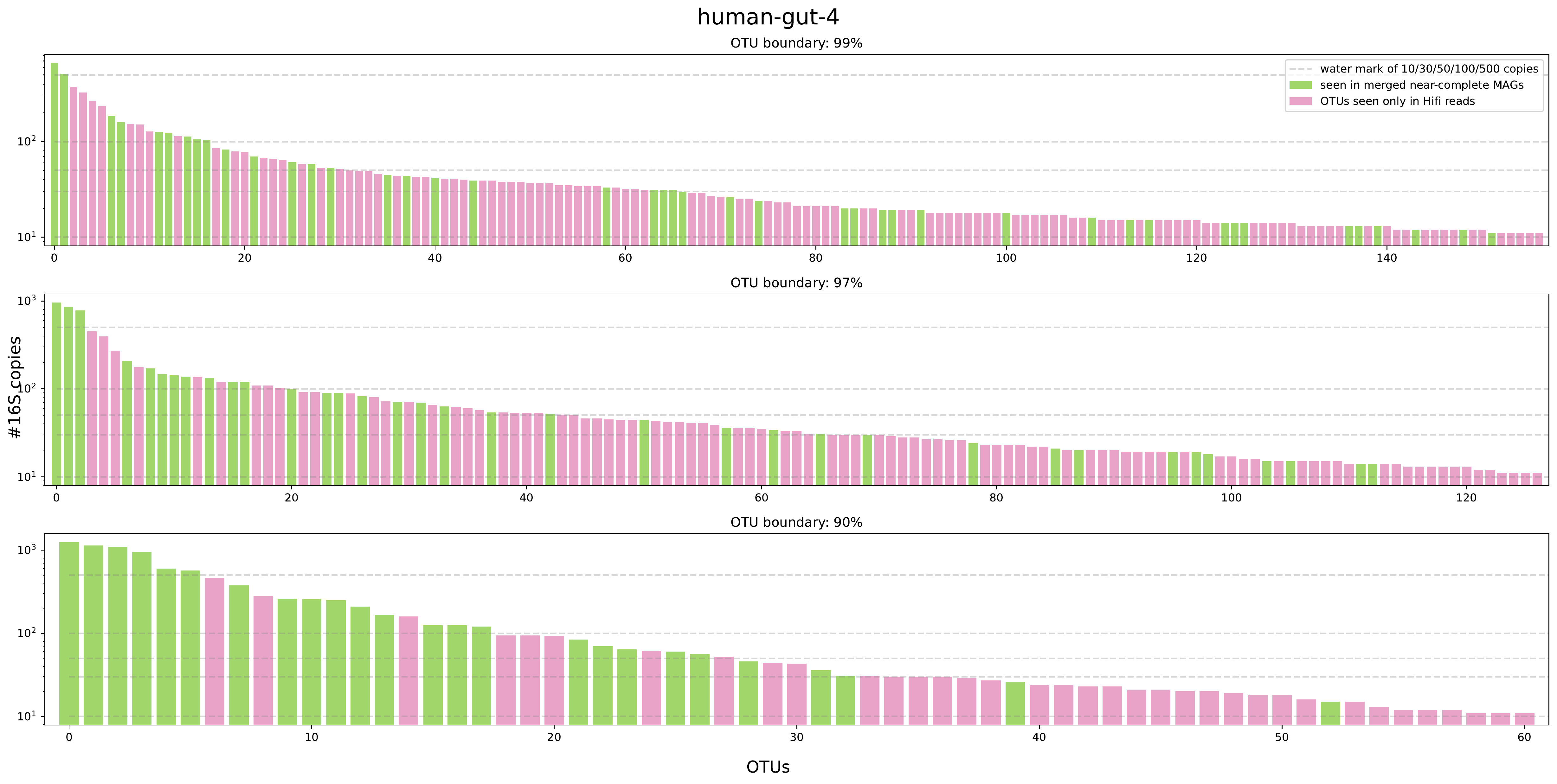}
\end{figure}
\begin{figure}[bp!]
  \includegraphics[width=0.95\linewidth]{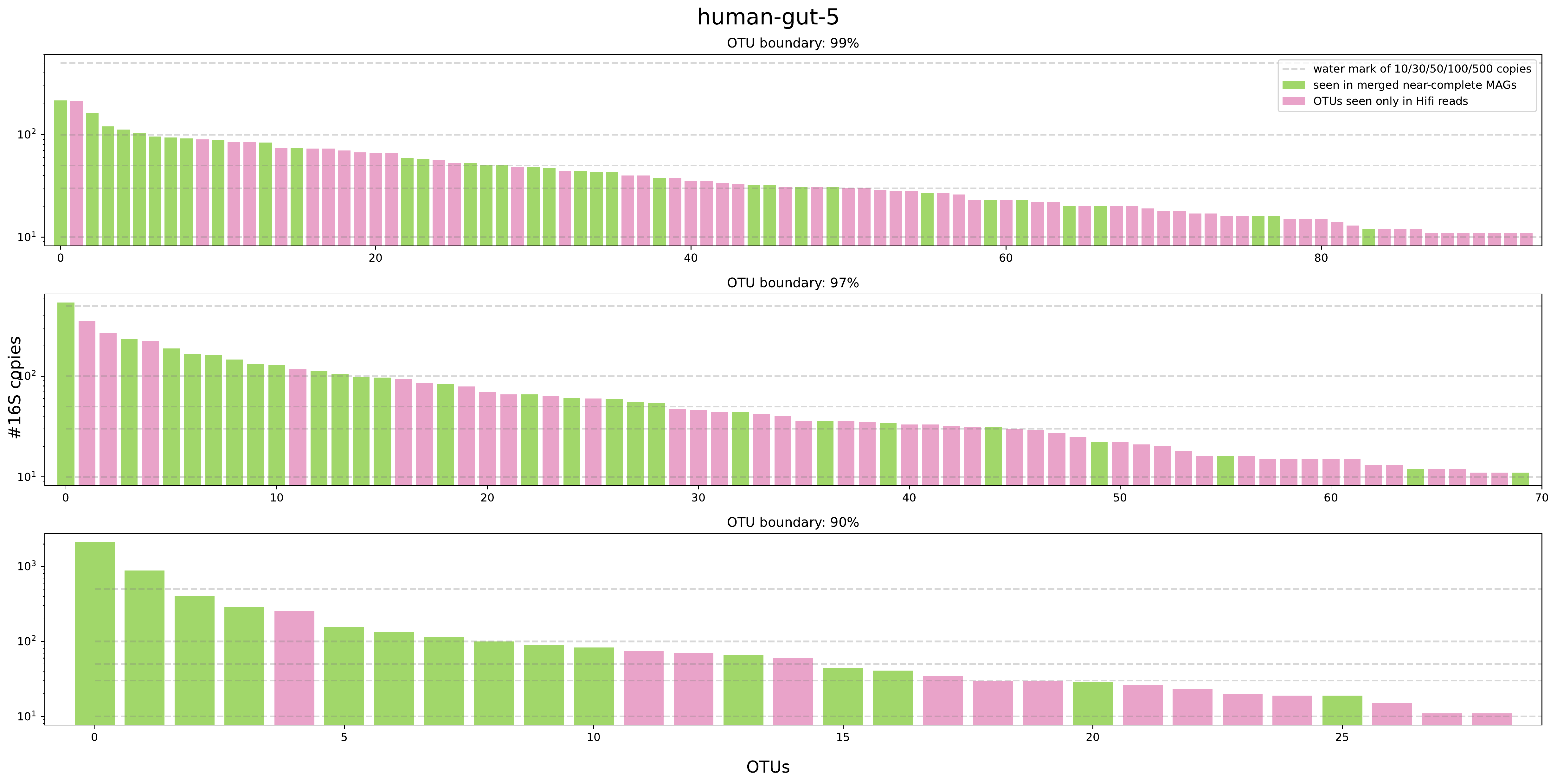}
  \includegraphics[width=0.95\linewidth]{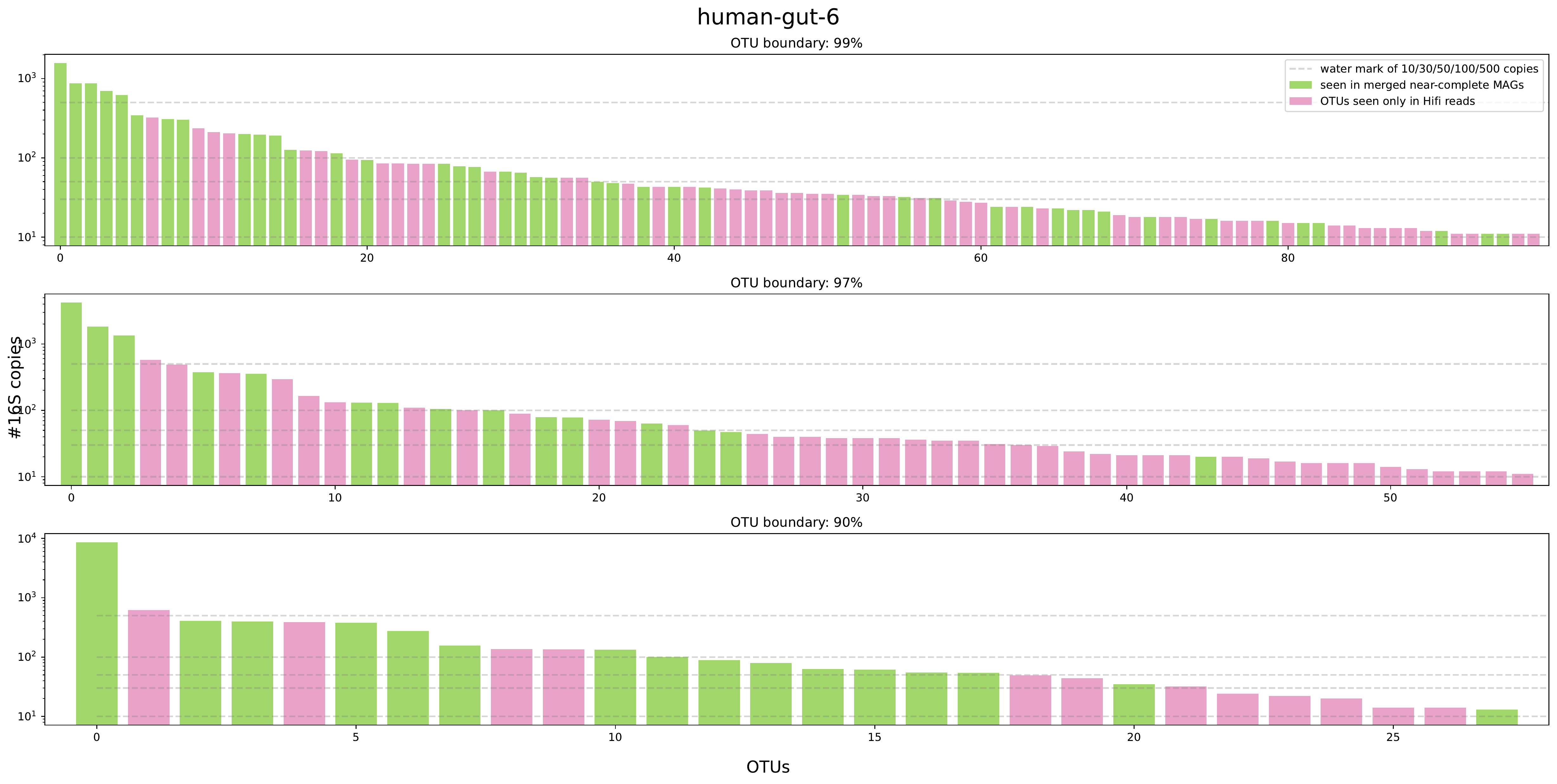}
\includegraphics[width=0.95\linewidth]{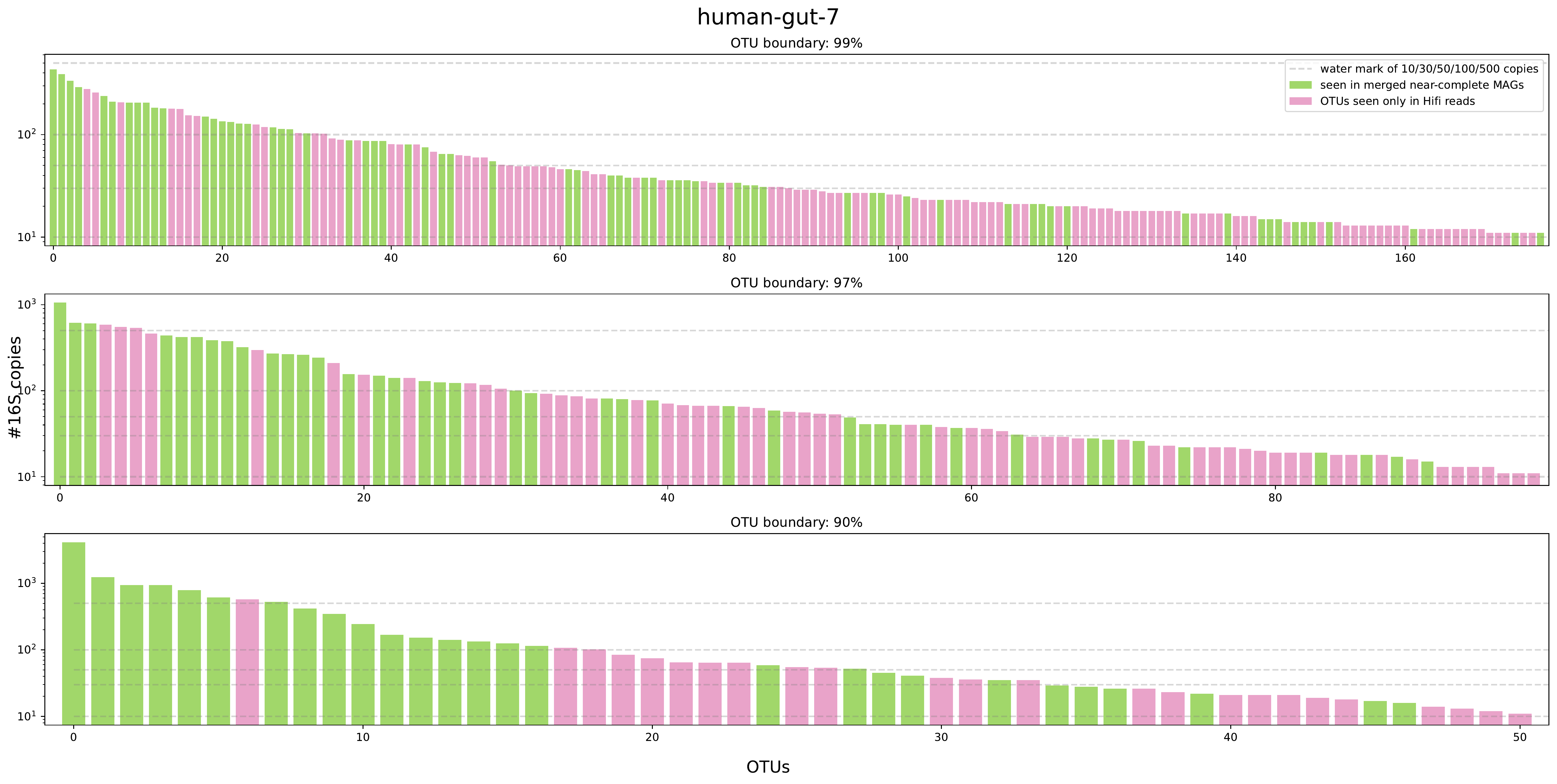}
  
\end{figure}  
\begin{figure}[bp!]
\includegraphics[width=0.95\linewidth]{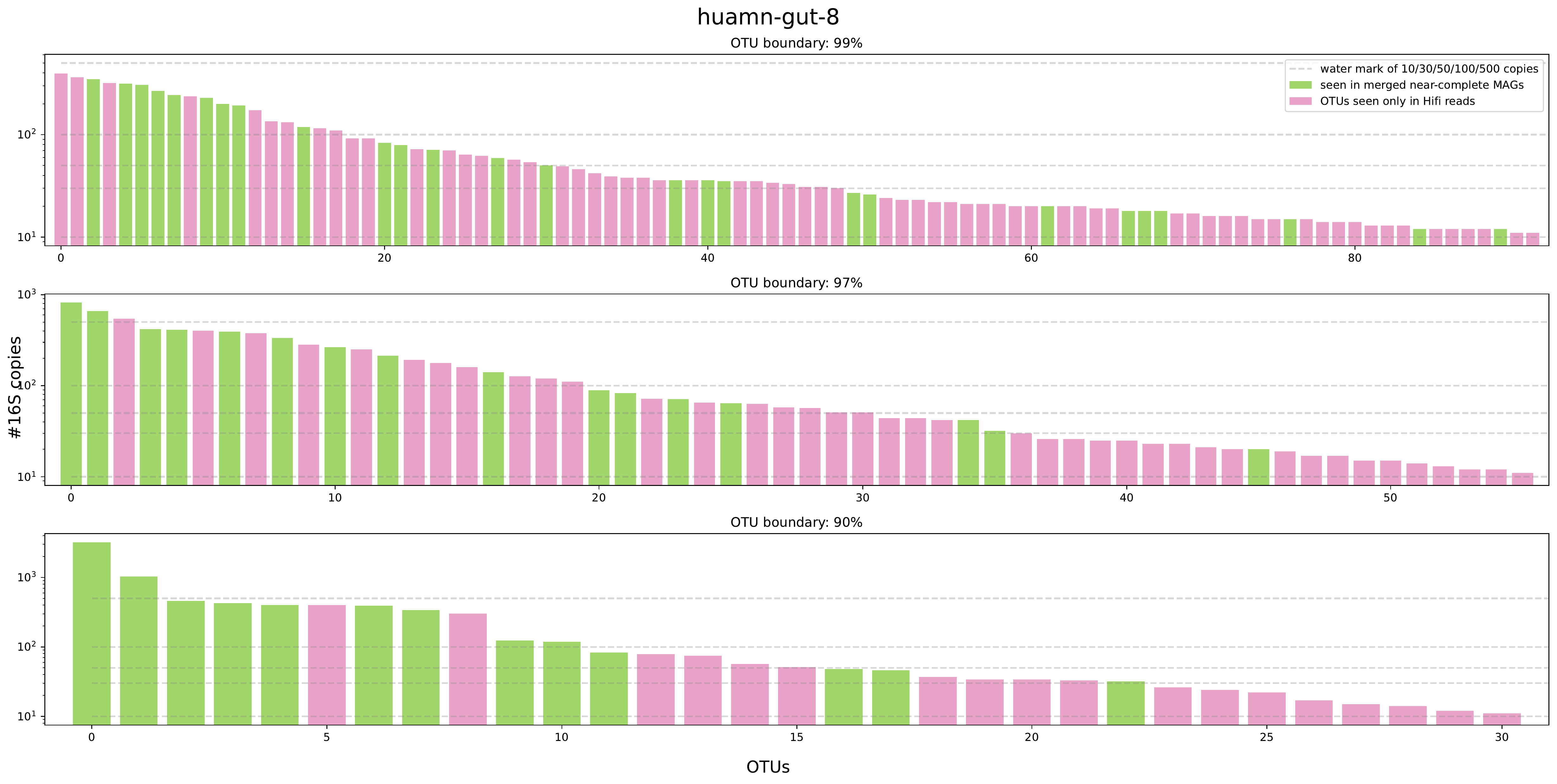}
\includegraphics[width=0.95\linewidth]{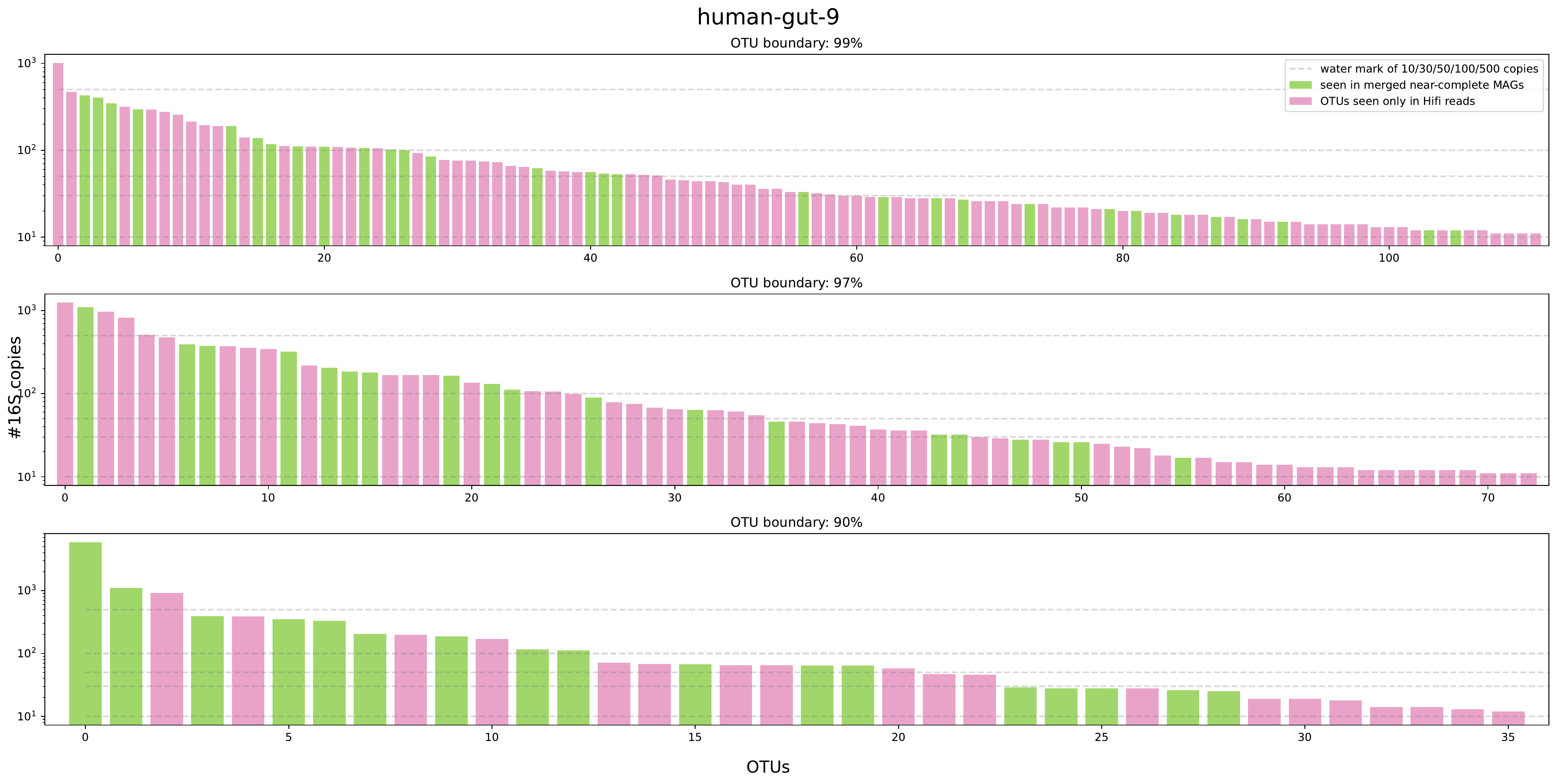}
\includegraphics[width=0.95\linewidth]{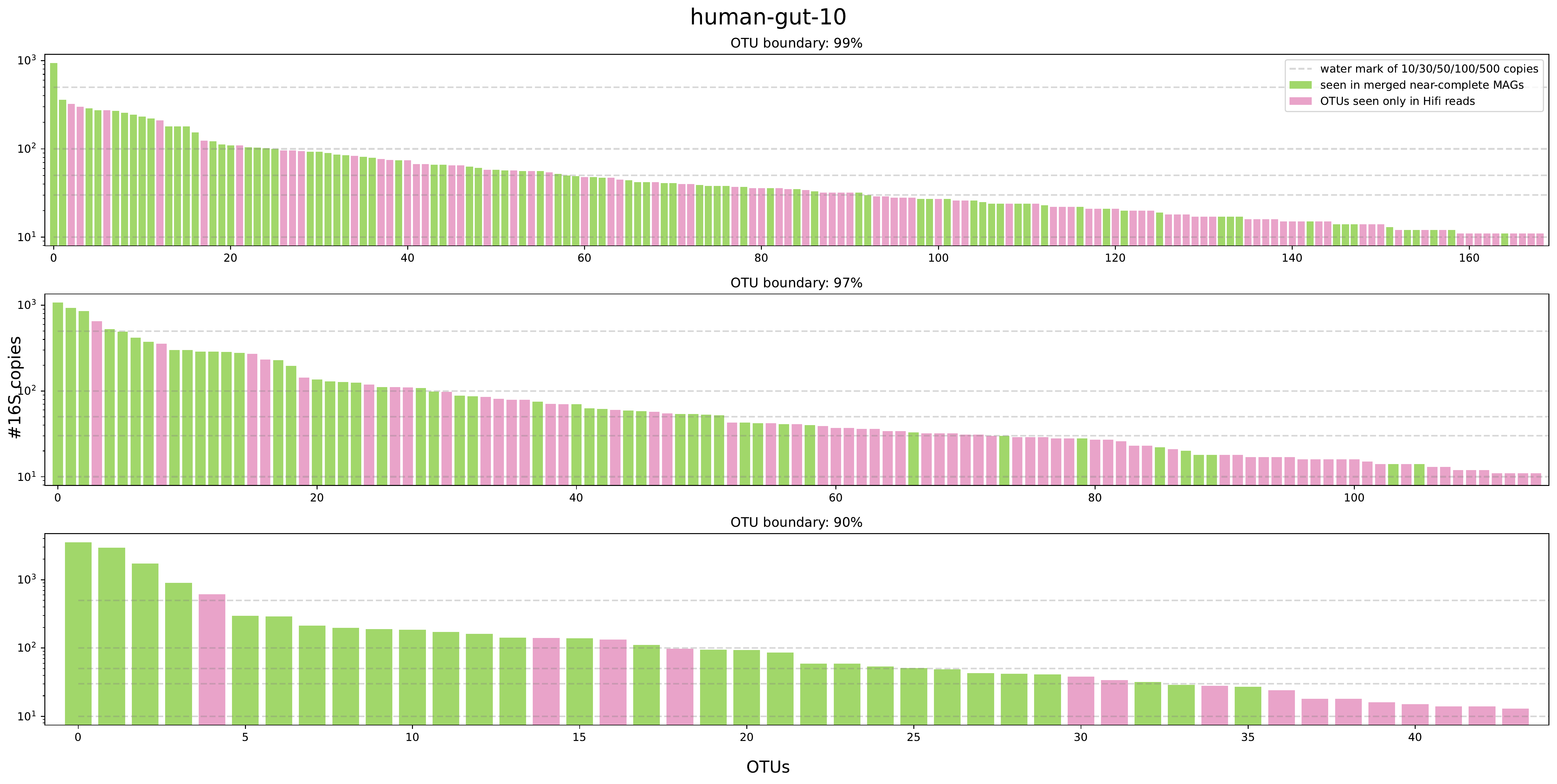}
  \caption{OTU recovery in samples, with OTU boundary set at 
99\%, 95\% and 90\%. The plots for human-gut-1 and human-gut-2
were omitted, because the each of these libraries are a pool
of four different samples. 
}
\end{figure}

%%%%%%%%%%%%%%%%%%%%%%%%%%%%%%%%%%%
%%                               %%
%% Tables                        %%
%%                               %%
%%%%%%%%%%%%%%%%%%%%%%%%%%%%%%%%%%%

%% Use of \listoftables is discouraged.
%%
\section*{Tables}

%%%%%%%%%%%%%%%%%%%%%%%%%%%%%%%%%%%
%%                               %%
%% Additional Files              %%
%%                               %%
%%%%%%%%%%%%%%%%%%%%%%%%%%%%%%%%%%%

\section*{Additional Files}
  \subsection*{Table S1. Supplementary tables and data release
used a different sample naming convention. This file provides
the name mapping.}
  
  \subsection*{Table S2. Binning information and evaluation of bins.}
A tab-delimited table.

  \subsection*{Table S3. The 3490 samples from Almeida et al. used
in this manuscript.}
List of files (sra ftq).

  \subsection*{Table S4. Sample and HiFi library information.}

  \subsection*{Table S5. List of tools their versions used in this manuscript.}

  \subsection*{Table S6. BLAST result summary of 20 high read- and contig- 
multiplicity k-mers intervals.}

\end{backmatter}
\end{document}